# Comprehensive Multiparty Session Types

Andi Bejleri[a,d], Elton Domnori[b], Malte Viering[a], Patrick Eugster[c], and Mira Mezini[a]

a TU Darmstadt, Darmstadt, Germany

b Canadian Institute of Technology, Tirana, Albania

c Universitá della Svizzera Italiana, Lugano, Switzerland

d IBM GBS, Frankfurt, Germany

**Abstract** *Multiparty session types* (MST) are a well-established type theory that describes the interactive structure of a fixed number of components from a global point of view and type-checks the components through projection of the global type onto the participants of the session. They guarantee communication-safety for a *language of multiparty sessions* (LMS), i.e., distributed, parallel components can exchange values without deadlocking and unexpected message types.

Several variants of MST and LMS have been proposed to study key features of distributed and parallel programming. We observe that the population of the considered variants follows from only one ancestor, i.e. the original LMS/MST, and there are overlapping traits between features of the considered variants and the original. These hamper evolution of session types and languages and their adoption in practice. This paper addresses the following question: *What are the essential features for MST and LMS, and how can these be modelled with simple constructs?* To the best of our knowledge, this is the first time this question has been addressed.

We performed a systematic analysis of the features and the constructs in MST, LMS, and the considered variants to identify the essential features. The variants are among the most influential (according to Google Scholar) and well-established systems that cover a wide set of areas in distributed, parallel programming. We used classical techniques of formal models such as BNF, structural congruence, small step operational semantics and typing judgments to build our language and type system. Lastly, the coherence of operational semantics and type system is proven by induction.

This paper proposes a set of essential features, a *language of structured interactions* and a type theory of *comprehensive multiparty session types*, including global types and type system. The analysis removes overlapping features and captures the shared traits, thereby introducing the essential features. The constructs of the language are simple and fundamental, based on the $\lambda$ and $\pi$ calculi. Analogously, our global types reflect what is omitted and introduced in the language. Our system covers all the features of the original and variants, with a better ratio of the number of language and type constructs over the number of covered features.

The features of the original, variants, and our system along with the number of constructs in the respective language and global types to model them are presented through a table. The syntax, operational semantics, meta-theory and type system of our system are given. We modelled all the motivating examples of the variants in our model, describing the reduction and typing steps. The work discusses how new features, in particular the non-essential ones (formerly excluded) and advanced ones can be either modelled atop the essential ones or added with minimal efforts, i.e. without modifying the existing ones. The fundamental properties of typed processes such as subject reduction, communication safety, and progress are established.

## 1 Introduction

*Multiparty session types* (MST) [22] provide a well-established type theory for statically *safe* and *deadlock free* interactions among a fixed number of processes. *Global types* specify the interaction structure between participants of a *session* from a global point of view. These types are defined by constructs modelling features of value exchange over shared channels, branching over labels, parallelism, recursion and inaction. The corresponding *language of multiparty sessions* (LMS) is a variant of Pict [38]– a programming language build over the asynchronous $\pi$-calculus [28], with queue that extends the features in the types with others that do not define an interaction structure: session initiation, conditional, and restriction.

Several variants of MST and LMS [5, 6, 7, 15, 32, 46] have been proposed to study key features of distributed programming. Bettini et al. [5] introduce value exchange over unshared channels to enable message-passing in distributed systems. Bhargavan et al. [6] introduce disjunction of prefixes to describe and verify cryptographic protocols. Mostrous et al. [32] introduce partial commutativity of interactions to describe and verify stream processing and multicore computing. Bocchi et al. [7] introduce asserted interactions to meet the needs of multi-organisational applications. Yoshida et al. [46] introduce primitive recursive sessions to describe and verify parallel computations. Chen et al. [15] introduce partial failures to meet the needs of distributed systems.

We observe[1] (1) that the population of the considered variants follows from only one ancestor, i.e., the original [22] and (2) overlapping features and redundancies in constructs between the considered variants and the original. This means that the common ancestor of these variants does not provide the shared traits of the overlapping features as a single one. Similarly, the constructs of the ancestor that model the features for both languages and types are complex and designed for a specific domain of distributed programming, i.e. web services. These hamper evolution of session types and languages and their adoption in practice.

To avoid the unraveling of efforts, we address the following questions:

*What are the essential features of MST and LMS (a), and how can these be modelled with simple constructs (b)?*

Part (a) of the question is addressed in two steps. First, we remove overlapping features introduced by the original system, and its variants respectively. Second, we add new features that capture the shared traits between the overlapping ones. Concretely, we propose features that define an interaction structure and features that do not. For the former we have: value exchange over unshared channels, disjunction, parallelism, recursion, assertions, functional abstraction, and inaction. For the latter we have: session initiation, restriction, and queue. Part (b) of the question is addressed by modelling the essential features using a concise set of well-known basic constructs from the $\pi$ and $\lambda$ calculi. Concretely, we propose the *Language of Structured Interactions*

[1] This observation came from a systematic analysis of the constructs in MST, LMS, and their respective variants under consideration.

(LoSI) and a consolidated type theory of *Comprehensive Multiparty Session Types* (CMST). As a result, LoSI and CMST improve on the solutions offered in literature by providing a better ratio of the number of language and type constructs (13, 9) over the number of covered features (16).

In particular, this paper makes the following contributions:

(I) We define the essential features of session programming and typing. By retaining only essential features, we provide a modular setting of study for session programming and typing. This means that new essential features can be added without modifying the existing ones (not as asserted interactions in [7]), and non-essential features can be layered atop (see Section 2).

(II) LoSI models the features by using the syntax of LMS and introducing new constructs from the $\pi$ and $\lambda$ calculi. By having simple, fundamental constructs, we ensure the modular nature of our framework and provide fewer operational semantics rules (see Section 3).

(III) Analogously, CMST's global types model the features by using the syntax of MST and introducing advanced types [2] in the same spirit as in the $\lambda$LF [20]. (see Section 4). All constructs are simple, resulting in less typing rules where reasoning for specific features is easier than in variants [15].

(IV) CMST checks the individual processes of a program by the end-point types, resulting from the global type projection onto participants of the session (see Sections 5 and 6). The verification of primitive recursive sessions (a.k.a. parameterised sessions, i.e., sessions of an arbitrary number of participants) presents a challenge due to the arithmetical expressions defining participants, i.e., the order of interactions specified in the global type may not hold in the end-point types returned from projection. To address this issue we sort the (input-output) prefixes on end-point types based on the participants that define the sender in an action, i.e., actions are placed according to the position of the sender on the topology of the primitive recursive session. We use arithmetic expressions as in Dependent ML (DML) [43, 44] to express the set of naturals typing the argument of a dependent product global type and to achieve decidability of the type system (when proving predicates over parameterised participants and assertions). The set of naturals range over infinite sets, in contrast to [46] that range over finite ones, not restricting the expressivity power, e.g., that of parallel algorithms (issue also addressed in [3, 4]). Our type system ensures types of messages exchanged: (1) well-assertedness of interactions, i.e., that assertions are defined over variables visible to the participants of the corresponding interaction, (2) well-defined applications of arithmetical expressions, i.e., that arithmetical expressions applied to a product global type conform to the type of its argument, (3) robustness, i.e., that participants affected by a failure occurrence are notified.

(V) We establish the fundamental properties of subject reduction, communication safety and progress for typed LoSI programs (see Section 7). Intuitively, subject reduction ensures that type annotations are well-defined; communication safety ensures that for some inputs (or outputs) in disjunction (summation) of a process, reciprocally, there are fewer outputs (or more inputs) of the respectively same sets of values in a parallel process; progress ensures that the reduction of a closed (no free variables), simple (one multiparty session), and well-typed process will never get stuck.

Finally, the paper shows how the removed, non-essential features can be layered atop the essential ones, discusses advanced features in our framework, surveys on related work and concludes with an outlook on future work. Omitted definitions, theorems, and proofs can be found in the appendix.

## 2 Analysis of the essential features

This section overviews the features of the original system and the considered variants along with their modelling and introduces the essential ones.

### 2.1 Features and modelling in the variants

*Session initiation* is introduced [22] to describe the hand-shaking of all processes of a session. It is modelled in LMS through two constructs $\bar{a}[2..n](\tilde{s}).P$ and $a[i](\tilde{s}).P_i$ where a denotes a shared channel, $\tilde{s}$ session channels, and $i$ process identity. The former construct sends an invitation to the latter to initiate a session. This feature does not define an interaction structure, therefore does not have a global type construct.

*Value exchange over unshared channels* are added [5] to provide a more efficient type system and to greatly facilitate the statement of properties and consequently their proofs since the linearity analysis of shared channels introduced in MST [22] is avoided. A channel is modelled as $s[p]$ where $s$ denotes the queue and $p$ a participant in a session. The output-input prefixes are $s[p]!\langle p', v\rangle$ and $s[p]?(p', x)$ with $p'$ denoting message receiver and sender respectively. In global types, an interaction is modelled as $p \to p' : \langle S\rangle$, denoting a message exchange of type $S$ between $p$ and $p'$.

*Parallelism, Inaction, Restriction* and *Recursion* are introduced [22] to model parallel composition of processes, end of interactions, scoping of channels and repetitive behaviour. LMS models them as in $\pi$-calculus, with the exception of recursion modelled as $\mathsf{def}\ \{X_i(\tilde{x_i}, tildes_i)\}_{i\in I}\ \mathsf{in}\ P$. In global types, they are modelled respectively as $G_1, G_2$, $\mathsf{end}$, $\mu X.G$; restriction is a feature that does not defines an interaction structure.

*Queue* is introduced [22] to describe messages in transmission in the medium of communication, i.e., to model asynchronicity. It is modelled in LMS as $s : \tilde{h}$ where $h$ denotes messages. This feature does not define an interaction structure.

*Branching labels* is introduced [22] to describe branching of processes. LMS models it as $\{l_i : P_i\}$ where $l_i$ is a label associated to a process branch. In global types, a label branch is modelled as $\{l_i : G_i\}$.

*Conditional* is introduced [22] and has a standard model in LMS. It is a feature that does not define an interaction structure.

*Primitive recursion* is introduced [46] to model and verify sessions of an arbitrary number of participants. Primitive recursion is modelled by the $\mathsf{R}$ operator ($\mathsf{R}\ P\ \lambda\mathtt{i}.\lambda X.Q$) from Gödel's theory $T$ [1] of primitive recursive functions. Symmetrically in global types, families of global types are expressed as $\mathsf{R}\ G\ \lambda i.\lambda X.G'$ types that can reduce to base types given a natural for parameter $i$. Parameters range over finite sets of naturals, restricting the computing power of programs.

*Asserted interactions* are added [7] to express and verify constraints on values exchanged between two participants. The output-input and branch-selection prefixes of LMS are extended with predicates of first-order logic, including equalities, and binders of variables, modelled as: $s!\langle\tilde{m}\rangle(\tilde{x})\{A\}.P$ denotes the sending of a message $m$ through channel $s$ guarded by predicate $A$ defined over variables $\tilde{x}$. In global types, interactions are specified as $p \to p' : \langle s(\tilde{x} : \tilde{U})\{A\}\rangle$, where $s$ denotes the channel, $\tilde{x}$ the interaction variables of type $U$ present in predicates $A$ which constrain the content $\tilde{x}$ exchanged between $p$ and $p'$. Assertions are studied in a non-modular setting. That is, if the system is extended with new features, say for the dynamic joining and leaving of a participant, then those features need to incorporate assertions.

*Disjunction of prefixes* is added [6] to model cryptographic protocols. At every prefix, every participant can express an internal or external choice; e.g., `send(Request(c, w, q); recv[Reply(x) | Fault])` and `recv[Request(c,w,q) -> send(Reply(x) + Fault)]`, where `Request` is a label, `c` and `w` participants, and `q` value. A global type is formally defined as a directed graph with a distinguished initial node, where each node identifies a participant (`c`, `w`) and each edge identifies a unique label (`Request`, `Reply`, `Fault`).

*Partial failures* are added [15] to express and verify failures on components or interactions while others continue respecting certain invariants. Failure is modelled as a label that is raised while performing a send or receive action as $c?/!(p,e)^F$ (send/receive or raise a failure $F$) and handled through a $\mathsf{try}\{P\}\mathsf{catch}(F : P')$ construct, i.e. $P$ is executed till handler $F : P$ is triggered. Symmetrically in global types, interactions that can raise a failure $f$ are expressed as $p \to p' : S \vee F$ and handled as $\mathsf{try}\{G\}\mathsf{catch}(F : G')$. The handle semantics is similar to that of a `go-to` statement, i.e. if a failure is triggered by one of the interactions in $P$ then the program flow shifts to $P'$. This makes reasoning about deadlock-freedom complex, i.e., as supported by seven dedicated operational semantics rules, six dedicated typing rules and a cumbersome meta-theory along with proofs of properties.

### 2.2 Essential features

Initially, we remove branching over labels [7, 22, 46], conditional features [6, 7, 15, 22, 46], and disjunction of prefixes [5] where the first is disjunction of processes guarded by labels, the second is disjunction of processes guarded by boolean expressions, and the third is disjunction over prefixes. Partial failure [15] is also a special instance of branching, i.e., disjunction over processes, prefixed by a *send* action in case of failure occurrence or a *receive* action in case of handling a failure. The same goes for primitive recursion [46], a special instance of recursion, i.e., finite iteration. We remove also asserted interactions [7]—a value exchange guarded by a boolean expression.

Next, we add three new features: disjunction, functional abstraction, and assertions. The first can express disjunction of processes where a process is independently chosen over others by sending a value along with its type (*internal choice*), and the dual process is chosen over others according to the type of the message received (*external choice*). The second feature can express reduction of arithmetical expression along with the reduction of processes, e.g. to express finite iterations. The third feature can express constraints of boolean expressions over processes.

**Table 1** Features of the original, variants, and our system. For each feature and system, the pair of numerals $[x, y]$ denotes respectively the number of constructs in the respective language and global types. † denotes that exchange of session channels is not considered. Non-essential features (highlighted in grey) can be layered atop of essential ones, labeled by letters (e.g., combining $a + b$).

| *Systems* / *Features* | Honda et al. [22] (Original) | Bettini et al. [5] Distrib. Systems | Bhargavan et al. [6] Cryptogr. Protocols | Mostrous et al. [32] Stream Process. | Bocchi et al. [7] Multiorg. Apps | Yoshida et al. [46] Parallel Comput. | Chen et al. [15] Distrib. Systems | LoSI/CMST |
|---|---|---|---|---|---|---|---|---|
| Session initiation | [2, 0] | [2, 0] | | [2, 0] | [2, 0] | [2, 0] | [1, 0] | [1, 0] |
| Value exchange[a] | [4, 1] | [4, 1] | [2, 1]† | [2, 1]† | [4, 1] | [2, 1]† | [2, 1] | [2, 1] |
| Parallelism | [1, 1] | [1, 0] | [1, 0] | [1, 1] | [1, 1] | [1, 0] | [1, 0] | [1, 1] |
| Inaction | [1, 1] | [1, 1] | | [1, 1] | [1, 1] | [1, 1] | [1, 1] | [1, 1] |
| Restriction | [1, 0] | [1, 0] | | [1, 0] | [1, 0] | [1, 0] | [1, 0] | [1, 0] |
| Recursion[b] | [2, 2] | [2, 2] | [2, 2] | [2, 2] | [2, 2] | [2, 2] | [2, 2] | [2, 2] |
| Queue | [1, 0] | [1, 0] | | [1, 0] | [1, 0] | [1, 0] | [1, 0] | [1, 0] |
| Branching labels[c] | [2, 1] | [2, 1] | (a+d) | [2, 1] | [2, 1] | [2, 1] | | (a+f) |
| Conditional | [1, 0] | [1, 0] | [1, 0] | [1, 0] | [1, 0] | [1, 0] | | (e+f) |
| Primitive recur. | | | | | | [2, 2] | | (b+e+f+g) |
| Asserted interac. | | | | | (a+b+c) | | | (a+b+e) |
| Disjunction pref.[d] | | | [1, 1] | | | | | (a+f) |
| Partial Failure | | | | | | | [3, 1]+a | (a+f) |
| Assertions[e] | | | | | | | | [1, 1] |
| Disjunction proc.[f] | | | | | | | | [1, 1] |
| Function abstr.[g] | | | | | | | | [2, 2] |
| **Total [16]** | **[15, 6]** | **[15, 5]** | **[7, 4]** | **[13, 6]** | **[15, 6]** | **[15, 7]** | **[12, 5]** | **[13, 9]** |

Table 1 summarises the features of the original system, its variants, and our system, along with the number of constructs used to model and verify a certain program feature. The features of the original system and variants are listed above the bold line. Those newly introduced in LoSI are shown below the bold line. LoSI models all the features shown in white background. Constructs between features of one system are disjoint. This is valid also between features among different systems (variants and our system), e.g., constructs used for primitive recursion and partial failure are disjoint. The table provides evidence that our approach contributes significantly to addressing points (a) and (b) of the question in the introduction, e.g., it covers all the 16 features of the original and main variants, with a better ratio of the number of language and type constructs over the number of covered features.

## 3 The LoSI: Language of Structured Interactions

Our language is based on small-step operational semantics, defined over inference rules and a relation of structural congruence. Examples from cryptographic protocols (Web service), multi-organisational applications (Financial protocol), parallel algorithms (Ring), and networking (Network) illustrate the expressive power of the language.

| $P ::=$ | Processes: | $\mid X$ | Variables | $\mid fn\ x{:}I \Rightarrow P$ | Abstraction |
|---|---|---|---|---|---|
| $\mid\ init(u{:}G,p).P$ | Initiation | $\mid 0$ | Inaction | $\mid P\ e$ | Application |
| $\mid\ u[p,q]!\langle m{:}S\rangle.P$ | Output | $\mid P \mid P$ | Parallel | $\mid (new\ a{:}G)P$ | Restriction |
| $\mid\ u[p,q]?(w{:}S).P$ | Input | $\mid P + P$ | Summation | $\mid a{:}h$ | Queues |
| $\mid\ rec\ X = P$ | Recursion | $\mid [b]P$ | Matching | | |

**Figure 1** Syntax of processes.

### 3.1 Syntax

Figure 1 provides the syntax of processes. The metavariables $P, Q$ stand for processes; $u$ for session identifiers $a$ and variables $x$; $w$ for variables $x$ and (session) channels $a[\hat{p}]$; $X$ ranges over process variables; $m$ over message expressions; $e$ over arithmetic expressions; $b$ over boolean expressions; $p, q$ over roles; $S$ over message types; $G$ over global types; $I$ over index sorts. The metavariable with subscripts and suffixes stands for the same class of terms.

Processes define a scope that includes the subsequent behaviours. The process $init(u{:}G,p).P$ represents the behaviour $P$ of participant $p$ in session $u$ typed as $G$ (locality). The session identifier $u$ serves as a public site for all participants of a session that perform the actions defined in $P$. Each process has a notion of locality in the site defined as $(u : G, p)$. Locality means that processes can be associated with a single site. The output-input behaviour in the session initiation constructs of MST [22] is unobservable; we delete the corresponding syntax, resulting in a silent initiation action similar to the $\tau$ action in the $\pi$-calculus. In our output and input prefixes $u[p,q]!\langle m : S\rangle.P$ and $u[p,q]?(w : S).P$, channel $u[p,q]$ denotes the sender $p$ and receiver $q$ of a message $m$ of type $S$ over site (session) $u$. Communications are defined over pairwise bidirectional channels to capture real-life programming abstractions. When $m$ and $w$ denote a session channel $a[p]$, the prefixes model session channel exchange, also known as session delegation. In static semantics, $a[p]$ abstracts all the actions that participant $p$ has yet to perform in site $a$ at the point of channel exchange. Our channels $u[p,q]$ are pairwise, bidirectional, and their restriction to a session is guaranteed by the $new$ operator at runtime.

The construct $rec\ X = P$ models recursion for the language as in ML [29]. Inact $0$ describes the end of a behaviour. Parallel composition is standard for composing the behaviour of two processes in parallel. Summation and matching from the $\pi$-calculus are introduced to model respectively disjunction and assertions. Summation is standard for composing alternative behaviours like $P$ or $Q$, i.e., if one process exerts a prefix, the other is rendered void. Matching $[b]P$ abstracts constraints $b$ over processes $P$. $fn\ x{:}I \Rightarrow P$ abstracts naturals of type $I$ over processes $P$ (introduced from the $\lambda$-calculus to model functional behaviour). Application $P\ e$ applies a process $P$ to an arithmetic expression $e$. Restriction *new* and message queues $a : h$ are part of the runtime syntax. The former restricts the actions' scope defined over site $a$ only to $P$ and the latter can be visualised as the network in distributed systems in the real world.

| | | | |
|---|---|---|---|
| $u ::= a \mid x$ | Identifiers | $N ::=$ Alice $\mid$ Worker $\mid \ldots$ | Participants |
| $w ::= a[\hat{p}] \mid x$ | Placeholders | $h ::= \epsilon \mid (\hat{q}, \hat{p}, v : S) \cdot h$ | Queues |
| $p ::= p[e] \mid N$ | Roles | $I ::= \{x{:}I \mid \Theta\} \mid$ nat | Index sorts |
| $\hat{p} ::= \hat{p}[c] \mid N$ | Value roles | $\Theta ::= e \leq e' \mid \Theta$ and $\Theta$ | Predicates |
| $m ::= e \mid b \mid a \mid a[\hat{p}] \mid \ldots$ | | | Messages expression |
| $e ::= x \mid c \mid e + e \mid c \times e$ | | | Arithmetical expression |
| $b ::=$ true $\mid$ false $\mid$ $b$ and $b$ $\mid$ $b$ or $b$ $\mid$ not $b$ $\mid$ $e$<$e$ $\mid$ $e$=$e$ | | | Boolean expressions |
| $v ::= a \mid a[\hat{p}] \mid c \mid$ true $\mid$ false $\mid \ldots$ | | | Values |
| $S ::= U \mid T \quad U ::=$ bool $\mid$ nat $\mid G \mid I \mid \ldots$ | | | Messages types |

**Figure 2** Auxiliary syntax.

Figure 2 shows the syntax of constructs that are part of the processes definition. Identifiers are names and variables. Placeholders include variables and session channels. Roles $p$ include parameterised participants defined over one or multiple arithmetic expressions (e.g., W[$e$] and W[$e$][$e'$]), and primitive participants $N$ such as Alice and Bob. Value roles $\hat{p}$ include non-parameterised participants, i.e., participants indexed by natural numbers (W[3], W[2][4], ...) or primitive participants. Messages $m$ include arithmetic and boolean expressions, sites, session channels, and other data. Arithmetic expressions $e$ are the same as index terms in DML [2, 44]: integers c, variables $x$, and complex expressions using addition and multiplication. Boolean expressions $b$ are standard: values true and false, complex expressions using boolean operators (and, or, not), and equalities and inequalities between arithmetic expressions. Queues $h$ are lists of messages which in turn are defined as triples: sender, receiver, and payload value along with its type (e.g, IP packet). Messages are run-time entities therefore they are defined over value roles. Values are sites, session channels, integers, boolean true and false. Message types $S$ include primitive types (bool, nat, and other data types), global types $G$, index sorts $I$, and end-point types $T$ to describe the exchange of session channels. The exchange of session $a$ is typed by $G$. The index sort $I$ is a set of naturals or a subset of index sorts. Subsets are formed by base predicates as inequalities between arithmetic expressions and complex predicates by conjunction. This is inspired by the definition of index sorts in DML[2].

[2] DML uses index sorts to ensure a correct use of API, e.g., "remove" and "delete" methods in arrays, lists, and stacks, through a dependent type system. Its constraint solver [45] uses Fourier methods [39] to solve index equations.

$$P \mid 0 \equiv P \qquad P \mid Q \equiv Q \mid P \qquad (P \mid Q) \mid R \equiv P \mid (Q \mid R)$$
$$P{+}0 \equiv P \qquad P{+}Q \equiv Q{+}P \qquad (P{+}Q){+}R \equiv P{+}(Q{+}R)$$
$$(\mathit{new}\ a{:}G)P \mid Q \equiv (\mathit{new}\ a{:}G)(P \mid Q) \quad \text{if } a \notin \mathsf{fn}(Q) \qquad (\mathit{new}\ a{:}G)0 \equiv 0$$
$$(\mathit{new}\ a{:}G)\ (\mathit{new}\ a'{:}G')P \equiv (\mathit{new}\ a'{:}G')\ (\mathit{new}\ a{:}G)P$$
$$\mathit{rec}\ X = P \equiv P[\mathit{rec}\ X = P/X] \qquad P\ e \equiv P'\ e \text{ if } P \equiv P'$$
$$a : (\hat{q}, \hat{p}, v{:}S) \cdot (\hat{q}', \hat{p}', v'{:}S') \cdot h \equiv a : (\hat{q}', \hat{p}', v'{:}S') \cdot (\hat{q}, \hat{p}, v{:}S) \cdot h \quad \text{if } \hat{p} \neq \hat{p}' \text{ or } \hat{q} \neq \hat{q}'$$

**Figure 3** Structural congruence.

$$\mathit{init}(a{:}G, \hat{p}_1).P_1 \mid \ldots \mid \mathit{init}(a : G, \hat{p}_n).P_n \longrightarrow (\mathit{new}\ a{:}G)(P_1 \mid \ldots \mid P_n \mid a{:}\emptyset)$$
$$\{\hat{p}_1, \ldots, \hat{p}_n\} = \mathsf{pid}(G) \qquad [\mathsf{Link}]$$
$$a[\hat{p}, \hat{q}]!\langle v : U\rangle.P{+}P' \mid a : h \longrightarrow P \mid a : h \cdot (\hat{p}, \hat{q}, v : U) \qquad [\mathsf{Send}]$$
$$a[\hat{p}, \hat{q}]?(x : U).P{+}P' \mid a : (\hat{q}, \hat{p}, v : U) \cdot h \longrightarrow P\{v/x\} \mid a : h \qquad [\mathsf{Recv}]$$
$$a[\hat{p}, \hat{q}]!\langle b[\hat{p}'] : T\rangle.P \mid a : h \longrightarrow P \mid a : h \cdot (\hat{p}, \hat{q}, b[\hat{p}'] : T) \qquad [\mathsf{SSend}]$$
$$a[\hat{p}, \hat{q}]?(b[\hat{p}'] : T).P \mid a : (\hat{q}, \hat{p}, b[\hat{p}'] : T) \cdot h \longrightarrow P \mid a : h \qquad [\mathsf{SRecv}]$$
$$[\mathsf{true}]P{+}P' \longrightarrow P \quad [\mathsf{MatchT}] \qquad [\mathsf{false}]P{+}P' \longrightarrow P' \quad [\mathsf{MatchF}]$$
$$P \longrightarrow P' \quad \Rightarrow \quad (\mathit{new}\ a{:}G)P \longrightarrow (\mathit{new}\ a{:}G)P' \qquad [\mathsf{Scop}]$$
$$P \longrightarrow P' \quad \Rightarrow \quad P \mid Q \longrightarrow P' \mid Q \quad [\mathsf{Par}] \qquad (\mathit{fn}\ x{:}I \Rightarrow P)\ \mathsf{c} \longrightarrow P\{\mathsf{c}/x\} \quad [\mathsf{App}]$$
$$P \equiv P' \text{ and } P' \longrightarrow Q' \text{ and } Q \equiv Q' \quad \Rightarrow \quad P \longrightarrow Q \qquad [\mathsf{Str}]$$

**Figure 4** Operational semantics.

Figure 3 defines all processes that have the same behaviour but different syntax. The first axiom defines 0 as the identity element of all processes with respect to parallel composition. The second and third axioms specify that parallel composition of processes is commutative and associative. The sum is analogous to the parallel composition (the other three axioms). The following three axioms express the scope of restriction and how it can move according to the term. The next axiom provides equi-recursion between a recursive function and its one-time unfolding. The subsequent rule defines congruence on the process in the application construct. The last rule defines commutativity of messages in the queue for different participants.

### 3.2 Operational semantics

Figure 4 shows the operational semantics described via the reduction relation $\longrightarrow$, written $P \longrightarrow P'$ and read "*process* $P$ *reduces to process* $P'$ *in one step*". A session is established among processes via public sites. Interactions are defined over value roles. Rule [Link] creates a session between $n$ different peers (handshake) by restricting

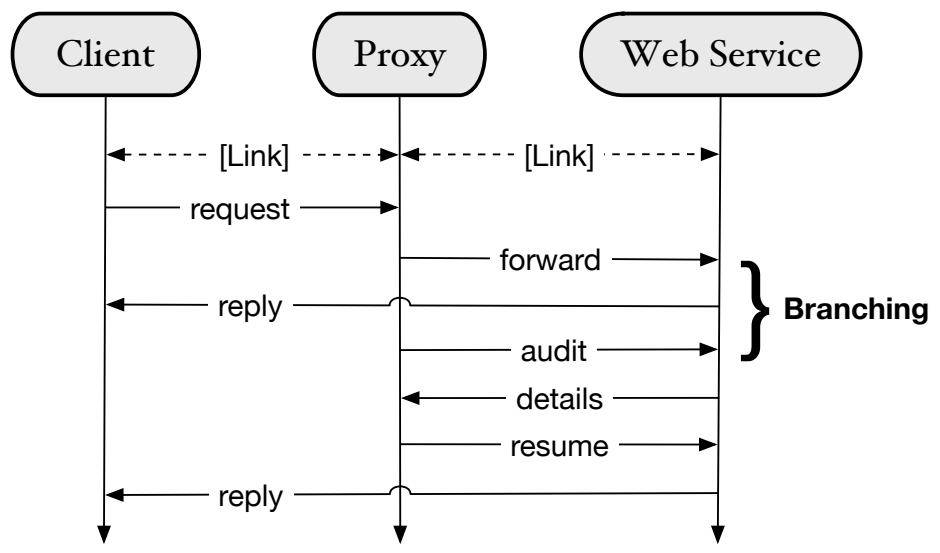

**(a)** Web service

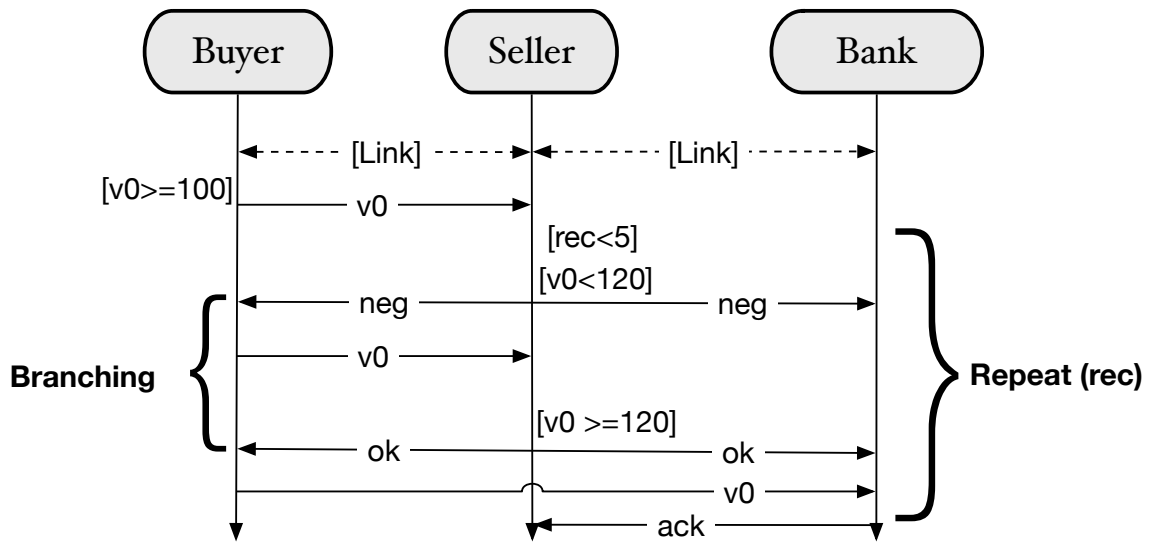

**(b)** Financial protocol

**Figure 5** Web service (a) and Financial protocol (b) diagram.

the scope of the channels on the site (public component), and generating a message queue. The side condition ensures that the participants of a session meet those in $G$; $\mathtt{pid}(G)$ denotes the set of participants occurring in $G$.

Rule [Send] inserts a message in the queue of the session by exerting one of its branches and rendering the others void. The result is the next term $P$ from the sender process and a modified queue with the message appended at the end. The receiving rule [Recv] removes a value message from the queue of the same sender, receiver and message type, as the one specified in the receiving construct by exerting one of the branches. The result is the next term of $P$ with value $v$ substituted in the scope and the modified queue $h$ without the message. Rules [SSend] and [SRecv] model the capability of participant $\hat{p}'$ to delegate its participation in session $b$ to another participant. Similar to the original semantics of session channels [22], the reduction rules for this feature are defined over bound outputs, i.e., the receiver does not receive the session channel $b[\hat{p}']$ during the exchange operation, but has it *a priori*. For simplicity, we do not consider the branch of interactions for session channels. Rules [MatchT] and [MatchF] exert the branch prefixed by a boolean expression that resolves true or false, rendering the others void. Rule [Scop] activates the reduction of the process inside the scope of the *new* operator. Rule [Par] defines computation in processes composed in parallel, first reducing the subprocesses. Rule [App] replaces argument $x$ in $P$ with integer c. Rule [Str] states that the reduction relation is defined on structurally congruent terms.

## 3.3 Running examples

We demonstrate the expressivity of LoSI by modelling the running examples of the considered variants [6, 7, 15, 46]. We do not model the examples of variants [5, 32], since through the other examples we illustrate modelling of value exchange on unshared channels.The global types of Web service, Financial protocol, Ring, and Network are denoted by respectively $G_W$, $G_F$, $G_R$ and $G_N$.

**Web service.** This example describes a session between a client, a proxy, and a web service [6] (see Figure 5a). The session starts with the client sending a *request* (*Req*) to the proxy. Next, the proxy chooses to transmit either a *forward* (*Fwd*) or an *audit* (*Aud*) message to the web service, which respectively *replies* to the client and sends *details* (*Dtls*) to the proxy, which in turn subsequently sends a *resume* (*Res*) message to the service. Finally, the web service *replies* to the client. In LoSI, the alternative sequences of the proxy, shown below, to send a *forward* or *audit* message to the web service can be modelled as a sum; similarly, the sequences prefixed by an input in the web service can be modelled as a sum.

$$\text{Proxy} \triangleq \mathit{init}(a{:}G_W, 2).a[1,2]?(x{:}Req).(a[2,3]!\langle f{:}Fwd\rangle.0 + a[2,3]!\langle a{:}Aud\rangle.a[3,2]?(x{:}Dtls).a[2,3]!\langle r{:}Res\rangle.0)$$

**Financial protocol.** An electronic commerce session between a buyer, a seller, and a bank [7] starts with the buyer sending an offer $v_0$ (which must be greater than 100) to the seller (see Figure 5b). The seller chooses to either negotiate the offer, here set to a limit of less than five times,or to accept it if the offer is greater than a base price set to 120. For coherence of the conversation, the seller notifies the bank at every stage of the negotiation. In the accept case, the buyer tells the bank to make a payment of the value negotiated. Then the bank acknowledges the payment. In LoSI, the condition that the initial offer has to be greater than 100 can be modelled by annotating the value in the Buyer with type $I = \{v' : \mathsf{nat} | v' \geq 100\}$ (ensured by the type system). The condition on the number of negotiations is defined by the matching construct, shared among all the participants. The other condition on the offer is expressed through matching for the seller and buyer (participants to whom the offer is visible). Next we give the definition of Buyer.

$$\text{Buyer} \triangleq \mathit{init}(a : G_F, 1).a[1,2]!\langle v_0 : I\rangle.(\mathit{rec}\ X = \mathit{fn}\ iter : \mathsf{nat} \Rightarrow [iter < 5] \ ([v_0 < 120]a[2,1]?(x : Neg).a[1,2]!\langle v_0 : I\rangle.X\ iter{+}1 + [\mathsf{not}\ v_0 < 120]a[2,1]?(x : Ok).a[1,3]!\langle v_0 : I\rangle.0))\ 1$$

The recursion variable $iter$ is initially set to 1 and it is increased by one on each iteration in the definition of each participant.

**Ring.** The ring describes communications of $n$ participants, where each participant receives a value from the neighbour on its left and then sends a value to the neighbour to its right [46] (see Figure 6a). To ensure the Ring structure, the number of participants should be $n{\geq}2$. It has three distinct processes: *Starter*, represented by $M[1]$, *Middle*, represented by $M[i]$, and *Last*, represented by $M[n]$. *Starter* and *Last* are parameterised by $n$ and *Middle* by $i$. The $\mathit{fn}, \mathit{rec}$ constructs are used to model iteration (primitive recursion) and matching models the recursive and base cases of a conditional. Below, we provide the process of Middle(i) and main program:

$$\text{Middle}(i) \triangleq \mathit{init}(a{:}G_R, M[i]).a[M[i{-}1], M[i]]?(z{:}U).a[M[i], M[i{+}1]]!\langle z{:}U\rangle.0$$
$$\text{Ring} \triangleq \mathit{fn}\ n : I \Rightarrow ((\mathit{rec}\ X = \mathit{fn}\ i : J \Rightarrow [i{=}n]\ (\text{Starter} \mid \text{Last}) + [i < n]\ (\text{Middle}(i) \mid X\ i{+}1))\ 2)$$

where $I{=}\{n' : \mathsf{nat} | n' \geq 2\}$ and $J = \{i' : \mathsf{nat} | i' \geq 2 \text{ and } i' \leq n\}$. *Middle* is composed in parallel with the process application $X\ i+1$. At each iteration, $X$ is substituted with *Middle*

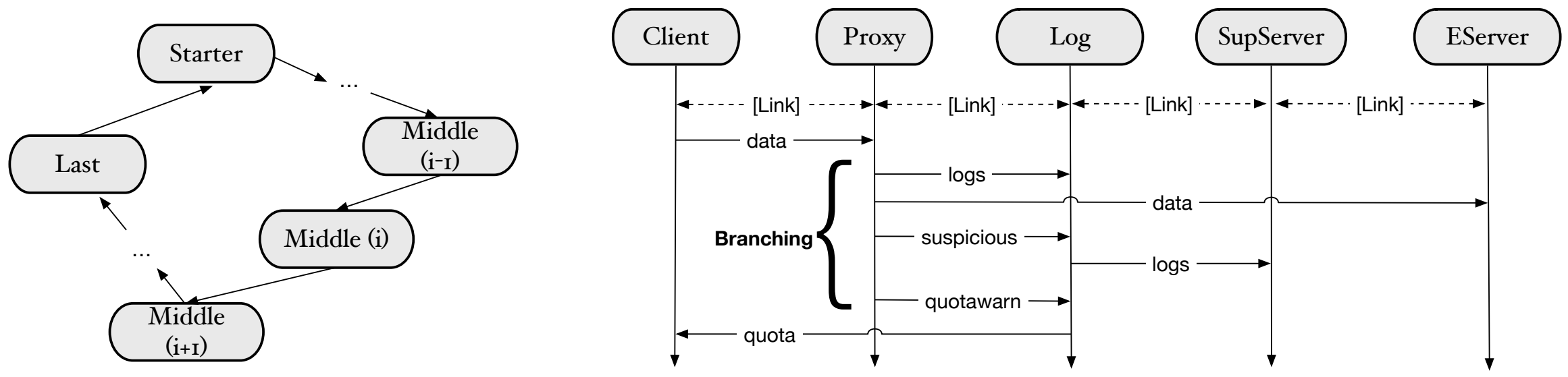

**(a)** Ring **(b)** Network

**Figure 6** Ring (a) and Network (b) diagram.

where parameter $i$ is substituted with 2 or a successor; in the last iteration for $i{=}n$, $X$ will be replaced with processes *Starter* and *Last*.

**Network.** In a network, traffic passes through a proxy that monitors the traffic and logs general information, e.g., consumed bandwidth [15] (see Figure 6b). The proxy sends this information to a log server and forwards the traffic from the client to an external server. If the proxy detects suspicious behaviour in the traffic, it raises a *Suspicious* failure and notifies the log server to handle the failure. In this case, the log server sends the logs to supervision server. Another scenario of failure is when the proxy detects that the quota of the client is low; in this case, the proxy raises a *QuotaWarn* failure and notifies the log server to handle the failure, subsequently the log sends the quota information to the client.

In LoSI, the raise of *Suspicious* and *QuotaWarn* failures by the proxy can be modelled as a branch prefixed by a send action. Symmetrically, handling of those failures by the log server can be modelled as a distinct branch prefixed by a receive action. Below, we provide the user defined process of the proxy server.

$$\begin{aligned}\mathrm{Proxy} \triangleq init(a{:}G_N, 1).a[0,1]?(x : Data).(&a[1,2]!(a : Log).a[1,4]!(x : Data).0\\ &+a[1,2]!(f : Suspicious).0\\ &+a[1,2]!(f : QuotaWarn).0)\end{aligned}$$

We can observe that the user-defined process does not express the notification of failures, e.g., suspicious, to affected participants, e.g., the external server. The system will enter such information in the run-time process similarly as in variant [15].

## 4 Global Types

This section provides the syntax of global types, including the definition of global types for the examples discussed in Section 3.3.

| | | | | | | | | |
|---|---|---|---|---|---|---|---|---|
| $G ::= p \to p' : \langle V \rangle.G$ | Interaction | $\mid$ end | End | $\mid [b]G$ | Matching |
| $\mid \; \mu X.G$ | Recursive | $\mid G, G$ | Parallel | $\mid \Pi x{:}I.G$ | Product |
| $\mid \; X$ | Variables | $\mid G + G$ | Summation | $\mid G\ e$ | Application |

$V ::= S \mid x{:}I$ Sorts

**Figure 7** Syntax of global types.

### 4.1 Syntax

Figure 7 shows the syntax of global types. Type $p \to p' : \langle V \rangle.G$ captures the sending of a message of sort $V$ from sender $p$ to receiver $p'$. We assume that in each interaction between $p$ and $p'$, we have $p \neq p'$. Sorts $V$ include message types $S$ and index-sort $I$ assignments to variables $x$ that describe assertions in the boolean expressions of matching, called interaction variables — visible only between $p$ and $p'$.

In the following, we give the definitions of a well-defined global type. The formal statement is given in the definition of projection.

**Definition 1** (Well-sorted). *A global type is* well-sorted *if the index sorts $I = \{i : I' | \Theta\}$ present in it define non-empty sets of naturals.*

Infinite behaviour is represented by recursively defined global types $\mu X.G$, where $\mu X.G \equiv G[\mu X.G/X]$. Type end signifies the end of a conversation and is used as a base type to build more complex global types. Type $G, G'$ represents two parallel sessions. Type $G{+}G'$ captures the branching of conversations, where $G{+}G' \equiv G'{+}G$ and $G{+}\mathsf{end} \equiv G$. Type $[b]G$ captures a boolean expression $b$ guarding conversation $G$. Such an expression in a matching construct must be consistent with the logic of participants, i.e., variables in $b$ must be visible to participants in G. Variables visible to every participant are called *global*. The definition of a well-asserted global type is given below:

**Definition 2** (Well-asserted). *A global type $G$ is* well-asserted *if variables of boolean constraints are visible to participants guarding those interactions.*

We relegate a more sophisticated analysis of assertions for future work as it is beyond the scope of this work.

The dependent product type $\Pi x{:}I.G$ defines a type-valued global type, mapping variables $x{:}I$ to global types $G$. We assume that in a global type $\Pi x : I.G$ we have $x \notin \mathsf{fv}(I)$. If the argument $(x)$ is a parameter of participants, then the product global type denotes a family of global types (parameterised sessions, e.g., Ring), otherwise it denotes an arrow type from global variables $x$ to $G$ as in the typed $\lambda$-calculus (e.g., Financial protocol). Application type $G\ e$ captures application of a product global type $(G)$ to an arithmetic expression $(e)$, which plays the role of an arrow type.

**Definition 3** (Well-dependent). *Global type $G$ is* well-dependent *if its product global types, denoting arrow type, are applied to arithmetical expressions which conform with the type of argument in that product global type.*

Participants affected by a failure or participating in the handling of a failure should be informed. In terms of our global types, such participants are present in the branches but not in the prefixes of each branch and define (1) an input or output dependency (as defined in MST [22]) with the prefix and (2) a different behaviour in more than one branch (similar to the compatibility property of branching [22]).

**Definition 4** (Robustness). *A global type $G$ is* robust *if participants of a branch construct that define a different behaviour in more than one branch are notified of the selection of a particular branch.*

Below we give a transformation definition that guarantees robustness. Firstly, we place interactions not related with the prefix branch before the branch to maintain asynchronicity, i.e. to not force all participants to be notified of a failure. Secondly, we add participants of branches in the recipients of the prefix for each branch in the form of a multicast interaction. For the sake of this definition, we assume to have multicast interaction in the syntax of global types as $p \rightarrow p_1, \ldots, p_n$ can be encoded as $p \rightarrow p_1.p \rightarrow p_2. \ \ldots \ .p \rightarrow p_n$.

**Definition 5** (Robustness Transformation). *Given a global type $G$, the transformation of $G$, written $\hat{t}(G)$, is defined:*
***First step:*** $\hat{t}(p \rightarrow p' : \langle V_i \rangle.G_1 + \ldots + p \rightarrow p' : \langle V_n \rangle.G_n) =$
$$q_i \rightarrow q_i' : \langle V_1 \rangle.(p \rightarrow p' : \langle V_1 \rangle.G_1' + \ldots + p \rightarrow p' : \langle V_n \rangle.G_n')$$
*where $q_i \rightarrow q_i' \in G_i$ and there is no input/output dependency from $p \rightarrow p'$ to $q_i \rightarrow q_i'$ in $G_i$.*
***Second step:*** $\hat{t}(p \rightarrow p' : \langle V_1 \rangle.G_1 + \ldots + p \rightarrow p' : \langle V_n \rangle.G_n) =$
$$p \rightarrow p', \tilde{p} : \langle V_1 \rangle.\hat{t}(G_1) + \ldots + p \rightarrow p', \tilde{p} : \langle V_n \rangle.\hat{t}(G_n)$$
*where $\tilde{p} = \bigcup_{i \in [1..n]} \mathsf{pid}(G_i) \setminus \{p, p'\}$. The other cases of global type are defined inductively on the subterms.*

We relegate a more sophisticated transformation definition for future work as it is beyond the scope of this work. The constructs showcase a declarative nature of global types, allowing our model to be extended with additional features.

### 4.2 Examples

**Web service.** The sequence of the proxy prefixed by sending $Fwd$ to the web service and the one prefixed by sending $Aud$ are composed together as a sum as demonstrated in the global type

$$G_W \triangleq C \rightarrow P : \langle Req \rangle.(P \rightarrow W : \langle Fwd \rangle.W \rightarrow C : \langle Rep \rangle.\mathsf{end}$$
$$+ P \rightarrow W : \langle Aud \rangle.W \rightarrow P : \langle Dtls \rangle.P \rightarrow W : \langle Res \rangle.W \rightarrow C : \langle Rep \rangle.\mathsf{end})$$

where $P$, $W$, and $C$ denote respectively the proxy, web service, and client.

**Financial protocol.** The global type of buyer $Bu$, seller $S$, and bank $Ba$ uses the product construct to bound variables that are visible to each participant to model global recursion. Those defined over variables exchanged are local to the sender and receiver, e.g., the one for checking the negotiation value between the buyer and seller, as demonstrated in the global type

$$G_F \triangleq Bu \to S:\langle v_0 : I\rangle.\mu X.\Pi iter : \mathsf{nat}.[iter < 5]$$
$$([v_0 < 120]\ S \to Bu, Ba:\langle Neg\rangle.Bu \to S:\langle v_0 : I\rangle.X\ iter + 1$$
$$+ [\mathsf{not}\ v_0 < 120]\ S \to Bu, Ba:\langle Ok\rangle.Bu \to Ba:\langle v_0 : I\rangle.Ba \to S:\langle Ack\rangle.\mathsf{end})\ 1$$

where $I = \{v' : \mathsf{nat} | v' \geq 100\}$, the guard $iter < 5$ applies to each participant since $iter$ is global, while $v_0 < 120$ applies only to seller and buyer since $v_0$ is visible only to them. We can observe that the global type is well-sorted, well-asserted and well-dependent.

**Ring.** The global type for the ring is straightforward, including (1) the parameterised interaction for $i$ and $i+1$ with $1 \leq i \leq n-1$ and $n \geq 2$, (2) the interaction from 1 to $n$, and (3) the family of ring instances parameterised by $i$, as demonstrated in

$$G_R \triangleq \Pi i : \{i' : \mathsf{nat} | i' < n \text{ and } i' \geq 1\}.M[i] \to M[i+1]:\langle U\rangle.M[n] \to M[1]:\langle U\rangle.\mathsf{end}$$

This means that e.g., $\Pi i : \{1,2\}.M[i] \to M[i+1]:\langle U\rangle$ captures the same behaviour as $1 \to 2:\langle U\rangle.2 \to 3:\langle U\rangle$. Variable $n$ is bound by the $fn$ in the definition of processes.

**Network.** The user-defined global type captures the normal and two exceptional behaviours as three distinct branches composed as a sum as demonstrated in the global type

$$G_N \triangleq C \to P:\langle Data\rangle.(P \to L:\langle Logs\rangle.L \to ES:\langle Data\rangle.\mathsf{end}$$
$$+ P \to L:\langle Suspicious\rangle.L \to SS:\langle Logs\rangle.\mathsf{end}$$
$$+ P \to L:\langle QuotaWarn\rangle.L \to C:\langle Quota\rangle.\mathsf{end})$$

where $C$, $P$, $L$, $ES$ and $SS$ denote respectively the client, proxy, and log along with external and supervision servers. We can observe that the above global type is not robust, e.g., the external server is not notified in case a suspicious behaviour, nor is a quota warning raised. Robustness is automatically achieved through Definition 5 that adds $C$, $ES$ and $SS$ in the recipients of the prefix for each branch in the form of a multicast interaction. Below we give its definition:

$$G_{RN} \triangleq C \to P:\langle Data\rangle.(P \to L, ES, SS, C:\langle Logs\rangle.L \to ES:\langle Data\rangle.\mathsf{end}$$
$$+ P \to L, ES, SS, C:\langle Suspicious\rangle.L \to SS:\langle Logs\rangle.\mathsf{end}$$
$$+ P \to L, ES, SS, C:\langle QuotaWarn\rangle.L \to C:\langle Quota\rangle.\mathsf{end})$$

# 5 Meta-theory

## 5.1 Projection

A projection of global type onto the participants of a session produces types (see Figure 8) that capture the respective behaviours of those processes. The constructs of end-point types read symmetrically to the constructs of the LoSI. Product and end types read the same as in global types, with $\mu X.T \equiv T[\mu X.T/X]$ and $T + \mathsf{end} \equiv T$.

| | | | | | |
|---|---|---|---|---|---|
| $T ::=$ | End-point types | $\mid X$ | Variable | $\mid \Pi x : I.T$ | Product |
| $\mid [p,q]!\langle V\rangle.T$ | Output | $\mid T{+}T$ | Summation | $\mid T\ e$ | Application |
| $\mid [p,q]?(V).T$ | Input | $\mid [b]T$ | Matching | $\mid$ end | End |
| $\mid \mu X.T$ | Recursion | | | | |

**Figure 8** Syntax of end-point types.

Initially we provide auxiliary definitions which capture the technical challenges underlying projection, starting with the equality between arithmetical expressions. The typing environment $\mathscr{C}$ consisting of interaction and index variables present in global types is defined as follows: $\mathscr{C} ::= \emptyset \mid x : I, \mathscr{C}$.

**Definition 6** (Equality of $e$). *The judgment $\mathscr{C} \vdash e = e'$ can be written in a subset of DML constraints, i.e., $\exists x_1 : I_1. \ldots .\exists x_n : I_n.e = e'$ where $x_i : I_i \in \mathscr{C}$. Hence, the constraint solver of DML can be used to check the judgment $\mathscr{C} \vdash e = e'$.*

The *equality between a role* p *of a global type and a role* q *of a process* is defined by using typing environment $\mathscr{C}$ to capture equality between parameterised roles.

**Definition 7** (Equality of p). *A global type role* p *is equal to a process role* q*, written* $\mathscr{C} \vdash p = q$ *and read "in the typing environment* $\mathscr{C}$*,* p *equals* q*", if*

1. $p = N$ *and* $q = N$ *then* $\mathscr{C} \vdash N = N$ *or*
2. $p = p'[e]$, $q = q'[e']$, $\mathscr{C} \vdash e = e'$ *and* $\mathscr{C} \vdash p' = q'$ *then* $\mathscr{C} \vdash p'[e] = q'[e']$.

*The parameters of roles in global types are different from the ones in processes*, i.e., they range over different sets of naturals. For instance, $i$ in the Ring example has different ranges of values for processes and global types and is therefore introduced by different binders ($fn$ in the definition of processes and $\Pi$ in global types). For this reason, we need to translate participants from terms of global type parameters into process ones. This will ensure the right end-point types to type-check processes.

**Definition 8** (Substitution on p). *Suppose that* p *and* p′ *have the same parameters. Substitution on roles, written* $p'\{q/p\}$*, is defined:*

$$W\{W/W\} = W \qquad p[e]\{q[e'']/p'[e']\} = p\{q/p'\}[e\{e''/e'\}]$$

Substitution on expressions is straightforward.

In the following, we give the definition of inequality between two boolean expressions:

**Definition 9** (Inequality of $b$). *The judgment $\mathscr{C} \vdash b_i \neq b_j$ can be written in a subset of DML constraints, i.e., $\neg(\exists x_1 : I_1. \ldots .\exists x_n : I_n.b_i = b_j)$ where $x_i : I_i \in \mathscr{C}$.*

Branches in an end-point type are compatible if for an internal and external choice, the first prefix of each branch has a different type from the others; otherwise, all prefixes per branch are the same. This ensures robustness of global types, i.e., a user-defined global type must be transformed in order for the projection to hold. Also branches are compatible if they are guarded by mutually disjoint boolean expressions to model (nested) conditionals, e.g., global *if-else* and *case-switch* over boolean expressions.

**Definition 10** (Compatibility of branching). *Compatibility is defined by $\asymp_P$ and $\asymp_{NP}$ for branches of prefixes and $\asymp_B$ for branches of boolean expressions:*

1. $\sum_{i\in I} T_i \asymp_P \sum_{j\in J} T_j$ if $\forall i \in I$ and $\forall j \in J$ such that $T_i = [p,p']!\langle V_i\rangle; T_i'$ and $T_j = [p,p']!\langle V_j\rangle; T_j'$ where $V_i \neq V_j$
2. $\sum_{i\in I} T_i \asymp_P \sum_{j\in J} T_j$ if $\forall i \in I$ and $\forall j \in J$ such that $T_i = [p,p']?(V_i); T_i'$ and $T_j = [p,p']?\langle V_j\rangle; T_j'$ where $V_i \neq V_j$
3. $\sum_{i\in I} T_i \asymp_{NP} \sum_{j\in J} T_j$ if $\forall i \in I$ and $\forall j \in J$ such that $T_i = T_j$
4. $\mathscr{C} \vdash \sum_{i\in I} T_i \asymp_B \sum_{j\in J} T_j$ if $\forall i \in I$ and $\forall j \in J$ such that $T_i = [b_i]T_i'$ and $T_j = [b_j]T_j'$ where $\mathscr{C} \vdash b_i \neq b_j$.

where $x : I \neq x' : I'$ is handled as $I \neq I'$ and for $S$ types inequality $\neq$ is handled in a standard manner.

The next definitions provide the meaning for set membership of an arithmetical expression into an index sort and meaning of an empty index sort.

**Definition 11** (Set membership). *The judgment $\mathscr{C}, x : I \vdash e \in I$ can be written in a subset of DML constraints, i.e., $\exists x' : I'.\Theta \wedge e = x'$, where $I = \{x' : I' \mid \Theta\}$.*

**Definition 12** (Empty set). *The judgment $\mathscr{C} \vdash I \neq 0$ can be written in a subset of DML constraints, i.e., $\mathscr{C}, i : I' \vdash \Theta$ where $I = \{i : I' \mid \Theta\}$.*

**Definition 13** (Projection). *Given global type $G$, role $q$, and context $\mathscr{C}$ of free variables in $G$ and $q$, the projection of $G$ onto $q$, $G\restriction_q^{\mathscr{C}}$, is defined inductively as:*

$$p \to p' : \langle S\rangle.G \restriction_q^{\mathscr{C}} = \begin{cases} [q,r']!\langle S\rangle.[r,q]?(S).(G\restriction_q^{\mathscr{C}}) & \text{if } \mathscr{C} \vdash p,p' = q,\\ [q,p']!\langle S\rangle.(G\restriction_q^{\mathscr{C}}) & \text{if } \mathscr{C} \vdash p = q,\\ [p,q]?(S).(G\restriction_q^{\mathscr{C}}) & \text{if } \mathscr{C} \vdash p' = q,\\ G\restriction_q^{\mathscr{C}} & \text{otherwise} \end{cases}$$

*where $r = p\{q/p'\}$ and $r' = p'\{q/p\}$.*

$$p \to p' : \langle x : I\rangle.G \restriction_q^{\mathscr{C}} = \begin{cases} [q,r']!\langle x{:}I\rangle.[r,q]?(x{:}I).(G\restriction_q^{\mathscr{C},x:I}) & \text{if } \mathscr{C} \vdash p,p' = q \text{ and } \mathscr{C} \vdash I{\neq}\emptyset,\\ [q,p']!\langle x{:}I\rangle.(G\restriction_q^{\mathscr{C},x:I}) & \text{if } \mathscr{C} \vdash p = q \text{ and } \mathscr{C} \vdash I{\neq}\emptyset,\\ [p,q]?(x{:}I).(G\restriction_q^{\mathscr{C},x:I}) & \text{if } \mathscr{C} \vdash p' = q \text{ and } \mathscr{C} \vdash I{\neq}\emptyset,\\ G\restriction_q^{\mathscr{C}} & \text{if } \mathscr{C} \vdash p,p' \neq q \text{ and } \mathscr{C} \vdash I{\neq}\emptyset \end{cases}$$

*where $r = p\{q/p'\}$ and $r' = p'\{q/p\}$.*

$$\mu X.G \restriction_q^{\mathscr{C}} = \mu X.(G\restriction_q^{\mathscr{C}}) \qquad X\restriction_q^{\mathscr{C}} = X \qquad \mathsf{end}\restriction_q^{\mathscr{C}} = \mathsf{end}$$

$$[b]G \restriction_q^{\mathscr{C}} = \begin{cases} [b](G\restriction_q^{\mathscr{C}}) & \text{if } fv(b) \in dom(\mathscr{C})\\ G\restriction_q^{\mathscr{C}} & \text{if } fv(b) = \emptyset \end{cases} \qquad (G,G')\restriction_q^{\mathscr{C}} = \begin{cases} G\restriction_q^{\mathscr{C}} & \text{if } q \in pid(G), q \notin pid(G'),\\ G'\restriction_q^{\mathscr{C}} & \text{if } q \in pid(G'), q \notin pid(G),\\ \mathsf{end} & \text{if } q \notin pid(G), q \notin pid(G') \end{cases}$$

$$(G{+}G')\restriction_q^{\mathscr{C}} = (G\restriction_q^{\mathscr{C}} + G'\restriction_q^{\mathscr{C}}) \text{ if } pid(\mathsf{n}(G)) = pid(\mathsf{n}(G')) \text{ and: } \begin{cases} G\restriction_q^{\mathscr{C}} \asymp_P G'\restriction_q^{\mathscr{C}} & \text{where } q \in pid(\mathsf{n}(G))\\ G\restriction_q^{\mathscr{C}} \asymp_{NP} G'\restriction_q^{\mathscr{C}} & \text{where } q \notin pid(\mathsf{n}(G))\\ \mathscr{C} \vdash G\restriction_q^{\mathscr{C}} \asymp_B G'\restriction_q^{\mathscr{C}} & \text{where } pid(\mathsf{n}(G)) = \emptyset) \end{cases}$$

$$\Pi x{:}I.G \upharpoonright_q^{\mathscr{C}} = \begin{cases} \Pi x : I.(G \upharpoonright_q^{\mathscr{C},x:I}) & \textit{if } \mathscr{C}\vdash I\neq\emptyset \textit{ and } x\notin \mathsf{dv}(\mathit{pid}(G)) \\ (G \upharpoonright_q^{\mathscr{C},x:I}) & \textit{if } \mathscr{C}\vdash I\neq\emptyset \textit{ and } x\in \mathsf{dv}(\mathit{pid}(G)) \end{cases}$$

$$G\ e \upharpoonright_q^{\mathscr{C}} = G \upharpoonright_q^{\mathscr{C}}\ e \quad \textit{if } G \equiv \Pi x : I.G' \textit{and } \mathscr{C}, x : I \vdash e \in I$$

Projection is largely intuitive, formalising the definitions in Section 4. When a side condition does not hold, the projection is undefined. In parameterised global types, a *parameterised role can appear on both sides of a parameterised interaction* for different values of the parameter. This occurrence is covered by the first case of projection for interaction. In case of an interaction holding an interaction variable, (1) the side conditions check for well-sortedness of the index sort, and (2) the typing environment $\mathscr{C}$ is extended with the type assignment of the variable only for $p$ and $p'$ in the projection of the remaining global type. Both (1) and (2) enforce the well-assertedness of global types. In matching, the boolean expression is kept in the end-point type only for participants that have visibility of the variables in $b$. That is, for some value of those variables $b$ can evaluate to true or false, i.e, to well-assertedness.

Coherence of matching ensures that the boolean expression guarding an interaction is visible to both participants of that interaction and not only one.

**Definition 14** (Matching coherence). *Given the typing environment $\mathscr{C}$ and $\mathscr{C}'$ of the projection of $[b]G' \upharpoonright_p^{\mathscr{C}}$ and $[b]G' \upharpoonright_{p'}^{\mathscr{C}'}$ respectively, we say that $G$ is* matching coherent *if for every $[b]G'$ in $G$ then $fv(b)\in dom(\mathscr{C})$ and $fv(b)\in dom(\mathscr{C}')$ hold for respectively $G' \upharpoonright_p^{\mathscr{C}}$ and $G' \upharpoonright_{p'}^{\mathscr{C}'}$ for some $p \to p'$ in $G'$.*

In parallel global types, $q$ can be in at most one global type, ensuring single-threading in the system. Projection for branching holds if the returned end-point types are compatible, according to one of the conditions. The function $\mathsf{n}(G)$ returns the first prefix of $G$ in case of an interaction branching or an empty set if the branching is guarded by a boolean expression. For product global type, projection is defined inductively for every role with an extended typing environment containing its argument assignment. The $\Pi$ prefix is kept in end-points only if it plays the role of an arrow type and is not kept if it abstracts families of global types. For application, projection is defined inductively only if the subsequent global type is (a) congruent to a product global type that plays the role of an arrow type and (b) the arithmetic expression applied conforms to the type of the argument in the product global type.

**Definition 15** (Coherence). *We say that $G$ is* coherent *if it is* matching coherent *and for each $p \in \mathit{pid}(G), G \upharpoonright_p^{\mathscr{C}}$ is defined for $dom(\mathscr{C}) = \mathsf{fv}(G)$.*

**Proposition 16** (Coherent end-point types). *Let $G$ be coherent. Then for each $p \in \mathit{pid}(G)$ if $G \upharpoonright_p^{\mathscr{C}}$ is defined then $G \upharpoonright_p^{\mathscr{C}}$ is also coherent.*

### 5.2 Sorting

We can observe from the first case in the projection of an interaction, that the order of prefixes is forced in parameterised global types, i.e., the output prefix is always placed before the input. For example, the end-point type returned from the projection of the global type for the Ring onto $\mathtt{W}[i]$ is

$$[\mathsf{W}[i],\mathsf{W}[i+1]]!\langle U\rangle.[\mathsf{W}[i-1],\mathsf{W}[i]]?(U).\mathsf{end}$$

where the output prefix is placed before the input. The interaction order in this participant does not mirror the one in the global type where the input prefix precedes the output (any participant between two others receives from the one to the right and then sends to the one to the left). However, the order of prefixes is not only broken in the projection of an interaction, but also when prefixes of sequential interactions are composed. The order of actions is based on the order of who sends first. In the following, we present how we model ordering.

The order of arithmetic expressions in participants is defined over the monotony of those expressions, i.e. if the expressions are increasing over the index, e.g. $i$+1, then they are monotonically increasing (see the first case below), otherwise, e.g. $n$-$i$ they are monotonically decreasing (see the second case below).

**Definition 17.** *The order relation* rel *between arithmetic expressions is defined as:*

1. $\mathsf{rel}(e,e')$ *iff* $\mathscr{C}\vdash e\leq e'$ *and* $\mathscr{C}\vdash e\{\mathsf{c}:I/x:I\}<e'\{\mathsf{c}':I/x:I\}$
2. $\mathsf{rel}(e,e')$ *iff* $\mathscr{C}\vdash e\geq e'$ *and* $\mathscr{C}\vdash e\{\mathsf{c}:I/x:I\}>e'\{\mathsf{c}':I/x:I\}$

*where* $\mathsf{c}<\mathsf{c}'$ *for* $\forall x\in\mathsf{fv}(e)\cup\mathsf{fv}(e')$, *the first relation after the iff defines the relation of expressions and the second defines their monotonicity accordingly.*

**Definition 18.** *The order relation* rel *between roles is defined as a lexicographical order over the arithmetic expressions that define them:*

$\mathsf{rel}(N[e_1]...[e_i]...[e_n],N[e'_1]...[e'_i]...[e'_n])$ *iff for* $1\leq i\leq n$ *such that* $\mathscr{C}\nvdash e_i=e'_i$ *then* $\mathsf{rel}(e_i,e'_i)$.

**Definition 19.** *The order relation* rel *between two prefixes is defined as the order of the senders on them:*

$\mathsf{rel}\ ([p_1,q_1]!/?^{3}(V),\ [p_2,q_2]!/?(V'))$ *iff* $\mathsf{rel}(p_1,p_2)$

Thus, prefixes in the end-point types returned by projection are sorted, through a sorting algorithm sort, to preserve the order as in global types in case of a parameterised global type. We can use a sorting algorithm (Mergesort, Quicksort, etc.) defined over this ordering relation.

### 5.3 Examples

**Web service.** The projection of global type $G_W$ onto $P$ is:

$$G_W\restriction^{\emptyset}_{P}=[Cc,P]?(Req).([P,W]!\langle Fwd\rangle.\mathsf{end}+[P,W]!\langle Aud\rangle.[W,P]?(Dtls).[P,W]!\langle Res\rangle.\mathsf{end})$$

Since no parameterised interaction is present, sorting does not affect the end-point type returned by projection, which is used to type-check the process.

[3] !/? denotes either ! or ?.

**Financial protocol.** The projection of global type $G_F$ onto $Bu$ is:

$$G_F \upharpoonright^{\emptyset}_{Bu} = [Bu,S]!\langle v_0 : I\rangle.(\mu X.\Pi iter : \mathsf{nat}.[iter < 5] \\ ([v_0 < 120][S,Bu]?(Neg).[Bu,S]!\langle v_0 : I\rangle.X\ iter+1 \\ + [\mathsf{not}\ v_0 < 120][S,Bu]?(Ok).[Bu,Ba]!\langle v_0 : I\rangle.\mathsf{end}))\ 1$$

**Network.** The projection of global type $G_{RN}$ onto $P$ is:

$$G_{RN} \upharpoonright^{\emptyset}_{P} = [C,P]?(Data).([P,\{L,ES,SS\}]!\langle Logs\rangle.\mathsf{end} \\ +[P,\{L,ES,SS\}]!\langle Suspicious\rangle.\mathsf{end} \\ +[P,\{L,ES,SS\}]!\langle QuotaWarn\rangle.\mathsf{end})$$

This end-point type also contains the participants that must be notified in case of failure or normal behaviour. We use this type to generate the run-time process from the user-defined one. The transformation for processes is the same as in variant [15], i.e., missing actions in the user-defined process are added according to the info in the type.

**Ring.** Given the type assignment $n : I$, the projection of global type $G_R$ onto $\mathtt{W}[i]$ is:

$$G_R \upharpoonright^{n:I}_{\mathtt{W}[i]} = [\mathtt{W}[i],\mathtt{W}[i+1]]!\langle U\rangle.[\mathtt{W}[i-1],\mathtt{W}[i]]?(U).\mathsf{end}.$$

We showcase projection steps. The global type index is renamed from $i$ to $j$ to not have the same name as in processes. Firstly, the equality $j$: $\{i'$:nat $\mid i' <$ n and $i' \geq 1\}$, n:$I$, $i$:$J \vdash \mathtt{W}[j], \mathtt{W}[j+1] = \mathtt{W}[i]$ holds, satisfying the first case of causality projection. Second, we perform role substitutions $\mathtt{W}[j+1]\{\mathtt{W}[i]/\mathtt{W}[j]\}$ and $\mathtt{W}[j]\{\mathtt{W}[i]/\mathtt{W}[j+1]\}$. From here, we achieve the end-point type $[\mathtt{W}[i],\mathtt{W}[i+1]]!\langle U\rangle.[\mathtt{W}[i-1],\mathtt{W}[i]]?(U)$. This end point type does not reflect the behaviour as in the global type, i.e., "sending" to the neighbour precedes "receiving" from the other. This is why we sort it to $[\mathtt{W}[i-1],\mathtt{W}[i]]?(U)$. $[\mathtt{W}[i],\mathtt{W}[i+1]]!\langle U\rangle$. The sender of the second prefix is smaller than the sender of the first, meaning that the second prefix occurs before the first. This end-point type well-types the behaviour of the $Middle$ role, which firstly receives from the neighbour on the right and then sends to the one on the left.

## 6 Typing Relation

Figure 9 describes the typing rules for processes. The typing judgement is of the form $\Gamma \vdash P \triangleright \Delta$, read "*in the context* $\Gamma$ *process* $P$ *has type* $\Delta$".

The formal definition of types and typing contexts is the following: $\Delta ::= \emptyset \mid \Delta, u[p]{:}T \quad \Gamma ::= \emptyset \mid \Gamma, a{:}G \mid \Gamma, X{:}\Delta \mid \Gamma, x : S$. The type $\Delta$ represents sequences of session channels along with their types $T$. The typing context $\Gamma$ maps sites, process variables, and variables to types. We write $dom(\Gamma)$ and $dom(\Delta)$ for the set of names, variables, and session channels bound in $\Gamma$ and $\Delta$ respectively. When the type $S$ has shape $G$ we assume that $G$ is coherent. The context is extended with new elements on the right

$$\Gamma, a : G \vdash a \triangleright G \quad \lfloor \text{TName} \rfloor \qquad \frac{\Gamma, u : G \vdash P \triangleright \Delta, u[p] : \mathsf{sort}(G\lceil_p^{\Gamma(\mathsf{fv}(G))})}{\Gamma \vdash \mathit{init}(u{:}G, p).P \triangleright \Delta} \; \lfloor \text{TSInit} \rfloor$$

$$\frac{\Gamma \vdash m \triangleright U \quad \Gamma \vdash P \triangleright \Delta, u[p] : T}{\Gamma \vdash u[p,q]!\langle m : U\rangle.P \triangleright \Delta, u[p] : [p,q]!\langle U\rangle.T} \; \lfloor \text{TSend} \rfloor$$

$$\frac{\Gamma, x : U \vdash P \triangleright \Delta, u[q] : T}{\Gamma \vdash u[p,q]?(x : U).P \triangleright \Delta, u[q] : [p,q]?(U).T} \; \lfloor \text{TRcv} \rfloor$$

$$\frac{\Gamma \vdash x \triangleright I \quad \Gamma \vdash P \triangleright \Delta, u[p] : T}{\Gamma \vdash u[p,q]!\langle x : I\rangle.P \triangleright \Delta, u[p] : [p,q]!\langle x : I\rangle.T} \; \lfloor \text{TISend} \rfloor$$

$$\frac{\Gamma, x : I \vdash P \triangleright \Delta, u[q] : T}{\Gamma \vdash u[p,q]?(x : I).P \triangleright \Delta, u[q] : [p,q]?(x : I).T} \; \lfloor \text{TIRcv} \rfloor$$

$$\frac{\Gamma \vdash P \triangleright \Delta, u[p] : T}{\Gamma \vdash u[p,q]!\langle b[p']{:}T'\rangle.P \triangleright \Delta, u[p] : [p,q]!\langle T'\rangle.T, b[p'] : T'} \; \lfloor \text{TSSend} \rfloor$$

$$\frac{\Gamma \vdash P \triangleright \Delta, u[q] : T, b[p'] : T'}{\Gamma \vdash u[p,q]?(b[p'] : T').P \triangleright \Delta, u[q] : [p,q]?(T').T} \; \lfloor \text{TSRcv} \rfloor$$

$$\frac{\Gamma, X : \Delta \vdash P \triangleright \Delta}{\Gamma \vdash \mathit{rec}\ X = P \triangleright \Delta} \; \lfloor \text{TRec} \rfloor \qquad \Gamma, X : \Delta \vdash X \triangleright \Delta \quad \lfloor \text{TVar} \rfloor$$

$$\frac{\Gamma \vdash \Delta \quad \Delta \text{ end only}}{\Gamma \vdash 0 \triangleright \Delta} \; \lfloor \text{TInact} \rfloor \qquad \frac{\Gamma \vdash P \triangleright \Delta \quad \Gamma \vdash Q \triangleright \Delta'}{\Gamma \vdash P \mid Q \triangleright \Delta, \Delta'} \; \lfloor \text{TPar} \rfloor$$

$$\frac{\Gamma \vdash P \triangleright \Delta \quad \Gamma \vdash Q \triangleright \Delta'}{\Gamma \vdash P{+}Q \triangleright \Delta{+}\Delta'} \; \lfloor \text{TSum} \rfloor \qquad \frac{\Gamma \vdash b \triangleright \mathsf{bool} \quad \Gamma \vdash P \triangleright \Delta}{\Gamma \vdash [b]P \triangleright [b]\Delta} \; \lfloor \text{TMatch} \rfloor$$

$$\frac{\Gamma, x : I \vdash P \triangleright \Delta}{\Gamma \vdash \mathit{fn}\ x : I \Rightarrow P \triangleright \Pi x{:}I.\Delta} \; \lfloor \text{TFun} \rfloor \qquad \frac{\Gamma \vdash P \triangleright \Pi x{:}I.\Delta \quad \Gamma \vdash e \in I}{\Gamma \vdash P\ e \triangleright (\Pi x{:}I.\Delta)\ e} \; \lfloor \text{TApp} \rfloor$$

**Figure 9** Static type system.

side through the "comma" operator. In "$\Gamma, a : G, x : S, X : \Delta$", we assume $a, x, X$ do not occur in $\Gamma$, and in "$\Delta, u[p] : T$", we assume $u[p]$ does not occur in $\Delta$.

Rule $\lfloor$TName$\rfloor$ assigns a global type to a session identifier. Other rules such as $\Gamma \vdash \mathsf{true}, \mathsf{false} \triangleright \mathsf{bool}$, $\Gamma \vdash \mathsf{c} \triangleright \mathsf{Int}$ and $\Gamma \vdash b_1 \mathsf{or}\ b_2 \triangleright \mathsf{bool}$ if $\Gamma \vdash b_i \triangleright \mathsf{bool}$ assign type to messages, e.g., the type bool to true, false and to an "$or$" expression based on the types of its subexpressions that both must evaluate to a Boolean.

Rule $\lfloor$TSInit$\rfloor$ assigns a type to a process prefixed by session initiation based on the typing of the subprocess: the subprocess $P$ must be typed by the projection of the global type onto the participant of the prefix (sorting of prefixes in end-point types is applied in case of parameterised sessions) under an extended environment containing the typing for the session identifier with global type $G$. Projection is defined over a restricted $\Gamma$, containing only the assumptions for free variables in $G$, written as $\Gamma(\mathsf{fv}(G))$. The type assigned $\Delta$ means that the process does not evaluate to any interactions at this point but denotes the typing of other possible sessions running.

Rules $\lfloor$TSend$\rfloor$ and $\lfloor$TRcv$\rfloor$ assign a type to a process respectively prefixed by an output and input action based on the typing of the subprocess: participants $p, q$ in the prefix must match the ones in the end-point type; $P$ must be well-typed by a type $T$ (for the input prefix over an extended context with the place-holder of the message); for the output prefix, the message $m$ must evaluate to a value of type $U$. Rules for variables in assertions read similarly with the difference that the type of variables is present along with the variable in the end-point type. These rules enforce the typing of interaction variables that may appear in assertions.

Rules ⌊TSSEND⌋ and ⌊TSRCV⌋ assign a type to processes sending and receiving a session channel (session delegation and reception) respectively based on the typing of the subprocesses. Rule ⌊TSSEND⌋ extends the conclusion type with the new type assignment for the participant delegating the session. Rule ⌊TSRCV⌋ decreases the premise type to maintain disjoint typings when the two processes (delegating and receiving) are composed in parallel.

Rules ⌊TREC⌋, ⌊TVAR⌋ and ⌊TINACT⌋ are standard, where "$\Delta$ end only" means that the type $\Delta$ contains only elements of shape $u[p]:\mathtt{end}$. Rule ⌊TPAR⌋ assigns a type to the parallel composition of two processes based on their respective typings: each process must be well-typed and the result is the union of the two typings if they are disjoint. Rule ⌊TSUM⌋ assigns a type to the disjunction composition of two processes based on their respective typing: each process must be well-typed with typing $\Delta_i$ for $i \in \{1,2\}$. The summation of two types is defined as

$$\begin{aligned}\Delta{+}\Delta' = \{a[p]: T{+}T' | a[p]: T \in \Delta, a[p]: T' \in \Delta', \text{ and } T \neq T'\}\\ \cup \{a[p]: T | a[p]: T \in \Delta \text{ and } a[p]: T \in \Delta'\}\end{aligned}$$

where two different sequences of a participant $p$ in session $a$ are composed in disjunction; otherwise sequences are the same.

Rule ⌊TMATCH⌋ assigns a type to a process prefixed by a boolean expression based on its respective typing: the subprocess and the boolean expression must be well-typed. The session type is defined as $[b]\Delta = \{a[p]:[b]T | a[p]: T \in \Delta\}$ where the boolean expression is carried in the end-point types.

Rule ⌊TFUN⌋ assigns a type to a $fn$-abstraction based on the typing of the subprocess: $P$ must be typed under an augmented $\Gamma$ with mapping for variable $x{:}I$. The session type is defined as $\Pi x : I.\Delta = \{a[p]: \Pi x{:}I.T | a[p]: T \in \Delta\}$ where the argument $\Pi$ is carried in the end-point types.

Rule ⌊TAPP⌋ assigns a type to an application based on the type of the $fn$ term: the $fn$ term must be well-typed by a product type and the expression must be of the type $I$. The type assigned is the one of the $fn$ along the expression as $(\Pi x{:}I.\Delta)\, e = \{a[p]: (\Pi x{:}I.T)\, e\, | a[p]: T \in \Delta\}$ where the expression $e$ is carried in the end-point types.

## 7 Properties of CMST

In this section, we describe the approach taken for proving subject reduction, communication safety, and the progress of CMST. The definition of properties and theorems are in the same spirit of Deniélou and Yoshida [17] which provide a latter and a more concise, intuitive presentation.

**Subject reduction** states that the type of a process is preserved or reduced. As a session runs, messages are sent to the queue and subsequently are received (and also consumed); branches guarded by boolean expressions are selected according to the values of those expressions; occurrences of an argument in a product type are substituted by naturals. We formalise the way session types can change through the reduction operation $\Delta \Rightarrow \Delta'$, assuming coherence of types. This judgment means that $\Delta'$ differs from $\Delta$ in one of five cases:

- $\{a[\hat{p}]{:}[\hat{p},\hat{q}]!\langle U_1\rangle.T_1{+}...{+}[\hat{p},\hat{q}]!\langle U_n\rangle.T_n, a[\hat{q}]{:}[\hat{p},\hat{q}]?(U_1).T_1'{+}...{+}\ [\hat{p},\hat{q}]?(U_m).T_m'\} \Rightarrow \{a[\hat{p}]{:}T_i, a[\hat{q}]{:}T_i'\} \quad (i{\in}[1..n],\ n{\leq}m)$
- $\{a[\hat{p}]{:}[\hat{p},\hat{q}]!\langle T''\rangle.T, a[\hat{q}]{:}[\hat{p},\hat{q}]?(T'').T'\}{\Rightarrow}\{a[\hat{p}]{:}T, a[\hat{q}]{:}T'\}$
- $\{a[\hat{p}] : [\mathsf{true}]T + T'\} \;\Rightarrow\; \{a[\hat{p}] : T\}$, $\{a[\hat{p}] : [\mathsf{false}]T + T'\} \;\Rightarrow\; \{a[\hat{p}] : T'\}$
- $\{a[\hat{p}] : (\Pi x : I.T)\mathrm{c}\} \;\Rightarrow\; \{a[\hat{p}] : T[\mathrm{c}/x]\}$
- $\{a[\hat{p}] : T_1, a[\hat{q}] : T_2\}, \Delta \Rightarrow \{a[\hat{p}] : T_1', a[\hat{q}] : T_2'\}, \Delta$ if $\{a[\hat{p}] : T_1, a[\hat{q}] : T_2\} \Rightarrow \{a[\hat{p}] : T_1', a[\hat{q}] : T_2'\}$

Intuitively, the rules represent: (1) the interaction of some values, (2) the interaction of a single session channel, (3) the action of choosing a branch guarded by a true or false value, and (4) the application of a natural to a $fn$ process.

Subject reduction ensures that the type annotations are well-defined, i.e., global types $G$ are coherent and types $S$ denote the correct sets of values associated to variables and values. This rules out terms such as $a[p,q]!\langle 5{:}\mathsf{bool}\rangle.P$ or $a{:}\langle G\rangle[p].P$ where $G$ is one of the non-coherent types given at Section 5.1.

**Theorem 20** (Subject congruence and reduction)**.**

1. *If* $\Gamma \vdash P \triangleright \Delta$ *and* $P \equiv P'$ *then* $\Gamma \vdash P' \triangleright \Delta$.
2. *If* $\Gamma \vdash P \triangleright \Delta$, *and* $P \longrightarrow P'$, *then* $\Gamma \vdash P' \triangleright \Delta'$ *where* $\Delta = \Delta'$ *or* $\Delta \Rightarrow \Delta'$.
3. *If* $\emptyset \vdash P \triangleright \emptyset$ *and* $P \longrightarrow P'$ *then* $\emptyset \vdash P' \triangleright \emptyset$.

Properties (1) and (2) are proved by induction on the typing judgment assuming respectively $P \equiv P'$ and $P \to P'$. Property (3) ensures subject reduction for closed processes and follows from the second property.

**Type safety** ensures that expressions present in a process $P$ are well-asserted and well-dependent according to the corresponding construct, i.e., boolean expressions are well-asserted for matching and arithmetic expressions are well-dependent for application. For an untyped LoSI, wrong terms such as $[x > 5]P$ and $(fn\ x{:}I \Rightarrow P)\,3$, where $I = \{x' : \mathsf{nat}|x' > 3\}$, can be defined.

**Corollary 21** (Type safety)**.** *If* $\Gamma \vdash P \triangleright \Delta$ *then* $P$ *is* type safe.

This is a corollary from subject reduction (Theorem 20) and well-asserted, well-dependent end-point types (Proposition 16).

**Communication safety** ensures that for some inputs (or outputs) in disjunction of a process, reciprocally, there are fewer outputs (or more inputs) of the respectively same sets of values in a parallel process. This is in the same line of logic with the one of subtyping by Gay and Hole [19]. We define the reduction context as $\mathscr{F} ::= \mathscr{F}\,|\,P \;\mid\; P|\mathscr{F} \;\mid\; \bullet$.

**Definition 22** (Communication safety). *Process $P$ is* communication safe *if* $P \equiv \mathscr{F}[Q]$

1. *with* $Q \equiv a[q,p]?(x{:}U_1).Q_1 + ... + a[q,p]?(x{:}U_n).Q_n$ *then there exists* $\mathscr{F}'$ *such that* $\mathscr{F}[a[q,p]?(x{:}U_1).Q_1 + ... + a[q,p]?(x{:}U_n).Q_n] \rightarrow^\star \mathscr{F}'[a[q,p]?(x{:}U_1).Q_1 + ... + a[q,p]?(x{:}U_n).Q_n \mid a:(q,p,v_i : U_i)\cdot h]$ *and* $i \in [1..m]$ *for* $m \leq n$
2. *with* $Q \equiv a[q,p]?(b[p'] : T).Q'$ *then there exists* $\mathscr{F}'$ *such that* $\mathscr{F}[a[q,p]?(b[p'] : T).Q'] \rightarrow^\star \mathscr{F}'[a[q,p]?(b[p'] : T).Q' \mid a:(q,p,b[p'] : T)\cdot h]$
3. *with* $Q \equiv a:(q,p,v_i : U_i)\cdot h$ *where* $i \in [1..n]$ *then there exists* $\mathscr{F}'$ *such that* $\mathscr{F}[a:(q,p,v_i : U_i)\cdot h] \rightarrow^\star \mathscr{F}'[a[q,p]?(x{:}U_1).Q_1 + ... + a[q,p]?(x{:}U_m).Q_m \mid a:(q,p,v:U)\cdot h']$ *where* $m \geq n$
4. *with* $Q \equiv a:(q,p,b[p'] : T)\cdot h'$ *then there exists* $\mathscr{F}'$ *such that* $\mathscr{F}[a:(q,p,b[p'] : T)\cdot h'] \rightarrow^\star \mathscr{F}'[a[q,p]?(b[p'] : T).Q' \mid a:(q,p,b[p'] : T)\cdot h']$

*or* $P \equiv 0$.

The following result is defined when a single session has started for an open process, i.e., a queue is created and the process is not restricted.

**Theorem 23** (Communication safety). *Suppose that* $a : G \vdash Q \mid a : h \triangleright a[p_1] : T_1, ..., a[p_n] : T_n$. *Then* $Q \mid a : h$ *is communication safe.*

This property is proved by the derivation of the typing judgment, analysing all possible cases for the session type reduction $\Delta \Rightarrow \Delta'$. The observation is that a session type reduction represents a processes reduction, e.g., a type reduction of a value will indeed be a value interaction; this is stated in an inversion lemma.

By Theorems 20 and 23, we have that a typed process specifying a multiparty session will not go wrong, i.e., (1) type annotations (global types and message types) in variables and values are well-defined and (2) an interaction will consist of outputs satisfying a subset of enabled inputs.

**Progress** asserts that the reduction of a closed, simple, and well-typed process denoting a multiparty session will never get stuck, i.e., the process is an inaction 0 or can make a reduction step. A *simple* process specifies only one multiparty session, say on $a$, and no other sessions.

**Definition 24** (Simple). *A process is* simple *if (1) it specifies the behaviour of each participant* $\hat{p}_i$ *present in the global type* $G$ *i.e.* $\{\hat{p}_1, \ldots, \hat{p}_n\} = pid(G)$*, (2) parallel composition of initial processes, i.e.* $init(a : G, [\hat{p}_1]).P_1 \mid ... init(a : G, [\hat{p}_n]).P_n$*, (3) summation of simple processes only prefixed by matching, i.e.* $[true]P + P'$, $[false]P + P'$.

**Theorem 25** (Progress). *If* $\emptyset \vdash P \triangleright \emptyset$ *and* $P$ *is simple then* $P \equiv 0$ *or* $P \rightarrow P'$.

## 8 Layering non-essential features

In the following, we provide how non-essential features that were removed can be layered atop of the essential ones, and the mapping of their constructs in LMS and MST and variants into the LoSI and CMST.

- *Branching over labels* can be layered atop summation and value exchange. Its modelling in the LMS (and some variants) can be mapped into the LoSI as: $s \triangleleft l.P \stackrel{\text{def}}{=} a[p,q]!\langle m{:}U\rangle.\ P$ and $s \triangleright \{l_i{:}P_i\}_{i\in I} \stackrel{\text{def}}{=} a[p,q]?(x_1{:}U_1).P_1 + \ldots + a[p,q]?(x_n{:}\ U_n).P_n$ where $I = [1..n]$ and for type constructs as: $\{l_i : G_i\}_{i\in J} \stackrel{\text{def}}{=} G_1 + G_2 + \ldots + G_n$ where $J = [1..n]$.
- *Primitive recursion* can be layered atop functional abstraction, recursion, summation and assertions as in Programming Computable Functions (PCF) [40]. Its modelling in [46] can be mapped in the LoSI as: $\mathbf{R}\ P\ \lambda i.\lambda X.Q \stackrel{\text{def}}{=} rec\ X = fn\ i : \mathsf{nat} \Rightarrow [i = 0]\ P + [i < n]\ Q$. Similarly, for type constructs as: (1) for parameterised interactions as $\mathbf{R}\ G\ \lambda i.\lambda X.G' \stackrel{\text{def}}{=} \Pi i : I.G'.G$ and for non parameterised as $\mathbf{R}\ G\ \lambda i.\lambda X.G' \stackrel{\text{def}}{=} \mu X.\Pi i : \mathsf{nat}.[i = 0]G + [i < n]G'$.
- *Conditional* can be layered atop assertions and summation where each guard is dual of the other. Its modelling in LMS and variants can be mapped into the LoSI as: $\mathsf{if}\ b\ \mathsf{then}\ P\ \mathsf{else}\ Q \stackrel{\text{def}}{=} [b]P + [\mathsf{not}\ b]Q$.
- *Partial failure* can be layered atop summation and value exchange. Its modelling in Chen et al. [15] can be mapped into the LoSI as: $c?/!(p,e)^F \stackrel{\text{def}}{=} a[p,q]!/?\langle f{:}U\rangle.P$ and $\mathsf{try}\{P\}\ \mathsf{catch}\{f_1 : P_1, \ldots f_n : P_n\} \stackrel{\text{def}}{=} P + a[p,q]?(f_1{:}U_1).P_1 + \ldots + a[p,q]?(f_n{:}U_n).P_n$. Similarly, for types as: $p \to p' : \langle U \vee f_1, \ldots f_n\rangle \stackrel{\text{def}}{=} p \to p' : \langle U\rangle + p \to p' : \langle U_1\rangle \ldots p \to p' : \langle U_n\rangle$ and $\mathsf{try}\ \{G\}\ \mathsf{catch}\{f_1 : G_1, \ldots f_n : G_n\} \stackrel{\text{def}}{=} G + p \to q : \langle U_1\rangle.P_1 + \ldots + p \to q : \langle U_n\rangle.P_n$.
- *Asserted interactions* can be layered atop assertions, value exchange or functional abstraction. Its modelling in Bocchi et al. [7] can be mapped into the LoSI as: $s!\langle m\rangle(x)\{A\}.P \stackrel{\text{def}}{=} fn\ x : I \Rightarrow [A]a[p,q]!\langle m{:}U\rangle.P$ or $s!\langle m\rangle(\tilde{x})\{A\}.P \stackrel{\text{def}}{=} a[p,q]?\langle x{:}I\rangle.[A]a[p,q]!\langle m{:}\ U\rangle.P$. Similarly, for types as: $p \to p' : \langle s(x : U)\{A\}\rangle \stackrel{\text{def}}{=} \Pi x : U.[A]p \to p' : \langle U\rangle$ or $p \to p' : \langle s(x : U)\{A\}\rangle \stackrel{\text{def}}{=} p \to p' : \langle x : U\rangle.[A]p \to p' : \langle U\rangle$ .
- Session initiation of the LMS (and some variants) can be mapped into the LoSI as: $a[p](\tilde{s}).P$ and $\bar{a}[0..n](\tilde{s}).P \stackrel{\text{def}}{=} init(a{:}G,p).P$.
- def recursion constructs of the LMS and variants can be mapped into the LoSI as: $\mathsf{def}\ \{X_i(x_i) = P_i\}_{i\in I}\ \mathsf{in}\ P \stackrel{\text{def}}{=} P[rec\ X_i = fn\ x_i : \mathsf{nat} \Rightarrow P_i/X_i]_{i\in I}$ and $X\langle e\rangle \stackrel{\text{def}}{=} X\ e$.

## 9 Advanced features

This section discusses advanced features coming from three influential works on session types [9, 10, 17]. Deniélou and Yoshida [17] introduce dynamic joining and leaving of participants in session types to meet the needs of peer-to-peer protocols and cloud systems. Carbone, Honda and Yoshida [9, 10] introduce independent choice and local variable assignment to describe and verify web services protocols, where a global protocol is projected onto the corresponding participants to obtain the running processes.

Initially we look at the language of [17]. The join operation is defined as our session initiation process in parallel to an ongoing session $a[\hat{p}].P \mid (\nu a : G)(P')$. The operation is defined only if the behaviour $P$ of participant $\hat{p}$ is not in $P'$. The input and output prefixes are annotated by labels motivated by the hard-wiring of branching into the input prefix. A new feature introduced by Deniélou and Yoshida is sequential composition, modelled as $P; P'$. The polling construct instantiates processes of the same "role", i.e., $c\forall(x : r) \setminus p.\{P\}$ instantiates processes from role $P$ for all participants in $r$ that are different from $p$. This is supported through a distinct reduction rule. Leaving is modelled through a new construct: $quit$.

A participant joining a session can be mapped in LoSI by (1) recording the participants that already joined in the restriction operator as $\nu a : G; \{\hat{p}_1, .., \hat{p}_n\}$, and (2) introducing a new joining rule that checks if $\hat{p} \in \mathsf{pid}(G)$ and $\hat{p} \notin \{\hat{p}_1, .., \hat{p}_n\}$, resulting in $(\nu a : G; \{\hat{p}_1, .., \hat{p}_n, \hat{p}\})(P' \mid P)$. As we have shown, label annotated messages can be mapped through our type annotated messages. We can model sequential composition by extending $\lambda$ abstractions from naturals to processes as $\lambda x : T.P$, thus mapping sequential composition as $(\lambda x : T.P)\, P'$. We believe that by using a functional operator, session types and their properties will benefit from the expressivity power of functional composition, missing so far in the literature. Polling can be modelled in LoSI through our semantics of instantiating processes, e.g., $Middle(i)$ role in the Ring, through $\lambda$ and $\mu$ recursion. We need to extend $\lambda$ expression with participants as $\lambda x : p.P$ to map the polling operation as $\mu X.\lambda x : p.P$ (i.e., with no new reduction rules). The new essential feature added is the leaving of one or more participants, modelled by a simple construct, quit, to model at both language and type definition.

Analogously, in global types we can model those features by extending our $\Pi i : I.G$ to $\Pi p : \{p\}.G$ where $\{p\}$ is a set of participants to model global specification of polling. Martin-Löf [27] identified a correspondence between universal quantification $\forall$ and the dependent product $\Pi$. Also, we must extend the dependent product type and application to $\Pi x : G'.G$ and $GG'$ to model the sequential composition of sessions. This is similar to the $\lambda$LF [20] type theory, where the $G' \to G$ type is replaced by the dependent product type $\Pi x : G'.G$. The meta-theory needs to be slightly adapted to reflect projection for the new version of the $\Pi$ construct.

The global calculus [9, 10] adds independent choice over global behaviours and assignment of local variables to the original global types. Independent choice is similar to Bhargavan et al. [6], where choice is defined by the operation in the signature of the interaction construct. For example in $A \to B{:}s\langle f, e, x\rangle.G + A \to B{:}s\langle g, e, x\rangle.G'$, $f$ and $g$ define which branch the conversation will follow, and $e$ represents the value sent by $A$ to $B$, captured by $x$. As shown above, independent choice can be modelled in out theory through the summation construct. Local variable assignment $x@A = e.G$, where $x$ is a variable local to $A$, can be easily represented by extending the LoSI with local variables as $(\lambda x@A.G)\, e$. A fundamental difference with our work is that the typing discipline is based on binary session types, lacking safety and progress for the MST.

## 10 Related Work

**Session types.** Binary session types [21, 41] are the fore-runners of multiparty session types. They capture the communication structure between two processes and, as global types, are part of the program annotations. Processes are verified by type-checking and reciprocity (duality) of session types. That is, session types of two processes must be reciprocal. The LMS is based on the session language introduced in [21]. Channels are shared only by the two complementary processes and are bidirectional.

A survey [18] gives an overview of the many extensions of binary session types and languages. In particular, our modelling of internal and external choice is in the same spirit as a version of binary session types [12].

The Scribble [47] framework provides a language to specify global protocols (our global types), relying on asynchronous and reliable message transports such as TCP. It checks well-formedness of a global specification, subsequently projecting them into the participants to achieve the local protocols (our end-point types). The features along with constructs that Scribble supports are the ones of MST. The end-point types are captured into the static type system of a programming language to benefit from the framework of session types. This is a well-established, practical work that certainly can be a very good starting point for implementing CMST into practice.

Montesi and Yoshida [31] provide a study on how to compose global description to ease the programming effort on roles (our processes), i.e., one does not have to implement a new role for each global type if that behaviour of role is similar in more than one global type. Their approach is based on a language of (partial) choreographies, that communicate through "send" and "receive" of end-point. The choreography language and the typing discipline, i.e., typing the language without using local types, extends the one of Montesi and Carbone [11] with end-point's "send" and "receive", evoking the direction of conversation types [8] (discussed below). A key feature is the mobility of shared channels, used to compose two protocols. An implementation is provided, unfortunately not in Scribble, but in Jolie [30]—a service-oriented programming language. Our work, like that of Montesi and Carbone, attempts to ease the programming effort; however, the respective strategies are on two different levels. Our work is focussed on making programming easy for a single protocol on a wide range of areas. Further we can investigate a language of choreographies in the presence of summation, matching and functional behaviour.

Interruptions (over processes) to Scribble specified in Python are introduced in [16, 23]. In contrast to our work, they do not capture partial failure handling, i.e., they notify also the participants not affected by a failure, and they request an agreement of all participants for the termination of the `try-catch`, i.e., reasoning about termination adds extra communications that in our system are not needed. Deadlock-freedom is supported through runtime monitoring to ensure conformance between end-point and global specification. Recently, Neykova and Yoshida [33] present an implementation of Erlang's crash recovery model into MST extended with parameterised sessions. In contrast to the other works [16, 23], the model of Neykova and Yoshida notifies only the affected participants. However, the lack of a failure construct allows handling of only crashes and not application or system failures.

Neykova, Yoshida and Hu [34] present runtime verification of distributed Python programs against Scribble specifications. Due to the presence of information at runtime, verification of assertions is more expressive than the one of DML, consequently of our system. Runtime monitoring requires less language design to incorporate session type into a programming language but does not provide the advantages of static type-checking as MST was originally defined.

Pabble [35] extends Scribble with parameterised roles and primitive recursion sessions. It implements the typing methodology adopted in [46] the pitfalls of which, relative to our system, are discussed in Sections 1 and 2. Prior to that work, a framework to describe session programming in C and global specification in Scribble is provided in [36]. The work is limited only to features of LMS and MST.

**Behavioural Types.** Contracts [13, 14], conversation types [8, 42] and generic types [24, 25, 26] type processes rather than channels as session types. Contracts are analogous to binary session types. Conversations typecheck processes through merging of global and process types rather than projection. Generic types type processes by keeping the frequency and order of use in the input-output actions of a process. Progress in these systems also guarantees dependencies between different sessions.

Furthermore, the type systems in the literature cannot describe and verify assertions on interactions, parameterised participants, primitive recursion and partial failure. They are not based on unshared, bidirectional channels, making these theories harder to apply to abstractions used in real world distributed systems.

## 11 Conclusion and Future Work

In years of productive research in session types and languages we have seen the emergence of many different systems. We reported observations about redundancies between these variants in terms of the features supported and the constructs used to model them, which lead to the question: *What are the essential features for MST and LMS and and how can these be modelled with simple constructs?*

We addressed this question by defining a set of essential features that can be extended with new ones by either adding new essential features with minimal efforts or layering non-essential ones atop. This setting allows the evolution of session types and languages, and prohibits efforts from unraveling. We introduced the *language of structured interactions* (LoSI) and the type theory of *comprehensive multiparty session types* (CMST), including global types and a type system. Both add a fundamental contribution to the formal study and verification of structured interactions for various kinds of distributed concurrent applications, with a better ratio of constructs (both language and types) versus the features supported than the original and its main variants. Conciseness is not limited to the language and types constructs, but also in fewer rules for operational semantics, typing, and proofs of properties. The key benefits of our system include not only consolidation, but also its ability to easily study advanced features such as local variables and the dynamic joining and leaving of participants. We are currently investigating a number of extensions and refinements: (a) extend abstract syntax trees of our global types with forking and merging of sessions in the presence of assertions and essential behaviour, (b) extend LoSI and CMST to support system failure, i.e., failures thrown by the underlying components of the system, e.g. network, operating system, and (c) implement a safe communication "library" of LoSI and global types constructs to devise a library for a mainstream class-based language.

**Acknowledgements** We are grateful to the anonymous reviewers of Programming for their comments on an earlier version of this paper. We would like to thank Timothy Hagen for his feedback on the presentation of this work.

## A Auxiliary Definitions

In this section, we give the definition of evaluation contexts of processes along with evaluation rules for arithmetical and boolean expressions, free names in processes, participants in a global type, free variables in a global type, and simplification of arithmetical expressions.

**Definition 26** (Contexts). *The evaluation contexts of expressions, written $\mathcal{E}$ with the hole "[]", are defined as:*

$$\begin{array}{rl} \mathcal{E}[_,\ldots,_] ::= & \mathtt{init}(u : G, [_]).\mathcal{E}[_,\ldots,_] \mid u[_,_]!\langle _ {:} S\rangle.\mathcal{E}[_,\ldots,_] \\ & \mid u[_,_]?(_ {:} S).\mathcal{E}[_,\ldots,_] \mid \mathtt{rec}\ X = \mathcal{E}[_,\ldots,_] \\ & \mid \mathtt{new}\ (a{:}G)\mathcal{E}[_,\ldots,_] \mid \mathcal{E}[_,\ldots,_]\mathcal{E}[_,\ldots,_] \\ & \mid \mathcal{E}[_,\ldots,_]{+}\mathcal{E}[_,\ldots,_] \mid [_]\mathcal{E}[_,\ldots,_] \\ & \mid \mathit{fn}\ x : I \Rightarrow \mathcal{E}[_,\ldots,_] \mid \mathcal{E}[_,\ldots,_]\ _ \end{array}$$

**Definition 27** (Reduction of expressions). *The reduction of arithmetic and boolean expressions in processes is defined as follows*

$$e_i \downarrow e_i' \Rightarrow \mathcal{E}[e_0, \ldots, e_i, \ldots, e_n] \downarrow \mathcal{E}[e_0, \ldots, e_i', \ldots, e_n]$$
$$b_i \downarrow b_i' \Rightarrow \mathcal{E}[b_0, \ldots, b_i, \ldots, b_n] \downarrow \mathcal{E}[b_0, \ldots, b_i', \ldots, b_n]$$

and named [ARed] and [BRed] respectively, and the relation $\downarrow$ denotes reduction of both arithmetic and boolean expressions.

**Definition 28** (Free names). *The set of free names of a process $P$, written $\mathit{fn}(P)$, is defined as follows:*

$$\begin{array}{rcl} \mathit{fn}(\mathtt{init}(u : G, P).P) &=& \mathit{fn}(u) \cup \mathit{fn}(P) \\ \mathit{fn}(u[p,q]!\langle m : S\rangle.P) &=& \mathit{fn}(u) \cup \mathit{dn}(m) \cup \mathit{fn}(P) \\ \mathit{fn}(u[p,q]?(w : S).P) &=& \mathit{fn}(u) \cup \mathit{fn}(P) \setminus \mathit{dn}(w) \\ \mathit{fn}(\mathtt{rec}\ X = P) &=& \mathit{fn}(P) \\ \mathit{fn}(X) &=& \emptyset \\ \mathit{fn}(0) &=& \emptyset \\ \mathit{fn}((\mathtt{new}\ a : G)P) &=& \mathit{fn}(P) \setminus \mathit{fn}(a) \\ \mathit{fn}(P \mid Q) &=& \mathit{fn}(P) \cup \mathit{fn}(Q) \\ \mathit{fn}(P{+}Q) &=& \mathit{fn}(P) \cup \mathit{fn}(Q) \\ \mathit{fn}([b]P,\ \mathit{fn}\ i : I \Rightarrow P,\ P\ e) &=& \mathit{fn}(P) \\ \mathit{fn}(a{:}h) &=& \mathit{fn}(h) \\ \mathit{dn}(x) &=& \emptyset \\ \mathit{dn}(a,\ a[p]) &=& \{a\} \end{array}$$

**Definition 29** (Participants in a global type). *The set of participants in a global type $G$ and not in its carried types, written $\mathsf{pid}(G)$, is defined as follows:*

$$\begin{array}{rcl} \mathsf{pid}(p \to p' : \langle V\rangle.G) &=& \{p,\ p'\} \cup \mathsf{pid}(G) \\ \mathsf{pid}(\mu X.G) &=& \mathsf{pid}(G) \\ \mathsf{pid}(X, \mathsf{end}) &=& \emptyset \\ \mathsf{pid}(G, G') &=& \mathsf{pid}(G) \cup \mathsf{pid}(G') \\ \mathsf{pid}(G{+}G') &=& \mathsf{pid}(G) \cup \mathsf{pid}(G') \\ \mathsf{pid}([b]G, \Pi x : I.G, G\ e) &=& \mathsf{pid}(G) \end{array}$$

**Definition 30** (Defined variables in a sort). *The set of defined variables in a sort, written $\mathsf{dv}(V)$, is defined as follows:*

$\mathsf{dv}(S) = \emptyset \qquad \mathsf{dv}(x : I) = \{x\}$

**Definition 31** (Defined variables in an arithmetic expression). *The set of defined variables in an arithmetic expression, written* $\mathsf{dv}(e)$*, is defined as follows:*
$\mathsf{dv}(x) = \{x\}$ $\quad\mathsf{dv}(\mathsf{c}) = \emptyset$ $\quad\mathsf{dv}(e + e') = \mathsf{dv}(e) \cup \mathsf{dv}(e')$ $\quad\mathsf{dv}(\mathsf{c} * e) = \mathsf{dv}(e)$

**Definition 32** (Defined variables in a boolean expression). *The set of defined variables in a boolean expression, written* $\mathsf{dv}(b)$*, is defined as follows:*
$\mathsf{dv}(\mathsf{true}, \mathsf{false}) = \emptyset$ $\quad\mathsf{dv}(b \text{ and } b', b \text{ or } b') = \mathsf{dv}(b) \cup \mathsf{dv}(b')$ $\quad\mathsf{dv}(\text{not } b) = \mathsf{dv}(b)$ $\quad\mathsf{dv}(e < e', e = e') = \mathsf{dv}(e) \cup \mathsf{dv}(e')$

**Definition 33** (Defined variables in a predicate expression). *The set of defined variables in a predicate expression, written* $\mathsf{dv}(\Theta)$*, is defined as follows:*
$\mathsf{dv}(e \leq e') = \mathsf{dv}(e) \cup \mathsf{dv}(e')$ $\quad\mathsf{dv}(\Theta \text{ and } \Theta') = \mathsf{dv}(\Theta) \cup \mathsf{dv}(\Theta')$

**Definition 34** (Free variables in an index sort). *The set of free variables in an index sort, written* $\mathsf{fv}(I)$*, is defined as follows:*
$\mathsf{fv}(\mathsf{nat}) = \emptyset$ $\quad\mathsf{fv}(\{x : I \mid \Theta\}) = \mathsf{dv}(\Theta) \cup \mathsf{fv}(I) \setminus \{x\}$

**Definition 35** (Free variables in a global type). *The set of free variables in a global type* $G$*, written* $\mathsf{fv}(G)$*, is defined as follows:*

$$\begin{array}{rcl}
\mathsf{fv}(p \to p' : \langle V \rangle.G) & = & \mathsf{dv}\ (p) \cup \mathsf{dv}\ (p') \cup \mathsf{fv}(G) \setminus \mathsf{dv}(V) \\
\mathsf{fv}(\mu X.G) & = & \mathsf{fv}(G) \setminus \{X\} \\
\mathsf{fv}(X) & = & \{X\} \\
\mathsf{fv}(\mathsf{end}) & = & \emptyset \\
\mathsf{fv}(G, G') & = & \mathsf{fv}(G) \cup \mathsf{fv}(G') \\
\mathsf{fv}(G + G') & = & \mathsf{fv}(G) \cup \mathsf{fv}(G') \\
\mathsf{fv}([b]G) & = & \mathsf{dv}(b) \cup \mathsf{fv}(G) \\
\mathsf{fv}(\Pi x : I.G) & = & \mathsf{fv}(G) \cup \mathsf{fv}(I) \setminus \{x\} \\
\mathsf{fv}(G\ e) & = & \mathsf{fv}(G) \cup \mathsf{dv}(e)
\end{array}$$

The defined names of a participant $p$ is defined as $\mathsf{dv}(\mathsf{Worker}) = \emptyset$ and $\mathsf{dv}(p[e]) = \mathsf{dv}(p) \cup \mathsf{dv}(e)$.

**Definition 36** (Simplification of arithmetic expressions). *Two arithmetic expressions are transformed, written* $e_1 := e_2 \mapsto e_1' := e_2'$ *according to the following rules:*

- $x := e$
- $\mathsf{c} * e := \mathsf{c} * \mathsf{c}' * e' \mapsto e := \mathsf{c}' * e'$
- $e + e' := e'' \mapsto e := e'' - e'$ *where* $\mathsf{dv}\ (e') = \emptyset$

In this case the function $\mathsf{dv}\ (e)$ returns the set of defined variables in arithmetic expressions. The simplification of two arithmetic expressions is written $e_1 := e_2 \mapsto e_1' := e_2'$ following arithmetic operations; $\mapsto^\star$ denotes a number of simplifications up to an irreducible expression, i.e., a variable $x$.

**Definition 37** (Substitution on $e$). *Suppose that* $x \in \mathsf{dv}(e')$ *and* $x \in \mathsf{dv}(e)$*. Substitution on arithmetic expressions, written* $e\{e''/e'\}$*, is defined as:*

$$e\{e''/e'\} = \begin{cases} e\{e'''/x\} & e' := e'' \mapsto^\star x := e''' \\ e & \textit{otherwise} \end{cases}$$

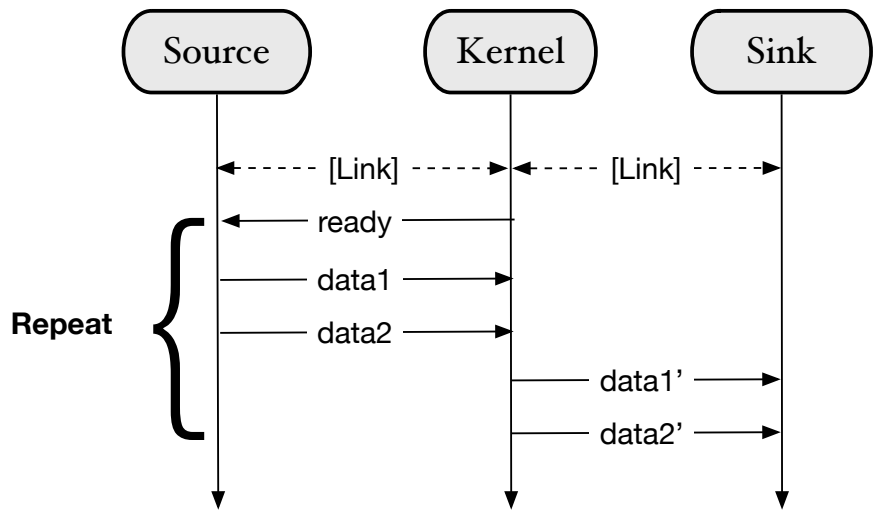

■ **Figure 10** Double-buffer protocol diagram

## B Double-buffering algorithm

A series of data chunks are transferred from a source (Source) to a transformer (Kernel) and subsequently delivered to a sink (Sink) [32] (see Figure 10). The data chunks are transformed in the kernel. The session starts with an infinite loop, where the kernel notifies the source that it is ready to receive data, and the source complies by sending two chunks of data in sequence. Then the kernel internally processes data and sends the two chunks of processed data to the sink.

The action of the kernel –sending the first chunk to the sink– can be placed before the reception of the second chunk from the source. This way, the kernel can process and send to the sink the first chunk that it has received while receiving the second chunk, thus reducing latency when large amounts of data are processed. Below we provide the process of Kernel; $G_D$ denotes the global type.

$$\text{Kernel} \triangleq \mathit{init}(a{:}G_D,2).\mathit{rec}\ X = a[2,1]!\langle \mathit{ready} : \mathit{Signal}\rangle.a[1,2]?(x : \mathit{Data}).\\ a[2,3]!\langle x' : \mathit{Data}\rangle.a[1,2]?(y : \mathit{Data}).a[2,3]!\langle y' : \mathit{Data}\rangle.X$$

By the subtyping relation, this pipelined implementation which optimises the efficiency of the algorithm is allowed.

## C Reduction Steps

This section gives the reduction steps of the examples given in Section 3.3.

**Web service** The reduction steps are given below where the label on the reduction relation $\longrightarrow$ is the name of the operational semantics rule applied. The first step establishes a session between the three participants, including the creation of a queue. In the following two steps, the client appends a message in the queue and reciprocally, the proxy removes and substitutes it in its processes scope. The other two consequent steps define the proxy choosing and sending the forward message to the web service, and the web service branching and receiving the forward message. To finish, the last message exchanged between the web service and client takes place, following the three processes that can do nothing.

$$
\begin{aligned}
&initialisation(a{:}G_W,1).a[1,2]!\langle req:Request\rangle.a[3,1]?(x:Reply).0 \mid\\
&initialisation(a{:}G_W,2).a[1,2]?(x:Request).(a[2,3]!\langle for:Forward\rangle.0 +\\
&\qquad a[2,3]!\langle audit:Audit\rangle.a[3,2]?(x:Details).0) \mid\\
&initialisation(a{:}G_W,3).(a[2,3]?(x:Forward).a[3,1]!\langle rep:Reply\rangle.0 +\\
&\qquad a[2,3]?(x:Audit).a[3,2]!\langle det:Details\rangle.\\
&\qquad a[3,1]!\langle rep:Reply\rangle.0) \rightarrow_{[\text{Link}]}\\
&(\textit{new}\ a:G_W)(a[1,2]!\langle req:Request\rangle.a[3,1]?(x:Reply).0 \mid\\
&\qquad (a[1,2]?(x:Request).(a[2,3]!\langle for:Forward\rangle.0 +\\
&\qquad a[2,3]!\langle audit:Audit\rangle.a[3,2]?(x:Details).0) \mid\\
&\qquad (a[2,3]?(x:Forward).a[3,1]!\langle rep:Reply\rangle.0 +\\
&\qquad a[2,3]?(x:Audit).a[3,2]!\langle det:Details\rangle.\\
&\qquad a[3,1]!\langle rep:Reply\rangle.0) \mid a:\emptyset) \rightarrow_{[\text{Send, Par, Scope}]}\\
&(\textit{new}\ a:G_W)(a[3,1]?(x:Reply).0 \mid\\
&\qquad (a[1,2]?(x:Request).(a[2,3]!\langle for:Forward\rangle.0 +\\
&\qquad a[2,3]!\langle aud:Audit\rangle.a[3,2]?(x:Details).0) \mid\\
&\qquad (a[2,3]?(x:Forward).a[3,1]!\langle rep:Reply\rangle.0 +\\
&\qquad a[2,3]?(x:Audit).a[3,2]!\langle det:Details\rangle.\\
&\qquad a[3,1]!\langle rep:Reply\rangle.0) \mid a:\langle 1,2,req:Request\rangle)\\
&\rightarrow_{[\text{Recv, Par, Scope}]}\\
&(\textit{new}\ a:G_W)(a[3,1]?(x:Reply).0 \mid\\
&\qquad (a[2,3]!\langle for:Forward\rangle.0 +\\
&\qquad a[2,3]!\langle aud:Audit\rangle.a[3,2]?(x:Details).0) \mid\\
&\qquad (a[2,3]?(x:Forward).a[3,1]!\langle rep:Reply\rangle.0 +\\
&\qquad a[2,3]?(x:Audit).a[3,2]!\langle det:Details\rangle.\\
&\qquad a[3,1]!\langle rep:Reply\rangle.0) \mid a:\emptyset) \rightarrow_{[\text{Send, Sum, Scope}]}\\
&(\textit{new}\ a:G_W)(a[3,1]?(x:Reply).0 \mid\\
&\qquad 0 \mid\\
&\qquad (a[2,3]?(x:Forward).a[3,1]!\langle rep:Reply\rangle.0 +\\
&\qquad a[2,3]?(x:Audit).a[3,2]!\langle det:Details\rangle.\\
&\qquad a[3,1]!\langle rep:Reply\rangle.0)\\
&\qquad a:\langle 2,3,for:Forward\rangle) \rightarrow_{[\text{Recv, Par, Scope}]}\\
&(\textit{new}\ a:G_W)(a[3,1]?(x:Reply).0 \mid 0 \mid a[3,1]!\langle rep:Reply\rangle.0 \mid\\
&\qquad a:\emptyset) \rightarrow_{[\text{Send, Recv, Par, Str, Scope}]} 0
\end{aligned}
$$

According to the structural congruence rule $P \mid 0 \equiv P$, three processes than can do nothing in parallel are congruent to the process that can do nothing. Also according to the congruence rule $(\textit{new}\ a{:}G_W)0 \equiv 0$, a session with a process that can do nothing is congruent to the process that can do nothing.

**Double-buffering algorithm** The reduction steps of the algorithm do not use any reduction rule not presented from above. So for presentation reasons, we omit them.

**Financial protocol** The reduction steps are given below. The first step establishes a session between the three participants, including the creation of a queue. After that the buyer sends an offer to the seller and the parameter $rec$ is replaced with the argument 1, and also the inner arithmetic, boolean expressions are evaluated ($1+1=2$ and $1<5$=true ). According to the reduction rules, we have that $[\text{true}]P \rightarrow P$. From here, the first case evaluates to true (100<120) and the other to false (not 100<120). According to reduction rules we have that $[\mathsf{false}]P[\mathsf{true}]P' \to P'$.

$$\mathit{Financial\ Protocol} \to_{[\text{Link}]}$$

$$(new\ a : G_F)(Buyer \mid Seller \mid Bank \mid a : \emptyset) \to_{[\text{Send, Rcv, Scope}]}$$

$$\begin{aligned}
&(new\ a : G_F)(recX = fn\ iter : \mathsf{nat} \Rightarrow [iter < 5]\\
&\qquad([100 < 120](a[2,1]?(x : Neg).a[1,2]!\langle v_0 : I\rangle.Buyer\ 2\ +\\
&\qquad[\mathsf{not}\ 100 < 120]a[2,1]?(x : Ok).a[1,3]!\langle v_0 : I\rangle.0))\ 1\\
&\qquad \mid rec\ X = fn\ iter : \mathsf{nat} \Rightarrow [\mathrm{iter} < 5]\\
&\qquad([100 < 120]a[2,1], a[2,3]!\langle neg : Neg\rangle.a[1,2]?(v_0 : I).\\
&\qquad\qquad Seller\ 2\ +\\
&\qquad[\mathsf{not}\ 100 < 120]a[2,1], a[2,3]!\langle ok : Ok\rangle.a[3,2]?(x : Ack).0))\ 1\\
&\qquad \mid rec\ X = fn\ iter : \mathsf{nat} \Rightarrow [\mathrm{iter} < 5](a[2,3]?(x : Neg).Bank\ 2\ +\\
&\qquad\quad a[2,3]?(x : Ok).a[1,3]?(v_0 : I).a[3,2]!\langle ack : Ack\rangle.0)\ 1\\
&\quad \mid a : \emptyset) \to_{[\text{Str, App, AEval, Scope}]}
\end{aligned}$$

$$\begin{aligned}
&(new\ a : G_F)([1 < 5]\\
&\qquad([100 < 120](a[2,1]?(x : Neg).a[1,2]!\langle v_0 : I\rangle.Buyer\ 2\ +\\
&\qquad[\mathsf{not}\ 100 < 120]a[2,1]?(x : Ok).a[1,3]!\langle v_0 : I\rangle.0))\\
&\qquad \mid [1 < 5]\\
&\qquad([100 < 120]a[2,1], a[2,3]!\langle neg : Neg\rangle.a[1,2]?(v_0 : I).\\
&\qquad\qquad Seller\ 2\ +\\
&\qquad[\mathsf{not}\ 100 < 120]a[2,1], a[2,3]!\langle ok : Ok\rangle.a[3,2]?(x : Ack).0))\\
&\qquad \mid [1 < 5](a[2,3]?(x : Neg).Bank\ 2\ +\\
&\qquad\quad a[2,3]?(x : Ok).a[1,3]?(v_0 : I).a[3,2]!\langle ack : Ack\rangle.0)\\
&\quad \mid a : \emptyset) \to_{[\text{MatchT, Scop}]}
\end{aligned}$$

$$\begin{aligned}
&(new\ a : G_F)(([100 < 120](a[2,1]?(x : Neg).a[1,2]!\langle v_0 : I\rangle.Buyer\ 2\ +\\
&\qquad[\mathsf{not}\ 100 < 120]a[2,1]?(x : Ok).a[1,3]!\langle v_0 : I\rangle.0))\\
&\qquad \mid ([v_0 < 120]a[2,1], a[2,3]!\langle neg : Neg\rangle.a[1,2]?(v_0 : I).\\
&\qquad\qquad Seller\ 2\ +\\
&\qquad[\mathsf{not}\ 100 < 120]a[2,1], a[2,3]!\langle ok : Ok\rangle.a[3,2]?(x : Ack).0))\\
&\qquad \mid (a[2,3]?(x : Neg).Bank\ 2\ +\\
&\qquad\quad a[2,3]?(x : Ok).a[1,3]?(v_0 : I).a[3,2]!\langle ack : Ack\rangle.0)\\
&\quad \mid a : \emptyset) \to_{[\text{BEval, MatchT, Scop}]}
\end{aligned}$$

$$\begin{aligned}
&(new\ a : G_F)(a[2,1]?(x : Neg).a[1,2]!\langle v_0 : I\rangle.Buyer\ 2\\
&\qquad \mid a[2,1], a[2,3]!\langle neg : Neg\rangle.a[1,2]?(v_0 : I).\\
&\qquad\qquad Seller\ 2\\
&\qquad \mid (a[2,3]?(x : Neg).Bank\ 2\ +\\
&\qquad\quad a[2,3]?(x : Ok).a[1,3]?(v_0 : I).a[3,2]!\langle ack : Ack\rangle.0)\\
&\quad \mid a : \emptyset) \to_{[\text{Send, Scop}]}
\end{aligned}$$

The remaining reductions regard choice and branching, which are illustrated in the Web Service example. So for presentation reason, we omit them in here.

**N-body problem (Ring)** The reduction steps of the Ring pattern for $n = 3$ are given below. The first step invokes the main program by replacing the parameter $n$ with the value 3 (rule [App]); the second step reduces to a natural number the participant expression in *Last*, namely in the receive action (rule [AEval]). At this point, *Starter* and *Last* are instantiated to running processes. In the next steps, $i$ is replaced by 2 and *Middle* is instantiated, replacing the process variable $X$.

$$
\begin{array}{l}
\mathit{Ring}\ 3 \rightarrow_{[\text{App, AEval}]} \\
(\mathit{rec}\ X = \mathit{fn}\ i : \mathsf{nat} \Rightarrow [i{=}3]\ \mathtt{init}(a{:}\langle G_R, \mathtt{W}[1]).a[\mathtt{W}[1],\mathtt{W}[2]]!\langle v : D[\,]\rangle. \\
\qquad a[\mathtt{W}[3],\mathtt{W}[1]]?(z : D[\,]).R \mid \\
\qquad \mathtt{init}(a{:}\langle G_R, \mathtt{W}[3]).a[\mathtt{W}[2],\mathtt{W}[3]]?(z : U).a[\mathtt{W}[3],\mathtt{W}[1]]!\langle z : D[\,]\rangle.S \\
\quad + [i < 3]\ \mathit{Middle(i)} \mid X\ i+1)\ 2 \rightarrow_{[\text{App, MatchT, AEval}]} \\
\mathtt{init}(a{:}\langle G_R, \mathtt{W}[2]).a[\mathtt{W}[1],\mathtt{W}[2]]?(z : D[\,]).a[\mathtt{W}[2],\mathtt{W}[3]]!\langle z : D[\,]\rangle.R' \mid \\
(\mathit{rec}\ X = \mathit{fn}\ i : \mathsf{nat} \Rightarrow [i{=}3]\ \mathtt{init}(a{:}\langle G_R, \mathtt{W}[1]).a[\mathtt{W}[1],\mathtt{W}[2]]!\langle v : D[\,]\rangle. \\
\qquad a[\mathtt{W}[3],\mathtt{W}[1]]?(z : D[\,]).R \mid \\
\qquad \mathtt{init}(a{:}\langle G_R, \mathtt{W}[3]).a[\mathtt{W}[2],\mathtt{W}[3]]?(z : D[\,]). \\
\qquad a[\mathtt{W}[3],\mathtt{W}[1]]!\langle z : D[\,]\rangle.S \\
\quad + [i < 3]\ \mathit{Middle(i)} \mid X\ i+1)\ 3 \rightarrow_{[\text{App, MatchT, AEval, Par}]} \\
\mathtt{init}(a{:}\langle G_R, \mathtt{W}[2]).a[\mathtt{W}[1],\mathtt{W}[2]]?(z : D[\,]).a[\mathtt{W}[2],\mathtt{W}[3]]!\langle z : D[\,]\rangle.R' \mid \\
\mathtt{init}(a{:}\langle G_R, \mathtt{W}[1]).a[\mathtt{W}[1],\mathtt{W}[2]]!\langle v : D[\,]\rangle.a[\mathtt{W}[3],\mathtt{W}[1]]?(z : D[\,]).R \mid \\
\mathtt{init}(a{:}\langle G_R, \mathtt{W}[3]).a[\mathtt{W}[2],\mathtt{W}[3]]?(z : D[\,]).a[\mathtt{W}[3],\mathtt{W}[1]]!\langle z : D[\,]\rangle.S \\
\rightarrow_{[\text{Link}]} \cdots
\end{array}
$$

The remaining reduction are similar to the ones in the Web Service, so we omit them for presentation reasons: a session of Ring of length three starts between the running processes, including the creation of a queue with the same identifier; the subsequent steps are intuitive: appending messages in the queue for a “send” and reciprocally, removing and substituting them in processes scope for a “receive”.

## D Double-buffering global type

The infinite loop is modelled through the $\mu$ operator and actions are composed in sequence as

$$
\begin{array}{l}
G_D \triangleq \mu X.K \to So : \langle Signal\rangle.So \to K : \langle Data\rangle.So \to K : \langle Data\rangle.K \to Si : \langle Data\rangle. \\
\qquad K \to Si : \langle Data\rangle.X
\end{array}
$$

where *K, So, Si* denote the kernel, source, and sink. Commutativity is achieved at the end-point types by subtyping to check the permutation of output prefixes.

## E 2D-Mesh: Global Type and End-point Types

To illustrate the expressive power of this system, we provide also the global type of the two-dimensional mesh (2D-Mesh). It organises workers in a grid (see diagram below) of size $n{\times}m$ where $n$ and $m$ represent respectively the number of rows and columns ($n, m \geq 1$). Variants of this pattern include toric meshes and hypercubes. The session starts with participant $\mathtt{W}[0][0]$ (top left corner of the diagram) sending to its neighbours to its left and below it. This behaviour along with the receives from the neighbours on right and top is repeated by all other processes , except the ones in the rightmost column and bottom row. Participants in the rightmost column only send to neighbours below, and participants in the bottom row only send to neighbours to the left.

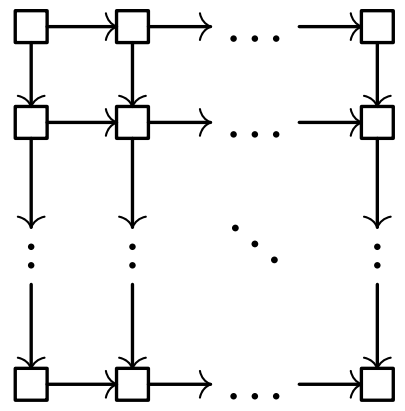

$$\begin{aligned} G \triangleq\ & \Pi i : J.\Pi j : K. \\ & \mathtt{W}[i][j] \to \mathtt{W}[i+1][j] : \langle \mathsf{nat} \rangle. \\ & \mathtt{W}[i][j] \to \mathtt{W}[i][j+1] : \langle \mathsf{nat} \rangle. \\ & \mathtt{W}[i][m] \to \mathtt{W}[i+1][m] : \langle \mathsf{nat} \rangle. \\ & \mathtt{W}[n][j] \to \mathtt{W}[n][j+1] : \langle \mathsf{nat} \rangle.\mathsf{end} \end{aligned}$$

In our system (right-hand side), we use two index variables through the $\Pi$ construct to express the family of all instances of the 2D mesh and, defining the lower bound of indices to 0 to specify participant $\mathtt{W}[0][0]$ starting the session and greater bound to $n-1$ and $m-1$, where $J = \{i' : \mathsf{nat} | i' < n\}$ and $K = \{j' : \mathsf{nat} | j' < m\}$. The adjunct global type specifies that the first message sent is by $\mathtt{W}[0][0]$ and the last one by $\mathtt{W}[n][m-1]$. Interactions $\mathtt{W}[i][j] \to \mathtt{W}[i+1][j] : \langle \mathsf{nat} \rangle.\mathtt{W}[i][j] \to \mathtt{W}[i][j+1] : \langle \mathsf{nat} \rangle$ specify the fact that each worker not situated in the bottom row or rightmost column sends a message to its neighbours situated below and to its left. Interactions $\mathtt{W}[i][m] \to \mathtt{W}[i+1][m] : \langle \mathsf{nat} \rangle$ and $\mathtt{W}[n][j] \to \mathtt{W}[n][j+1] : \langle \mathsf{nat} \rangle$ describe the messages sent in the rightmost column and bottom row. Given a value for $n$ and $m$, iteration starts by applying the value 0 to $i$ and $j$.

**End-point types** The end-point type for principal $\mathtt{W}[i][j]$ in the mesh global type is

$$\begin{aligned} G \upharpoonright^{n,m:I}_{\mathtt{W}[i][j]} = \ & [\mathtt{W}[i][j], \mathtt{W}[i+1][j]]!\langle \mathsf{nat} \rangle.[\mathtt{W}[i-1][j], \mathtt{W}[i][j]]?(\mathsf{nat}). \\ & [\mathtt{W}[i][j], \mathtt{W}[i][j+1]]!\langle \mathsf{nat} \rangle.[\mathtt{W}[i][j-1], \mathtt{W}[i][j]]?(\mathsf{nat}).\mathsf{end} \end{aligned}$$

where the prefixes are ordered according to the definition of projection: the output followed by the input composed in sequence with other prefixes. From sorting we have

$$\begin{aligned} \mathsf{sort}(G \upharpoonright^{n,m:I}_{\mathtt{W}[i][j]}) = \ & [\mathtt{W}[i-1][j], \mathtt{W}[i][j]]?(\mathsf{nat}).[\mathtt{W}[i][j-1], \mathtt{W}[i][j]]?(\mathsf{nat}). \\ & [\mathtt{W}[i][j], \mathtt{W}[i+1][j]]!\langle \mathsf{nat} \rangle.[\mathtt{W}[i][j], \mathtt{W}[i][j+1]]!\langle \mathsf{nat} \rangle.\mathsf{end} \end{aligned}$$

where the input prefixes are placed before the output; in particular, the first input is the one capturing the reception from the neighbour above, the second input is the one from the left; the output prefix captures the send to respectively to below and the right. This end-point type reflects the ordering of prefixes as in the global type.

## F Examples of coherence

The following global types are not coherent since their projections are not defined or they are not matching coherent. We consider cases where side conditions fail.

$$p \to p' : \langle x : \{x' : \mathsf{nat} | x' < 0\} \rangle.G$$

$$[p,q]!\langle V\rangle.[q',p]?(V').T \ll [q',p]?(V').[p,q]!\langle V\rangle.T \quad \lfloor\text{OI}\rfloor$$

$$[p,q]!\langle V\rangle.[p,q']!\langle V'\rangle.T \ll [p,q']!\langle V'\rangle.[p,q]!\langle V\rangle.T \quad \lfloor\text{OO}\rfloor$$

$$[p,q]?(V).[p',q]?(V').T \ll [p',q]?(V').[p,q]?(V).T \quad \lfloor\text{II}\rfloor$$

$$\frac{T \ll T'}{[p,q]!\langle V\rangle.T \ll [p,q]!\langle V\rangle.T'}\ \lfloor\text{Out}\rfloor \qquad \frac{T \ll T'}{[p,q]?(V).T \ll [p,q]?(V).T'}\ \lfloor\text{In}\rfloor$$

$$\frac{T_1 \ll T_1' \quad T_2 \ll T_2' \quad T_1' \asymp_{B,NP,P} T_2'}{T_1 + T_2 \ll T_1' + T_2'}\ \lfloor\text{Sum}\rfloor \qquad \frac{T \ll T'}{[b]T \ll [b]T'}\ \lfloor\text{Mat}\rfloor$$

$$\text{end} \ll \text{end}\ \lfloor\text{End}\rfloor \qquad \mu X.T \ll \mu X.T\ \lfloor\text{Rec}\rfloor$$

$$\frac{T \ll T' \quad I \ll I'}{\Pi x : I.T \ll \Pi x : I'.T'}\ \lfloor\text{Pro}\rfloor \qquad \frac{T \ll T'}{T\ e \ll T'\ e}\ \lfloor\text{App}\rfloor \qquad \frac{T_1 \ll T_2 \quad T_2 \ll T_3}{T_1 \ll T_3}\ \lfloor\text{Tra}\rfloor$$

**Figure 11** Prefix-subtype relation of end-point types.

Projection of the global type above onto $p$ and $p'$ is not defined, since the side condition $\emptyset \vdash \{x' : \text{nat}|x' < 0\} \neq \emptyset$ does not hold. Intuitively, an interaction value is neither defined nor null. This is similar also for product global types. In the following scenario, projection fails because, intuitively, an arithmetical expression is applied only to the product global types.

$p \to p' : \langle U\rangle.G\ 5$

The next global type assigns a value that is smaller than the expected one. Thus, the projection fails on any participant of $G$.

$\Pi x : \{x' : \text{nat}|x' > 3\}.G\ 3$

The next global type does not consist of sequences of interactions where the first prefix for each branch expresses the same event (sending or receiving) or sequences guarded by a boolean expression. Thus it is rejected by the compatibility operation when projecting it onto every participant.

$\Pi x : I.G + \Pi x : I'.G'$

Finally the next global type is not *matching coherent*. $x$ defines the boolean expression but is not visible to $p$, $p'$ and any $p''\in\text{pid}(G)$.

$[x > 5]p \to p' : \langle U\rangle.G$

# G Subtyping of Sessions

**Partial commutativity of actions** is added [32] for commuting asynchronous outputs to optimise the performance of stream processing and multicore computations. The language and types remain as in the original work. The subtyping relation on prefixes of end-point types is defined over named channels ($k$) and label branching following the traditional subtyping relation of binary session types [19]. However, the key contribution is the prefix asynchronous output subtyping of end-point types that allows the permutation of an input with an output prefixes of different channels. The rules do not only capture the combinations of output-input prefixes for values but also for labels (selection-branching) and mixed combinations.

Figure 11 provides the axioms and rules defining the prefix-subtyping relation for end-point types, written $T \ll T'$ and read "*end-point type $T$ is a prefix-subtype of $T'$*", meaning that $T$ is more optimised (asynchronous) than $T'$. The rules allow permutation of output-input, output-output, and input-input prefixes on different channels and types only if the permuted end-point type are coherent; e.g., the end-point type $[p,q]?(x : \mathsf{nat}).[p',q]?(y : z : \mathsf{nat}|z < x).T$ is not permuted to $[p',q]?(y : z : \mathsf{nat}|z < x).[p,q]?(x : \mathsf{nat}).T$ since the result is not coherent, i.e., the side condition $\emptyset \vdash z : \mathsf{nat}|z < x \neq \emptyset$ does not hold (well-formedness of end-point types follows that of global types, Definition 13; this is elided for brevity). Input-output permutation is prohibited to prevent well-typing of deadlocked processes: given three end-point types $[p,q]!\langle V\rangle.[p',p]?(V').T$, $[p,q]?(V).[q,p']!\langle V''\rangle.T$, and $[q,p']?(V'').[p',p]!\langle V'\rangle.T$, the permutation of the first type through a (IO) rule would result in $[p',p]?(V').[p,q]!\langle V\rangle.T$. This type in parallel with the others would result in a well-typed, deadlocked process. The rules also allow permutation of prefixes in branches only if they preserve the compatibility of the branches. This is done to prevent well-typing of wrong processes such as a branch over an input composed in disjunction with others prefixed by an output, e.g., $[q',p]?(V).T + [p,q]!\langle V_1\rangle.T_{11} + ... + [p,q]!\langle V_k\rangle.T_{1k}$. These rules allow typing of processes that permute actions to achieve better performance, while retaining typing soundness.

**Proposition 38** (Decidability). *Given end-point types $T_1$ and $T_2$, the relation $T_1 \ll T_2$ is decidable.*

Next, we generalise the prefix-subtyping relation $\ll$ in the presence of recursive types, i.e., defining $\ll$ for sequences of prefixes generated by unfolding recursive types. First, we define the unfolding up to a fixed level of nesting.

**Definition 39** (Unfolding). *The $n$-time unfolding function of an end-point type, written $\mathit{unfold}^n(T)$, is defined inductively as:*

$$\begin{array}{ll}
\mathit{unfold}^0(T) = T & \mathit{unfold}^1([p,q]!\langle V\rangle.T) = [p,q]!\langle V\rangle.\mathit{unfold}^1(T) \\
\mathit{unfold}^1(\mu X.T) = T[\mu X.T/X] & \mathit{unfold}^1([p,q]?(V).T) = [p,q]?(V).\mathit{unfold}^1(T) \\
\mathit{unfold}^1(X) = X & \mathit{unfold}^1(T_1 + T_2) = \mathit{unfold}^1(T_1) + \mathit{unfold}^1(T_2) \\
\mathit{unfold}^1([b]T) = [b]\mathit{unfold}^1(T) & \mathit{unfold}^1(\Pi x : I.T) = \Pi x : I.\mathit{unfold}^1(T) \\
\mathit{unfold}^1(T\ e) = \mathit{unfold}^1(T)\ e & \mathit{unfold}^{n+1}(T) = \mathit{unfold}^1(\mathit{unfold}^n(T)) \\
\mathit{unfold}^1(\mathsf{end}) = \mathsf{end} &
\end{array}$$

The coinductive definition of subtyping $\leq$ is based on traditional subtyping of mobile processes [37] and session types [19].

**Definition 40** (Subtyping). *The coinductive subtyping relation of end-point types, written $T_1 \leq T_2$ and read "end-point type $T_1$ is subtype of $T_2$", is defined:*

- *If $T_1 = [p,q]!\langle V\rangle.T_1'$ then $\mathit{unfold}^n(T_2) \geq [p,q]!\langle V\rangle.T_2'$ and $T_1' \leq T_2'$.*

- *If* $T_1 = [p,q]?(V).T_1'$ *then unfold*$^n(T_2) \geq [p,q]?(V).T_2'$ *and* $T_1' \leq T_2'$.
- *If* $T_1 = [p,q]!\langle V_1\rangle.T_{11}+...+[p,q]!\langle V_k\rangle.T_{1k}$ *then unfold*$^n(T_2) \gg [p,q]!\langle V_1\rangle.T_{21}+...+[p,q]!\langle V_l\rangle.T_{2l}$, $k \geq l$, *and* $\forall i \in [1..l]$ *such that* $T_{1i} \leq T_{2i}$.
- *If* $T_1 = [p,q]?(V_1).T_{11}+...+[p,q]?(V_k).T_{1k}$ *then unfold*$^n(T_2) \gg [p,q]?(V_1).T_{21}+...+[p,q]?(V_l).T_{2l}$, $k \leq l$, *and* $\forall i \in [1..k]$ *such that* $T_{1i} \leq T_{2i}$.
- *If* $T_1 = [b]T_1'$ *then unfold*$^n(T_2) \gg [b]T_2'$ *and* $T_1' \leq T_2'$.
- *If* $T_1 = \Pi x : I.T_1'$ *then unfold*$^n(T_2) \gg \Pi x : I'.T_2'$ *and* $T_1' \leq T_2'$.
- *If* $T_1 = T_1'\ e$ *then unfold*$^n(T_2) \gg T_2'\ e$ *and* $T_1' \leq T_2'$.
- *If* $T_1 = \mu X.T$ *then unfold*$^1(T_1) \leq T_2$.
- *If* $T_1 = \mathsf{end}$ *then unfold*$^n(T_2) = \mathsf{end}$.

Informally, an output of $T_1$ is a subtype of an unfolded $T_2$ after applying $\geq$; an input of $T_1$ is a subtype of an unfolded $T_2$ after applying $\geq$; in a selection, the capabilities of $T_1$ must be greater than in $T_2$ while in a branching, it is vice-versa; the remaining cases ensure subtyping on the subterms. We do not consider subtyping on sorts $V$ to maintain an independent operational semantics, i.e. sending and receiving, from the type system.

**Double-buffering algorithm.** The projection of global type $G_D$ onto $K$ is:

$$G_D \upharpoonright^{\emptyset}_{K} = \mu X.[K,So]!\langle Signal\rangle.[So,K]?(Data).[So,K]?(Data).\\ [K,Si]!\langle Data\rangle.[K,Si]!\langle Data\rangle.X$$

The end-point type $\mu X.[K,So]!\langle Signal\rangle.[So,K]?(Data).\ [K,Si]!\ \langle Data\rangle.[So,K]?(Data).$ $[K,Si]!\langle Data\rangle.X$ is a subtype of $G_D\upharpoonright^{\emptyset}_{K} = \mu X.[K,So]!\langle Signal\rangle.[So,K]?(Data).\ [So,K]?\ (Data).$ $[K,Si]!\ \langle Data\rangle.\ [K,Si]!\langle Data\rangle.X$ according to subtyping rules ⌊OI⌋, ⌊In⌋, ⌊Out⌋ and ⌊Rec⌋. The kernel process will be type-checked with the latter type.

## H Remaining static typing and runtime typing

Below, we provide the the remaining static and runtime typing rules: processes after session initiation, subtyping to session types and the queue

$$\frac{\Gamma \vdash P \triangleright \Delta \qquad \Delta' \leq \Delta}{\Gamma \vdash P \triangleright \Delta'}\ \lfloor \text{TSub} \rfloor$$

Rule ⌊TSub⌋ applies subtyping to session types, where $\Delta_1 \leq \Delta_2$ if (1) $\Delta_1 = \Delta', u : T$, $\Delta_2 = \Delta'', u : T'$ and $T \leq T'$ and $\Delta' \leq \Delta''$ and (2) $\emptyset \leq \emptyset$.

$$\frac{\Gamma, a : G \vdash P \triangleright \Delta, a[\hat{p}_1] : T_1, ..., a[\hat{p}_n] : T_n}{\Gamma \vdash (new\ a : G)P \triangleright \Delta}\ \lfloor \text{TRes} \rfloor$$

Rule ⌊TRes⌋ assigns a type to a session after initiation based on the typing of the subprocess: subprocess $P$ must be well-typed by the union of all end-points in parallel $a[\hat{p}_i] : T_i$ with $i$ in $[1..n]$. Types of queues are defined as follows:

$$\begin{array}{llll} \text{Queue} & T & ::= & [\hat{p},\hat{q}]!\langle S\rangle \quad \textit{message send} \\ & & \mid & T.T' \quad \textit{message sequence} \end{array}$$

where $[\hat{p},\hat{q}]!\langle S\rangle$ expresses the communication of $\hat{p}$ to $\hat{q}$ of a value of type $S$, and $T.T'$ represents sequencing of queue types (we assume associativity for '.'). The empty queue has an empty mapping from session channels to queue types: $\Gamma \vdash a : \emptyset \triangleright \emptyset$ $\lfloor \text{TQEmpt} \rfloor$.

A queue adds an output type to the type of the sender in the mapping. Below we present the rule for addition of a value, including session channel

$$\frac{\Gamma \vdash a : h \triangleright \Delta \quad \Gamma \vdash v : U}{\Gamma \vdash a : h \cdot (\hat{p},\hat{q},v : U) \triangleright \Delta.\{a[\hat{p}] :![\hat{p},\hat{q}]\langle U\rangle\}} \lfloor \text{TQSend} \rfloor$$

$$\frac{\Gamma \vdash a{:}h \triangleright \Delta}{\Gamma \vdash a{:}h \cdot (\hat{p},\hat{q},b[\hat{p}']{:}T') \triangleright \Delta.\{a[\hat{p}]{:}![\hat{p},\hat{q}]\langle T'\rangle, b[\hat{p}']{:}T'\}} \lfloor \text{TQSSend} \rfloor$$

where '.' is defined as: $\Delta.\{a[\hat{p}] : T\} = \begin{cases} \Delta', a[\hat{p}] : T'.T & \text{if } \Delta = \Delta', a[\hat{p}] : T' \\ \Delta, a[\hat{p}] : T & \text{otherwise} \end{cases}$

The typing of processes in parallel with queues is defined as

$$\frac{\Gamma \vdash P \triangleright \Delta \quad \Gamma \vdash Q \triangleright \Delta'}{\Gamma \vdash P \mid Q \triangleright \Delta \star \Delta'} \lfloor \text{TQPar} \rfloor$$

where $\star$ is defined as: $T \star T' = \begin{cases} T.T' & \text{if T is a queue type,} \\ T'.T & \text{if T}' \text{ is a queue type,} \\ \text{undefined} & \text{otherwise} \end{cases}$

The parallel composition of two types is defined as $\Delta \star \Delta' = \Delta \backslash dom(\Delta') \cup \Delta' \backslash dom(\Delta) \cup \{a[\hat{p}] : T \star T' | a[\hat{p}] : T \in \Delta \text{ and } a[\hat{p}] : T' \in \Delta'\}$ and $\Delta \star \emptyset = \Delta$.

## I Proofs

**Theorem 41** (Decidability of projection). *The projection of a global type onto principals is decidable.*

*Proof.* By definition of projection on interaction, branching, matching, product and application constructs, the decidability of projection follows from the decidability of the relations: $\mathscr{C} \vdash p = q$, $\mathscr{C} \vdash I \neq \emptyset$ and $\mathscr{C} \vdash b = \mathsf{true}, \mathsf{false}$, which follow from the decidability of DML constraint solving. For the other constructs, the statement of decidability holds by definition. This property ensures also the decidability of well-assertedness. □

**Theorem 42.** *Projection of G onto principals and sorting of end-point types returned are decidable at the worst case in $O(p \times d \times n \log n)$ time complexity.*

*Proof.* The projection algorithm has complexity $O(p \times d \times n)$ with $p$ the number of principals, $d$ the maximal number of mathematic expressions in principals, and $n$ the length of the global type. The sorting algorithm (Mergesort) requires $O(p \times$

$d \times n \log n$) computational steps. Thus, the sorting algorithm dominates the cost of computation[4]. □

**Proposition 43** (Decidability). *Given end-point types $T_1$ and $T_2$, the relation $T_1 \ll T_2$ is decidable.*

*Proof.* Straightforward from the axioms and rules in Figure 7. □

**Theorem 44** (Decidability of typing). *The typing relation $\Gamma \vdash P \rhd \Delta$ is decidable.*

*Proof.* By induction over the derivation $\Gamma \vdash P \rhd \Delta$. We consider each axiom and congruence rule of the type system that generates the typing $\Gamma \vdash P \rhd \Delta$.

Case: ⌊TFun⌋

| | |
|---|---|
| $\Gamma \vdash fn\ x : I \Rightarrow P \rhd \Pi x{:}I.\Delta$ | By assumption |
| $\Gamma, x : I \vdash P \rhd \Delta$ | By inversion |
| $\Gamma, x : I \vdash P \rhd \Delta$ is decidable | By i.h. |
| $\Gamma \vdash fn\ x : I \Rightarrow P \rhd \Pi x{:}I.\Delta$ is decidable | By rule ⌊TFun⌋ |

Case: ⌊TApp⌋

| | |
|---|---|
| $\Gamma \vdash P\ e \rhd \tau$ | By assumption |
| $\Gamma \vdash P \rhd \Pi x{:}I.\Delta \quad \Gamma \vdash e \in I$ | By inversion |
| $\Gamma \vdash P \rhd \Pi x{:}I.\Delta$ is decidable | By i.h. |
| $\Gamma \vdash e \in I$ is decidable | By DML constraint solver [45] |
| $\Gamma \vdash P\ e \rhd \Delta$ is decidable | By rule ⌊TApp⌋ |

Case: [SInit]

| | |
|---|---|
| $\Gamma \vdash init\ (u : G, p).P \rhd \Delta$ | By assumption |
| $\Gamma, u : G \vdash u : G$ | |
| $\Gamma, u : G \vdash P \rhd \Delta, u : \mathsf{sort}(G \upharpoonright_p^{\Gamma})$ | By inversion |
| $\Gamma, u : G \vdash u : G$ | |
| $\Gamma, u : G \vdash P \rhd \Delta, u : \mathsf{sort}(G \upharpoonright_p^{\Gamma})$ are decidable | By i.h., Theorem 42 |
| $\Gamma \vdash init\ (u : G, p).P \rhd \Delta$ is decidable | By rule [TSInit] |

Cases ⌊TSend⌋, ⌊TRcv⌋, ⌊TSSend⌋, ⌊TSRcv⌋, ⌊TISend⌋, ⌊TIRcv⌋, ⌊TRec⌋, ⌊TVar⌋, ⌊TInact⌋, ⌊TNu⌋, ⌊TPar⌋, ⌊Sum⌋, ⌊TPSum⌋, ⌊TMatch⌋, ⌊TPMatch⌋, and ⌊≤⌋ are trivial by induction. □

**Notation 45.** *"By inversion" denotes inversion on a rule. That is, a conclusion judgment that is achieved by applying a certain rule is true if the premises on that rule are true.*

[4] For the sake of this analysis, we do not consider the polynomial complexity of the DML constraints solver.

**Notation 46.** *"By rule" denotes applying a rule. That is, given the premises and side conditions of a rule then we can conclude the judgment by applying that rule.*

**Notation 47.** *"By i.h." or "By induction" denotes induction hypothesis. That is, a subterm holds the property we are proving.*

**Lemma 48** (Type Preservation Under Substitution).

1. *If* $\Gamma, x : U \vdash P \triangleright \Delta$ *and* $\Gamma \vdash v : U$ *then* $\Gamma \vdash P[v/x] \triangleright \Delta$.
2. *If* $\Gamma, x : I \vdash P \triangleright \Delta$ *and* $\Gamma \vdash \mathrm{c} \in I$ *then* $\Gamma \vdash P\{\mathrm{c}/x\} \triangleright \Delta\{\mathrm{c}/x\}$.
3. *If* $\Gamma, X : \Delta \vdash P \triangleright \Delta$ *and* $\Gamma \vdash Q \triangleright \Delta$ *then* $\Gamma \vdash P\{Q/X\} \triangleright \Delta$.

*Proof.* (1) is similar to (2) and the proof of (2) is given below. (2) is by induction on the typing judgement $\Gamma, x : I \vdash P \triangleright \Delta$. We present the most appealing cases, including those most difficult.

| | |
|---|---|
| $\Gamma, i : I \vdash fn\ j : I' \Rightarrow P \triangleright \Pi j : I'.\Delta$ and $\Gamma \vdash \mathrm{c} \in I$ | By assumption |
| $\Gamma, i : I, j : I' \vdash P \triangleright \Delta$ and $\Gamma \vdash \mathrm{c} \in I'$ | By inversion |
| $\Gamma, j : I' \vdash P\{\mathrm{c}/i\} \triangleright \Delta\{\mathrm{c}/i\}$ | By i.h. |
| $\Gamma \vdash fn\ j : I' \Rightarrow P\{\mathrm{c}/i\} \vdash \Pi j : I'.\Delta\{\mathrm{c}/i\}$ | By rule [TFun] |

where $(\Pi j : I'.\Delta)\{\mathrm{c}/i\} = \Pi j : I'.(\Delta\{\mathrm{c}/i\})$.

| | |
|---|---|
| $\Gamma, i : I \vdash P\ e \triangleright \Delta$ and $\Gamma \vdash \mathrm{c} \in I$ | By assumption |
| $\Gamma, i : I \vdash P \triangleright \Pi j : I' : \Delta$, $\Gamma \vdash e \in I'$, $\Gamma \vdash \mathrm{c} \in I$ | By inversion |
| $\Gamma \vdash P\{\mathrm{c}/i\} \vdash \Pi j : I' : \Delta\{\mathrm{c}/i\}$, $C \vdash e \in I'$ | By i.h. |
| $\Gamma \vdash P\{\mathrm{c}/i\}\ e \vdash \Delta\{\mathrm{c}/i\}$ | By rule [TApp] |

The remaining rules are similar. The proof of (3) is standard. □

**Lemma 49** (Weakening). *(1) If* $\Gamma \vdash P \triangleright \Delta$ *and* $\Delta'$ *is* end *only then* $\Gamma \vdash P \triangleright \Delta \star \Delta'$.
*(2) If* $\Gamma \vdash P \triangleright \Delta$ *and* $\Delta'$ *is* end *only then* $\Gamma \vdash P \triangleright \Delta + \Delta'$.
*(3) If* $\Gamma \vdash P \triangleright \Delta$ *and* $a \notin fn(P)$ *then* $\Gamma, a : G \vdash P \triangleright \Delta$.

*Proof.* Standard.

□

**Theorem 50** (Subject congruence). $\Gamma \vdash P \triangleright \Delta$ *and* $P \equiv P'$ *imply* $\Gamma \vdash P' \triangleright \Delta$.

*Proof.* By rule induction on the derivation of $\Gamma \vdash P \triangleright \Delta$ when assuming that $P \equiv P'$ and $\Gamma \vdash P \triangleright \Delta$. For each structural congruence axiom, we consider each session type system rule that can generate $\Gamma \vdash P \triangleright \Delta$.

**Case**

$P \mid 0 \equiv P$

$\Gamma \vdash P \mid \mathbf{0} \triangleright \Delta$ — By assumption
$\Delta = \Delta_1 \star \Delta_2$ — By ⌊TQPar⌋
$\Gamma \vdash P \triangleright \Delta_1$ and $\Gamma \vdash \mathbf{0} \triangleright \Delta_2$ — By inversion
$\Delta_2$ is only `end` — By rule ⌊TInact⌋
$\Gamma \vdash P \triangleright \Delta_1 \star \Delta_2$ — By weakening 49 (1)

$\Gamma \vdash P \triangleright \Delta$ — By assumption
$\Gamma \vdash \mathbf{0} \triangleright \emptyset$ — By rule ⌊TInact⌋
$\Gamma \vdash P \mid \mathbf{0} \triangleright \Delta \star \emptyset'$ — By rule ⌊TQPar⌋
$\Delta \star \emptyset = \Delta$

**Case**

$P \mid Q \equiv Q \mid P$

$\Gamma \vdash P \mid Q \triangleright \Delta$ — By assumption
$\Delta = \Delta_1 \star \Delta_2$ — By rule ⌊TQPar⌋
$\Gamma \vdash P \triangleright \Delta_1$ and $\Gamma \vdash Q \triangleright \Delta_2$ — By inversion
$\Gamma \vdash Q \mid P \triangleright \Delta_2 \star \Delta_1$ — By rule ⌊TQPar⌋
$\Delta_1 \star \Delta_2 = \Delta_2 \star \Delta_1$

The other case is symmetric to the above one.

**Case**

$(P \mid Q) \mid S \equiv P \mid (Q \mid S)$

$\Gamma \vdash (P \mid Q) \mid S \triangleright \Delta$ — By assumption
$\Delta = (\Delta_1 \star \Delta_2) \star \Delta_3$ — By rule ⌊TQPar⌋
$\Gamma \vdash P \triangleright \Delta_1$, $\Gamma \vdash Q \triangleright \Delta_2$ and $\Gamma \vdash S \triangleright \Delta_3$ — By inversion
$\Gamma \vdash P \mid (Q \mid S) \triangleright \Delta_1 \star (\Delta_2 \star \Delta_3)$ — By rule ⌊TQPar⌋
$(\Delta_1 \star \Delta_2) \star \Delta_3 = \Delta_1 \star (\Delta_2 \star \Delta_3)$

The other case is symmetric to the above one.

**Case**

$P + 0 \equiv P$

$\Gamma \vdash P + \mathbf{0} \triangleright \Delta$ — By assumption
$\Delta = \Delta_1 + \Delta_2$ — By ⌊TSum⌋
$\Gamma \vdash P \triangleright \Delta_1$ and $\Gamma \vdash \mathbf{0} \triangleright \Delta_2$ — By inversion

$\Delta_2$ is only end — By rule $\lfloor$TInact$\rfloor$
$\Gamma \vdash P \triangleright \Delta_1 + \Delta_2$ — By weakening 49 (2)

$\Gamma \vdash P \triangleright \Delta$ — By assumption
$\Gamma \vdash \mathbf{0} \triangleright \emptyset$ — By rule $\lfloor$TInact$\rfloor$
$\Gamma \vdash P + \mathbf{0} \triangleright \Delta + \emptyset'$ — By rule $\lfloor$TSum$\rfloor$
$\Delta + \emptyset = \Delta$

**Case**

$$P + Q \equiv Q + P$$

$\Gamma \vdash P + Q \triangleright \Delta$ — By assumption
$\Delta = \Delta_1 + \Delta_2$ — By rule $\lfloor$TSum$\rfloor$
$\Gamma \vdash P \triangleright \Delta_1$ and $\Gamma \vdash Q \triangleright \Delta_2$ — By inversion
$\Gamma \vdash Q + P \triangleright \Delta_2 + \Delta_1$ — By rule $\lfloor$TSum$\rfloor$
$\Delta_1 + \Delta_2 = \Delta_2 + \Delta_1$

The other case is symmetric to the above one.

**Case**

$$(P + Q) + S \equiv P + (Q + S)$$

$\Gamma \vdash (P + Q) + S \triangleright \Delta$ — By assumption
$\Delta = (\Delta_1 + \Delta_2) + \Delta_3$ — By rule $\lfloor$TSum$\rfloor$
$\Gamma \vdash P \triangleright \Delta_1$, $\Gamma \vdash Q \triangleright \Delta_2$ and $\Gamma \vdash S \triangleright \Delta_3$ — By inversion
$\Gamma \vdash P + (Q + S) \triangleright \Delta_1 + (\Delta_2 + \Delta_3)$ — By rule $\lfloor$TSum$\rfloor$
$(\Delta_1 + \Delta_2) + \Delta_3 = \Delta_1 + (\Delta_2 + \Delta_3)$

The other case is symmetric to the above one.

**Case**

$$rec\ X = P \equiv P[rec\ X = P/X]$$

$\Gamma \vdash rec\ X = P \triangleright \Delta$ — By assumption
$\Gamma, X : \Delta \vdash P \triangleright \Delta$ — By inversion
$\Gamma \vdash P[rec\ X = P/X] \triangleright \Delta$ — By substitution lemma 48(3)

**Case**

$$P\ e \equiv P'\ e \text{ if } P \equiv P'$$

$\Gamma \vdash P\ e \triangleright \Delta$ By assumption
$\Delta = (\Pi x : I.\Delta)\ e$ By rule $\lfloor$TApp$\rfloor$
$\Gamma \vdash P \triangleright \Pi x : I.\Delta$ and $\Gamma \vdash e \in I$ By inversion
$\Gamma \vdash P' \triangleright \Pi x : I.\Delta$ By i.h.
$\Gamma \vdash P'\ e \triangleright (\Pi x : I.\Delta)\ e$ By rule $\lfloor$TApp$\rfloor$

The other case is symmetric to the above one.

**Case**

$$(new\ a : G)P \mid Q \equiv (new\ a : G)(P \mid Q) \quad \text{if } a \notin \mathsf{fn}(Q)$$

$\Gamma \vdash (new\ a : G)P \mid Q \triangleright \Delta$ and $a \notin \mathsf{fn}(Q)$ By assumption
$\Delta = \Delta_1 \star \Delta_2$ By rule $\lfloor$TQPar$\rfloor$
$\Gamma \vdash (new\ a : G)P \triangleright \Delta_1$ and $\Gamma \vdash Q \triangleright \Delta_2$ By inversion
$\Gamma, a : G \vdash P \triangleright \Delta_1', a : [\hat{p}_1] : T_1, ..., a[\hat{p}_n] : T_n$ By inversion
$\Gamma, a : G \vdash Q \triangleright \Delta_2$ By weakening 49(3)
$\Gamma, a : G \vdash P \mid Q \triangleright (\Delta_1', a : [\hat{p}_1] : T_1, ..., a[\hat{p}_n] : T_n) \star \Delta_2$ By rule $\lfloor$TQPar$\rfloor$
Since $a \notin dom(\Delta_2)$ then $(\Delta_1', a : [\hat{p}_1] : T_1, ..., a[\hat{p}_n] : T_n) \star \Delta_2 =$
$(\Delta_1' \star \Delta_2), a : [\hat{p}_1] : T_1, ..., a[\hat{p}_n] : T_n$
$\Gamma \vdash (new\ a : G)(P \mid Q) \triangleright \Delta_1 \star \Delta_2 = \Delta$ By rule $\lfloor$TRes$\rfloor$

The other case is symmetric to the above one. The other axioms is trivial to proof.

**Case**

$$a : (\hat{q}, \hat{p}, v{:}S) \cdot (\hat{q}', \hat{p}', v'{:}S') \cdot h \equiv a : (\hat{q}', \hat{p}', v'{:}S') \cdot (\hat{q}, \hat{p}, v{:}S) \cdot h$$

$$\text{if } \hat{p} \neq \hat{p}' \text{ or } \hat{q} \neq \hat{q}'$$

$\Gamma \vdash a : (\hat{q}, \hat{p}, v{:}S) \cdot (\hat{q}', \hat{p}', v'{:}S') \cdot h \triangleright \Delta$ By assumption
$\Delta = \{a[\hat{q}] : [\hat{q}, \hat{p}]!\langle S\rangle\}.\{a[\hat{q}'] : [\hat{q}', \hat{p}']!\langle S'\rangle\}.\Delta' =$ By rule $\lfloor$TQSend$\rfloor$
$\quad \{a[\hat{q}] : [\hat{q}, \hat{p}]!\langle S\rangle, a[\hat{q}'] : [\hat{q}', \hat{p}']!\langle S'\rangle\}.\Delta' =$
$\quad \{a[\hat{q}'] : [\hat{q}', \hat{p}']!\langle S'\rangle, a[\hat{q}] : [\hat{q}, \hat{p}]!\langle S\rangle\}.\Delta' =$
$\quad \{a[\hat{q}'] : [\hat{q}', \hat{p}']!\langle S'\rangle\}.\{a[\hat{q}] : [\hat{q}, \hat{p}]!\langle S\rangle\}.\Delta'$
$\Gamma \vdash a : (\hat{q}', \hat{p}', v'{:}S') \cdot (\hat{q}, \hat{p}, v{:}S) \cdot h \triangleright$
$\quad \{a[\hat{q}'] : [\hat{q}', \hat{p}']!\langle S'\rangle\}.\{a[\hat{q}] : [\hat{q}, \hat{p}]!\langle S\rangle\}.\Delta'$ By rule $\lfloor$TQSend$\rfloor$

The other case is symmetric to the above one.

□

**Theorem 51** (Subject reduction). *If* $\Gamma \vdash P \triangleright \Delta$, *and* $P \rightarrow P'$, *then* $\Gamma \vdash P' \triangleright \Delta'$ *where* $\Delta = \Delta'$ *or* $\Delta \Rightarrow \Delta'$.

*Proof.* By induction over the derivation of $P \to P'$.

**Case**

$$init(a{:}G,\hat{p}_1).P_1 \mid ...init(a{:}G,\hat{p}_n).P_n \longrightarrow (new\ a : G)(P_1 \mid ... \mid P_n \mid a : \emptyset)$$
$$\{\hat{p}_1,...,\hat{p}_n\} = \mathsf{pid}(G)$$

$\Gamma \vdash init(a{:}G,\hat{p}_1).P_1 \mid ...init(a{:}G,\hat{p}_n).P_n \rhd \Delta$ By assumption
$\Gamma \vdash init(a{:}G,\hat{p}_1).P_1 \rhd \Delta_1,\ ...,\ \Gamma \vdash init(a{:}G,\hat{p}_n).P_n \rhd \Delta_n,$
where $\Delta = \Delta_1,...,\Delta_n$ By inversion
$\Gamma, a : G \vdash P_1 \rhd \Delta_1, a[\hat{p}_1] : \mathsf{sort}(G \upharpoonright_{\hat{p}_1}^{\Gamma(i:I)})$ By inversion
...
$\Gamma, a : G \vdash P_n \rhd \Delta_n, a[\hat{p}_n] : \mathsf{sort}(G \upharpoonright_{\hat{p}_n}^{\Gamma(i:I)})$ By inversion
$\Gamma, a : G \vdash P_1 \mid ... \mid P_n \rhd \Delta_1,...,\Delta_n, a[\hat{p}_1] : \mathsf{sort}(G \upharpoonright_{\hat{p}_1}^{\Gamma(i:I)}),...,$
$a[\hat{p}_n] : \mathsf{order}(G \upharpoonright_{\hat{p}_n}^{\Gamma(i:I)})$ By rule ⌊TPar⌋
$\Gamma, a : G \vdash a : \emptyset \rhd \emptyset$ By rule ⌊TQEmpt⌋
$\Gamma, a : G \vdash P_1 \mid ... \mid P_n \mid a : \emptyset \rhd (\Delta_1,...,\Delta_n, a[\hat{p}_1] : \mathsf{sort}(G \upharpoonright_{\hat{p}_1}^{\Gamma(i:I)}),$
$..., a[\hat{p}_n] : \mathsf{sort}(G \upharpoonright_{\hat{p}_n}^{\Gamma(i:I)})).\emptyset$ By rule ⌊TQPar⌋
$(\Delta_1,...,\Delta_n, a[\hat{p}_1] : \mathsf{sort}(G \upharpoonright_{\hat{p}_1}^{\Gamma(i:I)}),..., a[\hat{p}_n] : \mathsf{sort}(G \upharpoonright_{\hat{p}_n}^{\Gamma(i:I)})).\emptyset =$
$\Delta_1,...,\Delta_n, a[\hat{p}_1] : \mathsf{sort}(G \upharpoonright_{\hat{p}_1}^{\Gamma(i:I)}),..., a[\hat{p}_n] : \mathsf{sort}(G \upharpoonright_{\hat{p}_n}^{\Gamma(i:I)})$
$\Gamma \vdash (new\ a : G)(P_1 \mid ... \mid P_n \mid a : \emptyset) \rhd \Delta$ By rule ⌊TRes⌋

**Case**

$$a[\hat{p},\hat{q}]!\langle v : U\rangle.P{+}P' \mid a : h \longrightarrow P \mid a : h \cdot (\hat{p},\hat{q},v : U)$$

$\Gamma \vdash a[\hat{p},\hat{q}]!\langle v : U\rangle.P{+}P' \mid a : h \rhd \Delta$ By assumption
$\Delta = \Delta_1 \star \Delta_2$ By rule ⌊TQPar⌋
$\Gamma \vdash a[\hat{p},\hat{q}]!\langle v : U\rangle.P{+}P' \rhd \Delta_1 \quad \Gamma \vdash a : h \rhd \Delta_2$ By inversion
$\Delta_1 = (\Delta_1', a[\hat{p}] : [\hat{p},\hat{q}]!\langle U\rangle.T){+}\Delta_1''$ By rules ⌊Sum⌋, ⌊TSend⌋
$(\Delta_1', a[\hat{p}] : [\hat{p},\hat{q}]!\langle U\rangle.T){+}\Delta_1'' = \Delta_1', a[\hat{p}] : [\hat{p},\hat{q}]!\langle U\rangle.T{+}[\hat{p},\hat{q}]!\langle U_1\rangle.T_1...{+}[\hat{p},\hat{q}]!\langle U_n\rangle.T_n$
Where $\Delta = (\Delta_1', a[\hat{p}] : [\hat{p},\hat{q}]!\langle U\rangle.T{+}a[\hat{p}] : [\hat{p},\hat{q}]!\langle U_1\rangle.T_1...{+}\ [\hat{p},\hat{q}]!\langle U_n\rangle.T_n) \star \Delta_2$
$\Gamma \vdash P \rhd \Delta_1', a[\hat{p}] : T \quad \Gamma \vdash v : U$ By inversion
$\Gamma \vdash a : h \cdot (\hat{p},\hat{q},v : U) \rhd \Delta_2.\{a[\hat{p}] : [\hat{p},\hat{q}]!\langle U\rangle\}$ By rule ⌊TQSend⌋
$\Gamma \vdash P \mid a : h \cdot (\hat{p},\hat{q},v : U) \rhd (\Delta_1', a[\hat{p}] : T) \star (\Delta_2.\{a[\hat{p}] : [\hat{p},\hat{q}]!\langle U\rangle\}$ By rule ⌊TQPar⌋
$(\Delta_1', a[\hat{p}] : T) \star (\Delta_2.\{a[\hat{p}] : [\hat{p},\hat{q}]!\langle U\rangle\}) =$
$(\Delta_1', a[\hat{p}] : [\hat{p},\hat{q}]!\langle U\rangle.T) \star \Delta_2 \geq$
$(\Delta_1', a[\hat{p}] : [\hat{p},\hat{q}]!\langle U\rangle.T{+}[\hat{p},\hat{q}]!\langle U_1\rangle.T_1...{+}[\hat{p},\hat{q}]!\langle U_n\rangle.T_n) \star \Delta_2 = \Delta$
$\Gamma \vdash P \mid a : h \cdot (\hat{p},\hat{q},v : U) \rhd \Delta$ By rule ⌊TSubs⌋

**Case**

$a[\hat{p},\hat{q}]?(x:U).P{+}P' \mid a:(\hat{q},\hat{p},v:U)\cdot h \longrightarrow P\{v/x\} \mid a:h$

$\Gamma \vdash a[\hat{p},\hat{q}]?(x:U).P{+}P' \mid a:(\hat{q},\hat{p},v:U)\cdot h \rhd \Delta$ — By assumption

$\Delta = \Delta_1 \star \Delta_2$ — By rule $\lfloor$TQPar$\rfloor$

$\Gamma \vdash a[\hat{p},\hat{q}]?(x:U).P{+}P' \rhd \Delta_1$ and $\Gamma \vdash a:(\hat{q},\hat{p},v:U)\cdot h \rhd \Delta_2$ — By inversion

$\Delta_1 = \Delta_1', a[\hat{q}]:[\hat{p},\hat{q}]?(U).T{+}...{+}[\hat{p},\hat{q}]?(U_n).T_n$ — By rules $\lfloor$TSum$\rfloor$, $\lfloor$TRecv$\rfloor$

Where $\Delta = (\Delta_1', a[\hat{q}]:[\hat{p},\hat{q}]?(U).T{+}...{+}[\hat{p},\hat{q}]?(U_n).T_n) \star \Delta_2$

$\Gamma, x:U \vdash P \rhd \Delta_1', a[\hat{q}]:T$ — By inversion

$\Delta_2 = \{a[\hat{p}]:[\hat{p},\hat{q}]!\langle U\rangle\}.\Delta_2'$ — By rule [TQSend]

$\Gamma \vdash h \rhd \Delta_2' \quad \Gamma \vdash v:U$ — By inversion

Where $\Delta = (\Delta_1', a[\hat{q}]:[\hat{p},\hat{q}]?(U).T{+}...{+}[\hat{p},\hat{q}]?(U_n).T_n) \star \{a[\hat{p}]:[\hat{p},\hat{q}]!\langle U\rangle\}.\Delta_2'$

$\Gamma \vdash P\{v/x\} \rhd \Delta_1', s[\hat{q}]:T$ — By substitution lemma 48(1)

$\Gamma \vdash P\{v/x\} \mid a:h \rhd (\Delta_1', a[\hat{q}]:T) \star \Delta_2'$ — By rule $\lfloor$TQPar$\rfloor$

$\Delta = (\Delta_1', a[\hat{q}]:[\hat{p},\hat{q}]?(U).T{+}...{+}[\hat{p},\hat{q}]?(U_n).T_n) \star\{a[\hat{p}]:[\hat{p},\hat{q}]!\langle U\rangle\}.\Delta_2' =$
$(\Delta_1', \{a[\hat{q}]:[\hat{p},\hat{q}]?(U).T{+}...{+}[\hat{p},\hat{q}]?(U_n).T_n, a[\hat{p}]:[\hat{p},\hat{q}]!\langle U\rangle\} \star \Delta_2' \Rightarrow$
$(\Delta_1', a[\hat{q}]:T, a[\hat{p}]:\emptyset) \star \Delta_2'$ — By rule $\lfloor$TR-Ex, TR-Context$\rfloor$
$= (\Delta_1', a[\hat{q}]:T) \star \Delta_2'$

**Case**

$a[\hat{p},\hat{q}]!\langle b[\hat{p}']:T'\rangle.P \mid a:h \longrightarrow P \mid a:h\cdot(\hat{p},\hat{q},b[\hat{p}']:T')$

$\Gamma \vdash a[\hat{p},\hat{q}]!\langle b[\hat{p}']:T'\rangle.P \mid a:h \rhd \Delta$ — By assumption

$\Delta = \Delta_1 \star \Delta_2$ — By rule $\lfloor$TQPar$\rfloor$

$\Gamma \vdash a[\hat{p},\hat{q}]!\langle b[\hat{p}']:T'\rangle.P \rhd \Delta_1 \quad \Gamma \vdash a:h \rhd \Delta_2$ — By inversion

$\Delta_1 = \Delta_1', a[\hat{p}]:[\hat{p},\hat{q}]!\langle T'\rangle.T, b[\hat{p}']:T'$ — By rules $\lfloor$TSSend$\rfloor$

Where $\Delta = (\Delta_1', a[\hat{p}]:[\hat{p},\hat{q}]!\langle T'\rangle.T, b[\hat{p}']:T') \star \Delta_2$

$\Gamma \vdash P \rhd \Delta_1', a[\hat{p}]:T$ — By inversion

$\Gamma \vdash a:h\cdot(\hat{p},\hat{q},b[\hat{p}']:T') \rhd \Delta_2.\{a[\hat{p}]:[\hat{p},\hat{q}]!\langle T'\rangle, b[\hat{p}']:T'\}$ — By rule $\lfloor$TQSSend$\rfloor$

$\Gamma \vdash P \mid a:h\cdot(\hat{p},\hat{q},b[\hat{p}']:T') \rhd$
$(\Delta_1', a[\hat{p}]:T) \star (\Delta_2.\{a[\hat{p}]:[\hat{p},\hat{q}]!\langle T'\rangle, b[\hat{p}']:T'\}$ — By rule $\lfloor$TQPar$\rfloor$

$(\Delta_1', a[\hat{p}]:T) \star (\Delta_2.\{a[\hat{p}]:[\hat{p},\hat{q}]!\langle T'\rangle, b[\hat{p}']:T'\} = (\Delta_1', a[\hat{p}]:[\hat{p},\hat{q}]!\langle T'\rangle.T) \star \Delta_2.\{b[\hat{p}']:T'\}$
Since $\Delta_1 = \Delta_1', a[\hat{p}]:[\hat{p},\hat{q}]!\langle T'\rangle.T, b[\hat{p}']:T'$
then $b[\hat{p}'] \notin dom(\Delta_1')$
Hence $(\Delta_1', a[\hat{p}]:[\hat{p},\hat{q}]!\langle T'\rangle.T) \star \Delta_2.\{b[\hat{p}']:T'\} = (\Delta_1', a[\hat{p}]:[\hat{p},\hat{q}]!\langle T'\rangle.T, b[\hat{p}']:T') \star \Delta_2$

**Case**

$$a[\hat{p},\hat{q}]?(b[\hat{p}']:T').P \mid a:(\hat{q},\hat{p},b[\hat{p}']:T')\cdot h \longrightarrow P \mid a:h$$

$\Gamma \vdash a[\hat{p},\hat{q}]?(b[\hat{p}']:T').P \mid a:(\hat{q},\hat{p},b[\hat{p}']:T')\cdot h \rhd \Delta$ By assumption

$\Delta = \Delta_1 \star \Delta_2$ By rule $\lfloor$TQPar$\rfloor$

$\Gamma \vdash a[\hat{p},\hat{q}]?(b[\hat{p}']:T').P \rhd \Delta_1 \quad \Gamma \vdash a:(\hat{q},\hat{p},b[\hat{p}']:T')\cdot h \rhd \Delta_2$ By inversion

$\Delta_1 = \Delta_1', a[\hat{q}]:[\hat{p},\hat{q}]?(T').T$ By rules $\lfloor$TSRecv$\rfloor$

Where $\Delta = (\Delta_1', a[\hat{q}]:[\hat{p},\hat{q}]?(T').T) \star \Delta_2$

$\Gamma \vdash P \rhd \Delta_1', a[\hat{q}]:T, b[\hat{p}']:T'$ By inversion

$\Delta_2 = \{a[\hat{p}]:[\hat{p},\hat{q}]!\langle T'\rangle, b[\hat{p}']:T'\}.\Delta_2'$ By rule [TQSSend]

$\Gamma \vdash h \rhd \Delta_2'$ By inversion

Where $\Delta = (\Delta_1', a[\hat{q}]:[\hat{p},\hat{q}]?(T').T) \star\{a[\hat{p}]:[\hat{p},\hat{q}]!\langle T'\rangle, b[\hat{p}']:T'\}.\Delta_2'$

$\Gamma \vdash P \mid a:h \rhd (\Delta_1', a[\hat{q}]:T, b[\hat{p}']:T') \star \Delta_2'$ By rule $\lfloor$TQPar$\rfloor$

$\Delta = (\Delta_1', a[\hat{q}]:[\hat{p},\hat{q}]?(T').T) \star\{a[\hat{p}]:[\hat{p},\hat{q}]!\langle T'\rangle, b[\hat{p}']:T'\}.\Delta_2' =$
$(\Delta_1', a[\hat{q}]:[\hat{p},\hat{q}]?(T').T, a[\hat{p}]:[\hat{p},\hat{q}]!\langle T'\rangle, b[\hat{p}']:T') \star \Delta_2' \Rightarrow$
$(\Delta_1', a[\hat{q}]:T, a[\hat{p}]:\emptyset, b[\hat{p}']:T') \star \Delta_2'$ By rule $\lfloor$TR-SEx, TR-Context$\rfloor$
$= (\Delta_1', a[\hat{q}]:T, b[\hat{p}']:T') \star \Delta_2'$

**Case**

$$[\mathsf{true}]P{+}P' \longrightarrow P$$

$\Gamma \vdash [\mathsf{true}]P{+}P' \rhd \Delta$ By assumption

$\Delta = \Delta_1{+}\Delta_2$ By rule $\lfloor$TSum$\rfloor$

$\Gamma \vdash [\mathsf{true}]P \rhd \Delta_1$ and $\Gamma \vdash P' \rhd \Delta_2$ By inversion

$\Delta_1 = [\mathsf{true}]\Delta_1'$ By rule $\lfloor$TSum$\rfloor$

Where $\Delta = [\mathsf{true}]\Delta_1'{+}\Delta_2$

$\Gamma \vdash P \rhd \Delta_1'$ By inversion

$\Delta = \{a_1[\hat{p}_1]:[\mathsf{true}]T_1, ...., a_n[\hat{p}_n]:[\mathsf{true}]T_n\} + \{a_1[\hat{p}_1]:T_1', ...., a[n][\hat{p}_n]:T_n'\} =$
$\{a_1[\hat{p}_1]:[\mathsf{true}]T_1 + T_1', ...., a_n[\hat{p}_n]:[\mathsf{true}]T_n + T_n'\} \Rightarrow$
$\{a_1[\hat{p}_1]:T_1, ...., a_n[\hat{p}_n]:T_n\}$ By rules $\lfloor$TR-MatchT$\rfloor$, $\lfloor$TR-Context$\rfloor$

**Case**

$$[\mathsf{false}]P{+}P' \longrightarrow P$$

$\Gamma \vdash [\mathsf{false}]P{+}P' \rhd \Delta$ By assumption

$\Delta = \Delta_1{+}\Delta_2$ By rule ⌊TSum⌋

$\Gamma \vdash [\mathsf{true}]P \rhd \Delta_1$ and $\Gamma \vdash P' \rhd \Delta_2$ By inversion

$\Delta_1 = [\mathsf{false}]\Delta_1'$ By rule ⌊TSum⌋

Where $\Delta = [\mathsf{false}]\Delta_1'{+}\Delta_2$

$\Gamma \vdash P \rhd \Delta_1'$ By inversion

$\Delta = \{a_1[\hat{p}_1] : [\mathsf{false}]T_1, ...., a_n[\hat{p}_n] : [\mathsf{false}]T_n\} + \{a_1[\hat{p}_1] : T_1', ...., a[n][\hat{p}_n] : T_n'\} =$
$\{a_1[\hat{p}_1] : [\mathsf{false}]T_1 + T_1', ...., a_n[\hat{p}_n] : [\mathsf{false}]T_n + T_n'\} \Rightarrow$
$\{a_1[\hat{p}_1] : T_1', ...., a_n[\hat{p}_n] : T_n'\}$ By rules ⌊TR-MatchF⌋, ⌊TR-Context⌋

**Case**

$$P \longrightarrow P' \quad \Rightarrow \quad (\mathit{new}\ a : G)P \longrightarrow (\mathit{new}\ a : G)P'$$

$\Gamma \vdash (\mathit{new}\ a : G)P \rhd \Delta$ By assumption

$\Gamma, a : G \vdash P \rhd \Delta, a[\hat{p}_1] : T_1, ..., a[\hat{p}_n] : T_n$ By inversion

$\Gamma, a : G \vdash P' \rhd \Delta', a[\hat{p}_1] : T_1', ..., a[\hat{p}_n] : T_n'$

where $\Delta, a[\hat{p}_1] : T_1, ..., a[\hat{p}_n] : T_n = \Delta', a[\hat{p}_1] : T_1', ..., a[\hat{p}_n] : T_n'$

or $\Delta, a[\hat{p}_1] : T_1, ..., a[\hat{p}_n] : T_n \Rightarrow \Delta', a[\hat{p}_1] : T_1', ..., a[\hat{p}_n] : T_n'$ By induction

$\Gamma \vdash (\mathit{new}\ a : G)P' \rhd \Delta'$ By rule ⌊TRes⌋

where $\Delta = \Delta'$ or $\Delta \Rightarrow \Delta'$

**Case**

$$P \longrightarrow P' \quad \Rightarrow \quad P \mid Q \longrightarrow P' \mid Q$$

$\Gamma \vdash P \mid Q \rhd \Delta$ By assumption

$\Delta = \Delta_1 \star \Delta_2$ By rule ⌊TQPar⌋

$\Gamma \vdash P \rhd \Delta_1$ and $\Gamma \vdash Q \rhd \Delta_2$ By inversion

$\Gamma \vdash P' \rhd \Delta_1'$ where $\Delta_1 = \Delta_1'$ or $\Delta_1 \Rightarrow \Delta_1'$ By induction

$\Gamma \vdash P' \mid Q \rhd \Delta_1' \star \Delta_2$ By rule ⌊TQPar⌋

$\Delta_1 \star \Delta_2 = \Delta_1 \setminus dom(\Delta_2) \cup \Delta_2 \setminus dom(\Delta_1) \cup \{a[\hat{p}] : T_1 \star T_2 | a[\hat{p}] : T_1 \in \Delta_1$ and $a[\hat{p}] : T_2 \in \Delta_2\}$

**Firstsubcase** $\Delta_1 \Rightarrow \Delta_1'$ for some $a[\hat{p}'] \notin dom(\Delta_2)$.

Then $\Delta_1 \star \Delta_2 \Rightarrow \Delta_1' \star \Delta_2$ By rule ⌊TR-Context⌋

**Secondsubcase** $\Delta_1 \Rightarrow \Delta_1'$ for some $a[\hat{p}'] \in dom(\Delta_2)$.

Then, $\Delta_2$ is not a queue message type. So we have $\Delta_1'.\Delta_2$ by definition of $\star$. Hence $a[\hat{p}] : T_1 \Rightarrow a[\hat{p}] : T_1'$ implies that $a[\hat{p}] : T_1.T_2 \Rightarrow a[\hat{p}] : T_1'.T_2$.

Then $\Delta_1 \star \Delta_2 \Rightarrow \Delta_1' \star \Delta_2$ By rule ⌊TR-Context⌋

**Case**

$(fn\ x : I \Rightarrow P)\ c \longrightarrow P\{c/x\}$

| | |
|---|---|
| $\Gamma \vdash (fn\ x : I \Rightarrow P)\ c \triangleright \Delta$ | By assumption |
| $\Delta = (\Pi x : I.\Delta')\ c$ | By rule ⌊TApp⌋ |
| $\Gamma \vdash fn\ x : I \Rightarrow P \triangleright \Pi x : I.\Delta'$ and $\Gamma \vdash c \in I$ | By inversion |
| $\Gamma, x : I \vdash P \triangleright \Delta'$ and $\Gamma \vdash c \in I$ | By inversion |
| $\Gamma \vdash P\{c/x\} \triangleright \Delta'\{c/x\}$ | By substitution lemma 48(2) |
| $(\Pi x : I.\Delta')\ c \Rightarrow \Delta'\{c/x\}$ | By rule ⌊TR-App, TR-Context⌋ |

**Case**

$P \equiv P'$ and $P' \longrightarrow Q'$ and $Q \equiv Q' \quad \Rightarrow \quad P \longrightarrow Q$

| | |
|---|---|
| $\Gamma \vdash P \triangleright \Delta$, $P \equiv P'$, $P' \to Q'$ and $Q' \equiv Q$ | By assumption |
| $\Gamma \vdash P' \triangleright \Delta$, $P \equiv P'$, $P' \to Q'$ and $Q' \equiv Q$ | By subject cong. theo. 50 |
| $\Gamma \vdash Q' \triangleright \Delta'$, where $\Delta \leq \Delta''$ and $\Delta'' = \Delta'$ or $\Delta'' \Rightarrow \Delta'$ | By induction |
| $\Gamma \vdash Q \triangleright \Delta'$ | By subject cong. theo. 50 |

□

**Theorem 52** (Subject reduction for closed processes). *If $\emptyset \vdash P \triangleright \emptyset$ and $P \to P'$ then $\emptyset \vdash P' \triangleright \emptyset$.*

*Proof.* Follows directly from Subject reduction theorem 51. □

**Lemma 53** (Value inversion) *(a) If $a : G \vdash P \triangleright a[\hat{p}_1] : T_1, ..., a[\hat{p}_n] : T_n$ then $P \equiv P' \mid a : h$.*

*(b) If $\Gamma \vdash P \triangleright \Pi x : I.\Delta$ then $P \equiv fn\ x : I \Rightarrow P'$.*

*(c) If $\Gamma \vdash P \triangleright u[q] : [\hat{p}, \hat{q}]?(U_1).T_1 + ... + [\hat{p}, \hat{q}]?(U_n).T_n$ then $P \equiv u[\hat{p}, \hat{q}]?(x : U_1).P'_1 + ... + u[\hat{p}, \hat{q}]?(x : U_m).P'_m$ and $n \leq m$.*

*(d) If $a : G \vdash P \mid a : h \triangleright [\hat{p}, \hat{q}]!\langle U_1\rangle.T_1 + ... + [\hat{p}, \hat{q}]!\langle U_n\rangle.T_n$ then $h \equiv h' \cdot (\hat{p}, \hat{q}, v_i : U_i)\ i \leq n$ or $P \equiv u[\hat{p}, \hat{q}]!\langle v_1 : U_1\rangle.P'_1 + ... + u[\hat{p}, \hat{q}]!\langle v_m : U_m\rangle.P'_m$ and $n \geq m$.*

*(e) If $\Gamma \vdash P \triangleright u[q] : [\hat{p}, \hat{q}]?(T').T$ then $P \equiv a[\hat{p}, \hat{q}]?(b[\hat{p}'] : T').P'$.*

*(f) If $a : G \vdash P \mid a : h \triangleright [\hat{p}, \hat{q}]!\langle T'\rangle.T$ then $h \equiv h' \cdot (\hat{p}, \hat{q}, b[\hat{p}'] : T')$ or $P \equiv a[\hat{p}, \hat{q}]!\langle b[\hat{p}'] : T'\rangle.P'$.*

*Proof.* Straightforward from the typing rules.

□

**Theorem 54** (Communication-safety). *Suppose that $a : G \vdash Q \mid a : h \triangleright a[\hat{p}_1] : T_1, ..., a[\hat{p}_n] : T_n$. Then $Q \mid a : h$ is communication-safe.*

*Proof.* By Theorem 51, we have that $a : G \vdash Q' \mid a : h' \rhd a[\hat{p}_1] : T'_1, ..., a[\hat{p}_n] : T'_n$ where $a[\hat{p}_1] : T_1, ..., a[\hat{p}_n] : T_n \Rightarrow a[\hat{p}_1] : T'_1, ..., a[\hat{p}_n] : T'_n$.

**Case:**

$\{a[\hat{p}]:[\hat{p},\hat{q}]!\langle U_1\rangle.T_1+...+[\hat{p},\hat{q}]!\langle U_n\rangle.T_n,$
$a[\hat{q}]:[\hat{p},\hat{q}]?(U_1).T'_1+...+[\hat{p},\hat{q}]?(U_m).T'_m\} \Rightarrow \{a[\hat{p}]:T_i, a[\hat{q}]:T'_i\}$

$a : G \vdash Q \mid a : h \rhd \{a[\hat{p}]:[\hat{p},\hat{q}]!\langle U_1\rangle.T_1+...+[\hat{p},\hat{q}]!\langle U_n\rangle.T_n,$
$a[\hat{q}]:[\hat{p},\hat{q}]?(U_1).T'_1+...+[\hat{p},\hat{q}]?(U_m).T'_m\}, \Delta$ — By assumption

**First subcase:** $\{a[\hat{q}]:[\hat{p},\hat{q}]?(U_1).T'_1+...+[\hat{p},\hat{q}]?(U_m).T'_m\}$
$Q \equiv \mathscr{F}[a[\hat{q},\hat{p}]?(x : U_1).Q'_1+...+a[\hat{q},\hat{p}]?(x : U_l).Q'_l],\ l \geq m$ — By Lemma 53(c)
**First subsubcase:**
$Q \mid a : h \equiv \mathscr{F}[a[\hat{q},\hat{p}]?(x : U_1).Q'_1+...+a[\hat{q},\hat{p}]?(x : U_l).Q'_l]$
$\mid a : (\hat{p},\hat{q},v_i : U_i)\cdot h' \quad i \leq n$ — By Lemma 53(d)
$Q \mid a : h \rightarrow Q' \mid a : h'$ — By rule ⌊Rcv⌋
**Second subsubcase:**
$Q \mid a : h \equiv \mathscr{F}[a[\hat{q},\hat{p}]?(x : U_1).Q'_1+...+a[\hat{q},\hat{p}]?(x : U_l).Q'_l \mid$
$a[\hat{q},\hat{p}]!\langle v_1 : U_1\rangle.P'_1+...+a[\hat{q},\hat{p}]!\langle v_k : U_k\rangle.P'_k] \mid a : h$
$k \leq n$ — By Lemma 53(d)
$Q \mid a : h \rightarrow Q' \mid a : h'$ — By rule ⌊Send⌋
**Second subcase:** $\{a[\hat{p}]:[\hat{p},\hat{q}]!\langle U_1\rangle.T_1+...+[\hat{p},\hat{q}]!\langle U_n\rangle.T_n\}$
Similarly to above

**Case:** $\{a[\hat{p}]:[\hat{p},\hat{q}]!\langle T'\rangle.T, a[\hat{q}]:[\hat{p},\hat{q}]?(T').T''\} \Rightarrow \{a[\hat{p}]:T, a[\hat{q}]:T''\}$

**First subcase:** $\{a[\hat{q}]:[\hat{p},\hat{q}]?(T').T''\}$
$Q \equiv \mathscr{F}[a[\hat{q},\hat{p}]?(b[\hat{q}'] : T').Q']$ — By Lemma 53(e)
**First subsubcase:**
$Q \mid a : h \equiv \mathscr{F}[a[\hat{q},\hat{p}]?(b[\hat{q}'] : T').Q'] \mid a : (\hat{p},\hat{q},b[\hat{q}'] : T')\cdot h'$
$Q \mid a : h \rightarrow Q' \mid a : h'$ — By rule ⌊SRecv⌋
**Second subsubcase:**
$Q \mid a : h \equiv \mathscr{F}[a[\hat{q},\hat{p}]?(b[\hat{q}'] : T').Q' \mid a[\hat{q},\hat{p}]!(b[\hat{q}'] : T').P'] \mid a : h$ — By Lemma 53(f)
$Q \mid a : h \rightarrow Q' \mid a : h'$ — By rule ⌊SSend⌋
**Second subcase:** $\{a[\hat{p}]:[\hat{p},\hat{q}]!\langle T'\rangle.T\}$
Similar to above.

The other cases are vacuous. ☐

**Theorem 55** (Progress). *If $\emptyset \vdash P \rhd \emptyset$ and $P$ is simpler then either $P \equiv 0$ or $P \rightarrow^\star P'$.*

*Proof.* By induction on the derivation of the typing judgment, analysing all possible cases.

Case ⌊TSInit, TSend, TRcv, TSSend, TSRcv, TISend,⌋ ⌊TIRcv, TRec, TVar⌋ are vacuous. Case ⌊TInact⌋ is given.

**Case:**

$$\frac{\emptyset \vdash P \triangleright \emptyset \quad \emptyset \vdash Q \triangleright \emptyset}{\emptyset \vdash P \mid Q \triangleright \emptyset} \lfloor \text{TPar} \rfloor$$

| | |
|---|---|
| $\emptyset \vdash P \mid Q \triangleright \emptyset$ | By assumption |
| $P \mid Q = \mathit{init}(a{:}G, \hat{p}_1).P_1 \mid \ldots \mid \mathit{init}(a{:}G, \hat{p}_n).P_n$ | First subcase |
| $\{\hat{p}_1, \ldots, \hat{p}_n\} = \mathsf{pid}(G)$ | By assumption |
| $\mathit{init}(a{:}G, \hat{p}_1).P_1 \mid \ldots \mid \mathit{init}(a{:}G, \hat{p}_n).P_n \longrightarrow (\mathit{new}\ a : G)(P_1 \mid \ldots \mid P_n \mid a{:}\emptyset)$ | By rule ⌊Link⌋ |
| $\emptyset \vdash P \triangleright \emptyset \quad \emptyset \vdash Q \triangleright \emptyset$ | By inversion |
| $P \to P'$ | First subsubcase |
| $P \longrightarrow P' \quad \Rightarrow \quad P \mid Q \longrightarrow P' \mid Q$ | By rule ⌊Par⌋ |
| $P = 0$ | Second subsubcase |
| $Q \to Q'$ | Third subcase |
| $0 \mid Q \to 0 \mid Q'$ | By rule ⌊Str⌋ |
| $Q = 0$ | Forth subcase |
| $0 \mid 0 \equiv 0$ | By structural congruence rules |

**Case:**

$$\frac{\emptyset \vdash P_1 \triangleright \emptyset \quad \emptyset \vdash P_2 \triangleright \emptyset}{\emptyset \vdash P_1 + P_2 \triangleright \emptyset} \lfloor \text{TSum} \rfloor$$

| | |
|---|---|
| $\emptyset \vdash P_1 + P_2 \triangleright \emptyset$ | By assumption |
| $\emptyset \vdash P_1 \triangleright \emptyset \quad \emptyset \vdash P_2 \triangleright \emptyset$ | By inversion |
| $P_1 \to P_1'$ | First subcase |
| $P_1 = [b]P_1'$ | By assumption |
| $b = \mathsf{true}$ | First subsubcase |
| $[\mathsf{true}]P_1' + P_2 \to P_1'$ | By rule |
| $b = \mathsf{false}$ | Second subsubcase |
| $[\mathsf{false}]P_1' + P_2 \to P_2$ | By rule |
| $P_1 = 0$ | Second subcase |
| $P_1 + P_2 \equiv P_2$ | By structural congruence |
| $P_2 \to P_2'$ | First subcase |
| $0 + P_2 \to 0 + P_2'$ | By rule ⌊Str⌋ |
| $P_2 = 0$ | Second subcase |
| $P_1 + P_2 \equiv 0$ | |

**Case:**

$$\frac{\emptyset \vdash b \triangleright \mathsf{bool} \quad \emptyset \vdash P \triangleright \emptyset}{\emptyset \vdash [b]P \triangleright \emptyset} \lfloor \text{TMatch} \rfloor$$

| | |
|---|---|
| $\emptyset \vdash [b]P \rhd \emptyset$ | By assumption |
| $\emptyset \vdash b \rhd \mathsf{bool} \quad \emptyset \vdash P \rhd \emptyset$ | By inversion |
| $b = \mathsf{true}$ | First subcase |
| $[\mathsf{true}]P \equiv [\mathsf{true}]P{+}0$ | By congruence rules |
| $[\mathsf{true}]P{+}0 \rightarrow P$ | By rule ⌊MatchT⌋ |
| $[\mathsf{true}]P \rightarrow P$ | By rule ⌊Str⌋ |
| $b = \mathsf{false}$ | Second subcase |
| $[\mathsf{false}]P \equiv [\mathsf{false}]P{+}0$ | By congruence rules |
| $[\mathsf{false}]P{+}0 \rightarrow 0$ | By rule ⌊MatchF⌋ |
| $[\mathsf{false}]P \rightarrow [\textsc{Inact}]$ | By rule ⌊Str⌋ |

Case ⌊TFun⌋ is vacuous.
**Case:**

$$\frac{\emptyset \vdash P \rhd \Pi x : I.\emptyset \qquad \emptyset \vdash e \in I}{\emptyset \vdash P\ e \rhd \emptyset}\ \lfloor\textsc{TApp}\rfloor$$

| | |
|---|---|
| $\emptyset \vdash P\ e \rhd \emptyset$ | By assumption |
| $\emptyset \vdash P \rhd \Pi x : I.\emptyset \quad \emptyset \vdash e \in I$ | By inversion |
| $P \equiv fn\ x : I \Rightarrow P'$ | By Lemma 53(b) |
| $e \downarrow \mathsf{c}$ | |
| $(fn\ x{:}I \Rightarrow P')\ \mathsf{c} \longrightarrow P'\{\mathsf{c}/x\}$ | By rule ⌊App⌋ |

Case ⌊TSub⌋ is given.
**Case:**

$$\frac{a : G \vdash P \rhd a[\hat{p}_1] : T_1, \ldots, a[\hat{p}_n] : T_n}{\emptyset \vdash (new\ a : G)P \rhd \emptyset}\ \lfloor\textsc{TRes}\rfloor$$

| | |
|---|---|
| $\emptyset \vdash (new\ a : G)P \rhd \emptyset$ | By assumption |
| $a : G \vdash P \rhd a[\hat{p}_1] : T_1, \ldots, a[\hat{p}_n] : T_n$ | By inversion |
| $P \equiv P' \mid a : h$ | By Lemma 53(a) |
| $P \rightarrow P''$ or $P = 0$ | By Corollary 23 and Theorem 25 |
| $P \longrightarrow P'' \quad \Rightarrow \quad (new\ a : G)P \longrightarrow (new\ a : G)P''$ | By rule ⌊Scop⌋ |

Cases ⌊TQSend, TQSSend, TQPar⌋ are vacuous. □

**Proposition 56** (Properties of the Web service example). *Client* | *Proxy* | *W. Service constitutes a simple process. Since* $\emptyset \vdash$ *Client* | *Proxy* | *W. Service* $\rhd \emptyset$ *then there is no interaction mismatches, no type errors, and Client* | *Proxy* | *W. Service* $\rightarrow^* P'$ *for some* $P'$.

Similarly for all other examples.

## About the authors

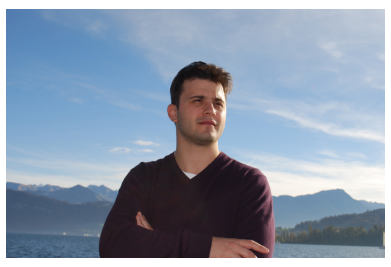

**Andi Bejleri** is Big Data architect at IBM. His research consists of addressing practical software-engineering challenges through design and implementation of type theories and programming concepts. He has contributed to session-types theory, a library for session-programming, a new agent-oriented model for the Cloud, static verification of dynamic, object-oriented models and predicates, object-oriented models. He studied computer science from 2001 to 2005 at the University of Pisa. In 2007, he obtained a Master in Advanced Computing from Imperial College London. Where at the same university, he was awarded a doctorate in 2012. He worked for one year at the HP Labs in Bristol and currently is a postdoctoral researcher at TU Darmstadt. Contact email bejleriandi@gmail.com.

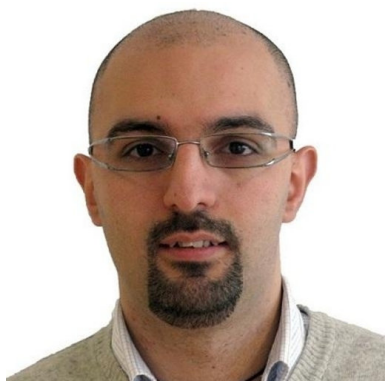

**Elton Domnori** is lecturer at the Canadian Institute of Technology, Tirana (Albania). On 2012 he was awarded with the PhD degree in Computer Science and Engineering from the University of Modena and Reggio Emilia. Since October 2015 he is member of the ICT COST Action IC1302. His research interests is focused on distributed systems and in particular on fully distributed coordination algorithms. His contribution are mainly towards intelligent and self coordination among peer nodes that aim to achieve a common goal. Make use of multi-agent systems platforms and paradigms to develop new algorithms and simulate scenarios. Contact email elton.domnori@cit.edu.al.

**Malte Viering** Contact email viering@dsp.tu-darmstadt.de

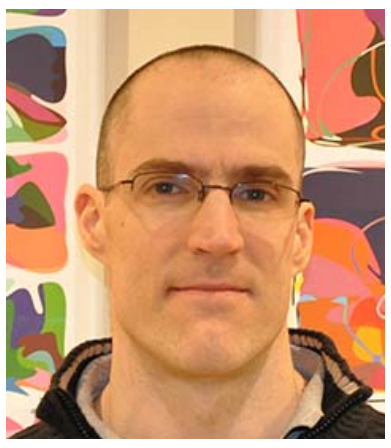

**Patrick Eugster** is an Associate Professor in Computer Science at Purdue University. He joined the faculty at Purdue University CS in Spring 2006, after having held associate researcher positions with both Swiss Federal Institutes of Technology in Zurich (ETHZ) and Lausanne (EPFL), and having worked for Sun Microsystems (now Oracle). Patrick holds M.S. and Ph.D. degrees from EPFL. Patrick is interested in programming support for distributed systems. Specific topics of interest include programming abstractions, program analysis, middleware, and distributed algorithms. Patrick has been recognized for his teaching and mentoring by a Price of Excellence for an Exceptional Teaching Contribution from EPFL (1998) and by an Undergraduate Student Advising Award from the College of Science of Purdue University (2010). His research has been awarded by a Price of Excellence for an Exceptional Research Contribution from EPFL (2001) and recognized by a Professional Achievement Award from the College of Science of Purdue University (2012), and has been supported by an Advanced Researcher Fellowship from the Swiss National Science Foundation (2001), a Postdoctoral Researcher Fellowship from the Swedish Research Council (2002), a CAREER Award from the US National Science Foundation (2007), and an Experienced Researcher Fellowship from the Alexander von Humboldt Foundation (2011). Patrick is also a member of the DARPA 2011 Computer Science Study Group. His research has been funded by NSF, DARPA, Northrop Grumman, Google, and Cisco Systems. Contact email patrick.thomas.eugster@usi.ch

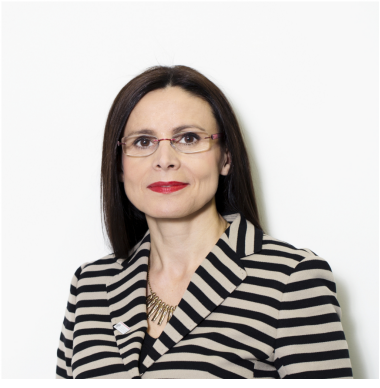

**Mira Mezini** is full professor at TU Darmstadt. She studied computer science from 1984 to 1989 at the University of Tirana. From 1992 until her doctorate in 1997, she was a research associate at the University of Siegen. From 1999, she taught at the Northeastern University of Boston in the USA for three years. From 2012, she was the dean of the Computer Science department and since January 1, 2014, she has been vice president of TU Darmstadt. Professor Mezini research focuses on software development paradigms and tools. On the paradigms side, she develops programming languages to enable the visions of “software as a service (SaaS)” and “software product-lines” by providing large-scale module concepts with built-in support for adaptability and extensibility. On the tools side, she works on intelligent software-development environments that guide developers to increase the development productivity and the software quality. In 2012, Mezini received the prestigious Advanced Grant from the ERC in the amount of EUR 2.3 million. Contact email mezini@st.informatik.tu-darmstadt.de.